\documentclass[12pt, a4paper]{article}
\usepackage{amsmath,amssymb,amsfonts}
\usepackage{graphicx}
\usepackage[colorlinks]{hyperref}

\renewcommand{\theequation}{\arabic{section}.\arabic{equation}}

\begin{document}

\title{Induced vacuum magnetic flux in quantum spinor matter in the background of a topological defect in
two-dimensional space}

\author{ Yurii A. Sitenko${}^{1}$ and Volodymyr M. Gorkavenko${}^{2}$\\
\it \small ${}^{1}$Bogolyubov Institute for Theoretical Physics,
 \it \small National Academy of Sciences of Ukraine,\\
 \it \small 14-b Metrologichna str., Kyiv 03143,
 Ukraine\\\phantom{bhh}\\
 \it \small ${}^{2}$Department of Physics, Taras Shevchenko National
 University of Kyiv,\\ \it \small 64 Volodymyrs'ka str., Kyiv
 01601, Ukraine}
 \date{}

\maketitle

\begin{abstract}
A topological defect in the form of the Abrikosov-Nielsen-Olesen vortex is considered as a gauge-flux-carrying tube that is impenetrable for quantum matter. The relativistic spinor matter field is quantized in the vortex background in $2+1$-dimensional conical spacetime which is a section orthogonal to the vortex axis; the most general set of boundary conditions ensuring the impenetrability of the vortex core is employed. We find the induced vacuum current circulating around the vortex and the induced vacuum magnetic field strength pointing along the vortex axis. The requirement of finiteness and physical plausibility for the total induced vacuum magnetic flux allows us to restrict the variety of admissible boundary conditions. The dependence of the results on the transverse size of the vortex, as well as on the vortex flux and the parameter of conicity, is elucidated. We discuss a significant distinction between the cases of massive and massless quantum spinor matter.


\end{abstract}

\section{Introduction}
\setcounter{equation}{0}

Spontaneous breakdown of continuous symmetries can give rise to
topological defects with rather interesting properties. A
topological defect in three-dimensional space, which is
characterized by the nontrivial first homotopy group, is known as
the Abrikosov-Nielsen-Olesen (ANO) vortex \cite{Abr,NO}. The vortex
is described classically in terms of a spin-zero (Higgs) field that
condenses and a spin-one field corresponding to the spontaneously
broken gauge group; the former field is coupled to the latter one in
the minimal way with constant ${\tilde e}_{\rm cond}$. Single valuedness
of the condensate field and finiteness of the vortex energy
implement that the vortex flux is related to ${\tilde e}_{\rm cond}$,
\begin{equation}\label{a1.1}
\Phi=\oint d \textbf{x} \textbf{V}(\textbf{x})=2\pi /{\tilde
e}_{\rm cond},
\end{equation}
where $\textbf{V}(\textbf{x})$ is the vector potential of the spin-one gauge field,
and the integral is over a path enclosing the vortex tube once (natural units
$\hbar=c=1$ are used). As some amount of energy (mass) is stored in the core of a
topological defect, this core is a source of gravity. Such a source in the case of
the linear ANO vortex makes the spatial region outside the vortex core to be conical,
i.e., with the deficit angle equal to $8 \pi \c{G} M$: the squared length element in the
outer region is
\begin{equation}\label{1.2}
    ds^2= dr^2+\nu^{-2} r^2 d\varphi^2+dz^2,
\end{equation}
where
\begin{equation}\label{1.3}
\nu=(1-4\c{G} M)^{-1},
\end{equation}
$\c{G}$ is the gravitational constant, and $M$ is the linear density of
mass stored in the core. The transverse size of the vortex core is
determined by the correlation length, and quantity $M$ is of order
of the inverse correlation length squared. Since constant $\c{G}$ is of the
order of the Planck length squared, the effects of conicity, which
are characterized by the value of the deficit angle, are negligible
for vortices in ordinary superconductors. However topological defects of
the type of ANO vortices may arise in a field that is seemingly
rather different from condensed matter physics -- in cosmology and high-energy physics.
This was realized by Kibble \cite{Ki1,Ki2} and Vilenkin
\cite{Vil1,Vil2} (see also \cite{Zel}), and, from the beginning of
the 1980s, such topological defects are known under the
name of cosmic strings. Cosmic strings with the thickness of the
order of the Planck length are definitely ruled out by astrophysical
observations, but there remains a room for cosmic strings with the
thickness that is more than 3 orders larger than the Planck
length (see, e.g., \cite{Bat}), although the direct evidence for
their existence is lacking.

A recent development in material science also provides an unexpected link
between condensed matter and high-energy physics, which is caused to a large
extent by the experimental discovery of graphene -- a two-dimensional crystalline
allotrope formed by a monolayer of carbon atoms \cite{Nov}. Low-energy electronic
excitations in graphene are characterized by dispersion law which is the same as
that for Dirac fermions in relativistic field theory, with the only distinction
that the velocity of light is changed to the Fermi velocity, see \cite{Ge,Kats}. It
is well established by now that a sheet of graphene is always corrugated and
covered by ripples which can be either intrinsic or induced by roughness of a
substrate. A single topological defect (disclination) warps a sheet of graphene,
rolling it into a nanocone which is similar to the transverse section of a spatial region out of a cosmic
string; carbon nanocones with deficit angles equal to $N_d \pi/3$ ($N_d = 1, 2, 3, 4, 5$,
i.e. $\nu = \frac 65, \frac 32, 2, 3, 6$) were observed experimentally, see \cite{Hei,Nae}
and references therein. Moreover, theory also predicts saddle-like nanocones with the
deficit angle taking negative values unbounded from below, $N_d =-1,-2,-3,...,-\infty$,
i.e. $\nu = \frac 67, \frac 34, \frac 23,..., 0$, which can be regarded as corresponding
to cosmic strings with negative mass density. Note that nanoconical structures may arise
as well in a diverse set of condensed matter systems known as the two-dimensional Dirac
materials, ranging from honeycomb crystalline allotropes (silicene and germanene \cite{Cah},
phosphorene \cite{Liu}) to high-temperature cuprate superconductors \cite{Tsu} and topological
insulators \cite{Qi}.

While considering the effect of the ANO vortex on the vacuum of
quantum matter, the following two circumstances should be kept in
mind.  First, the phase with broken symmetry exists outside the
vortex core and the vacuum is to be defined there.  Hence, the
quantum  matter field does not penetrate inside  the core, obeying a
boundary condition at its side edge.  Second, the impact of the
ANO vortex on quantum matter is through a vector potential of the
vortex-forming spin-one field, and the quantum matter field is assumed
to couple to this vector potential in the minimal way with coupling
constant $\tilde e$. Hence, the ANO vortex flux has no effect on the surrounding matter in the
framework of classical theory, and such an effect, if it exists, is of a
purely quantum nature. This phenomenon should be understood as a
quantum-field-theoretical manifestation of the famous Aharonov-Bohm
effect \cite{Aha} and is characterized by the periodic dependence on
the value of the vortex flux, $\Phi$ \eqref{a1.1}, with the period
equal to London flux quantum $2\pi /{\tilde e}$.

A crucial task in the study of the effect of the ANO vortex on the vacuum of quantum matter is to elucidate
the dependence on a boundary condition at the edge of the vortex core. It seems reasonable to start from
the most general set of mathematically admissible boundary conditions and then, after obtaining the outcoming effect, to restrict this set by
physically motivated arguments. Another task is to elucidate the dependence on the transverse size of the
vortex core. These two tasks will be thoroughly scrutinized and solved in the course of the present study by
considering a somewhat simplified case of two-dimensional space
\footnote{Quantum-field-theoretical models in $2+1$-dimensional spacetime play a role of toy models in particle physics and may be relevant to real systems in condensed matter physics. They exhibit a number of interesting features, such as fermion numder fractionization, parity violation, and flavor symmetry breaking; for a review see \cite{Dun}.}
being the transverse section of a three-dimensional spatial region out of the ANO vortex.

It should be noted that the current, the condensate, and the energy-momentum tensor that are induced in the vacuum of quantum relativistic spinor matter were considered in the above-described context in \cite{Bez1,Bel,Bez2}. However, a particular boundary condition was employed, and the issue of a dependence of the results on the choice of admissible boundary conditions remained undisclosed.

The current and the magnetic field strength, as well as the energy density and the Casimir force, that are induced in the vacuum of quantum relativistic scalar matter at $\nu=1$ in a space of arbitrary dimension were considered in the above-described context in \cite{Gor1,Gor2,Gor3,Gor4}. In these studies the Dirichlet boundary condition was employed; a physical motivation herein is in the assumption of a perfect reflection of quantum matter from the edge of the vortex core.

In the case of quantum relativistic spinor matter, neither the Dirichlet nor the Neumann boundary condition is admissible. A physically motivated demand is the absence of the matter flux across the boundary. In $2+1$-dimensional spacetime with a  connected boundary, this demand yields a one-parameter family of boundary conditions; see Section 4 below. Employing such boundary conditions, we shall find the induced vacuum current and the induced vacuum magnetic field strength; further physical arguments will be shown to remove an ambiguity in the choice of boundary conditions.

In the next section we define the current and the magnetic field that are induced in the vacuum of quantum relativistic spinor matter in the background of the ANO vortex of nonzero transverse size. In Section 3 we present the complete set of solutions to the Dirac equation that is relevant to the problem considered. In Section 4 we choose boundary conditions ensuring the absence of the matter flux across the edge of the vortex core. The induced vacuum current is obtained in Section 5. In Section 6 we consider the induced vacuum magnetic field and its total flux with the use of both analytical and numerical methods. Finally, the results are summarized and discussed in Section 7. Some details in the derivation of the expression for the induced vacuum current are given in Appendices A and B. The case of the infinitely thin vortex is reviewed in Appendix C. The results for massless quantum spinor matter are presented in Appendix D.

\section{Preliminaries and definition of physical characteristics of the vacuum}
\setcounter{equation}{0}

The operator of the second-quantized spinor field is presented as
\begin{equation}\label{1.1}
 \Psi({\textbf{x}},t)=\sum\hspace{-1.6em}\int\limits_{E>0} {\rm e}^{-{\rm
i}E t}\psi_{E}({\bf x})\,a_{E}+
 \sum\hspace{-1.6em}\int\limits_{E<0}
  {\rm e}^{-{\rm i}Et}\psi_{E}({\bf
  x})\,b^\dag_{E},
\end{equation}
where $a^\dag_E $ and $a_E$ ($b^\dag_E $ and $b_E$) are the spinor
particle (antiparticle) creation and destruction operators obeying
the anticommutation relations, $\psi_{E}({\bf x})$ is the solution
to the stationary Dirac equation,
\begin{equation}\label{1.5}
 H \psi_{E}({\bf x})=E \psi_{E}({\bf x}),
\end{equation}
and symbol \mbox{$\displaystyle \sum\hspace{-1.4em}\int $} denotes
summation over the discrete part and integration (with a certain
measure) over the continuous part of the energy spectrum; ground state $|{\rm vac} \rangle$ is conventionally defined by relation
\begin{equation}\label{1.5a}
a_E|{\rm vac}\rangle = b_E|{\rm vac}\rangle = 0.
\end{equation}
In the case of the ANO vortex background, the Dirac hamiltonian takes form
\begin{equation}\label{1.6}
H=-{\rm i} \mbox{\boldmath $\alpha$}\cdot \left(\mbox{\boldmath
$\partial$} - {\rm i}\tilde e\, \textbf{V}+\frac{{\rm i}}2
\mbox{\boldmath $\omega$}\right)+\beta m,
\end{equation}
where $\mbox{\boldmath $\omega$}$ is the spin connection
corresponding to conical space \eqref{1.2}. The current that is
induced in the vacuum is given by expression
\begin{equation}\label{1.7}
\textbf{j}(\textbf{x})=\langle {\rm vac}| \Psi^\dag(\textbf{x},t)
\mbox{\boldmath $\alpha$}   \Psi(\textbf{x},t) |{\rm vac}
\rangle=-\frac12\sum\hspace{-1.4em}\int \rm{sgn}(E)
\psi^\dag_E(\textbf{x}) \mbox{\boldmath $\alpha$}   \psi_E(\textbf{x})
\end{equation}
($\rm{sgn}(u)$ is the sign function, $\rm{sgn}(u)=\pm 1$ at $u
\gtrless 0$). The magnetic field strength, $\textbf{B}_{\rm
I}(\textbf{x})$, is also induced in the vacuum, as a consequence of
the Maxwell equation,
\begin{equation}\label{1.8}
\mbox{\boldmath $\partial$}\times \textbf{B}_{\rm I}(\textbf{x}) =
e\, \textbf{j}(\textbf{x}),
\end{equation}
where the electromagnetic coupling constant, $e$, differs in general
from $\tilde e$. The total flux of the induced vacuum magnetic field
is
\begin{equation}\label{1.9}
\Phi_I=\int d \mbox{\boldmath $\sigma$} \cdot \textbf{B}_{\rm
I}(\textbf{x}).
\end{equation}

Since the vacuum of quantum matter exists outside the ANO vortex
core, as was already  emphasized, an issue of the choice of boundary
conditions at the edge of the core is of primary  concern.
Turning to this issue, let us note first, that \eqref{1.6} is not
enough to define the hamiltonian operator rigorously in a
mathematical sense. To define an operator in an unambiguous way, one
has to specify its domain of definition. Let the set of functions $\psi$
be the domain of definition of operator $H$ and the set of functions
$\tilde \psi$ be the domain of definition of its adjoint, operator
$H^\dag$. Then the operator is Hermitian (or symmetric in
mathematical parlance),
\begin{equation}\label{1.10}
\int\limits_X d^3x \sqrt{g} \,{\tilde\psi}^\dag (H\psi)=
\int\limits_X d^3x \sqrt{g}\,(H^\dag \tilde \psi)^\dag \psi,
\end{equation}
if relation
\begin{equation}\label{1.11}
-{\rm i} \int\limits_{\partial X} d \mbox{\boldmath $\sigma$}
\cdot{\tilde \psi}^\dag \mbox{\boldmath $\alpha$} \psi =0
\end{equation}
is valid; here functions $\psi(\textbf{x})$ and $\tilde
\psi(\textbf{x})$ are defined in space $X$ with boundary $\partial
X$. It is evident that condition \eqref{1.11} can be satisfied by
imposing different boundary conditions for $\psi$  and $\tilde
\psi$.  But, a nontrivial task is to find a possibility that a
boundary condition for $\tilde \psi$ is the same as that for $\psi$;
then the domain of definition of $H^\dag$ coincides with that of
$H$, and operator $H$ is self-adjoint (for a review of the Weyl-von
Neumann theory of self-adjoint operators see \cite{Neu,Ree}). The action of a self-adjoint operator on functions belonging to its domain of definition results in functions of the same kind only, and a multiple action and functions of
such an operator, for instance, the resolvent and evolution operators,
can consistently be defined. Thus, the requirement of the
self-adjointness of operator $H$ \eqref{1.6} renders the most
general boundary condition at the edge of the vortex core for the
solution to the Dirac equation, $\psi_E(\textbf{x})$.

Note also that relation \eqref{1.11}, when applied to the solution to the Dirac equation, yields
\begin{equation}\label{1.11a}
-{\rm i} \int\limits_{\partial X} d \mbox{\boldmath $\sigma$}
\cdot\psi^\dag_E \mbox{\boldmath $\alpha$} \psi_E = 0,
\end{equation}
or
\begin{equation}\label{1.11b}
\int\limits_{\partial X} d \mbox{\boldmath $\sigma$}\cdot
\textbf{j} = 0
\end{equation}
with $\textbf{j}(\textbf{x})$ given by \eqref{1.7}.
If boundary $\partial X$ is connected, then \eqref{1.11b} is reduced to
\begin{equation}\label{1.11c}
\boldsymbol{n}\cdot\textbf{j}|_{\textbf{x}
\in
\partial X} = 0,
\end{equation}
where $\boldsymbol{n}$ is the unit normal that may be chosen as pointing outward to $X$. The last relation signifies the impenetrability of $\partial X$; i.e., the matter field is confined to $X$.

In the present paper we consider the vacuum polarization effects in
$2+1$-dimensional spacetime, which is a section orthogonal to the
ANO vortex axis, i.e., at $z={\rm }$. The irreducible representation
of the Clifford algebra is chosen in such a way that the Dirac
matrices in flat $2+1$-dimensional spacetime take form
\begin{equation}\label{1.12}
\alpha^{1}_{(0)}=-\sigma^2,\quad
\quad\alpha^{2}_{(0)}=\sigma^1,\quad\beta=\sigma^3,
\end{equation}
where $\sigma^1$, $\sigma^2$, and $\sigma^3$ are the Pauli matrices
(a transition to another inequivalent irreducible representation can
be made by changing the sign of $\beta$).
In the background of the ANO vortex, the only one component of the
vector potential and the spin connection is nonvanishing:
\begin{equation}\label{1.13}
    V_\varphi=\frac{\Phi}{2\pi}, \quad w_\varphi={\rm
    i}\frac{\nu-1}r\, \alpha_\varphi \alpha_r,
\end{equation}
and the Dirac hamiltonian takes form
\begin{equation}\label{1.14}
H=-{\rm i}
\left[\alpha^r\left(\partial_r+\frac{1-\nu}{2r}\right)+\alpha^\varphi\left(\partial_\varphi-{\rm
i}\frac{\tilde e \Phi}{2\pi}\right)\right]+\beta m,
\end{equation}
where
\begin{equation}\label{1.15}
\alpha^r=\alpha_r=\left(\begin{array}{cc}
                    0& {\rm i}  e^{-{\rm i}\varphi}\\
                   - {\rm i}  e^{{\rm i}\varphi} &0
                     \end{array}\right),\quad
\alpha^\varphi=\frac{\nu}r\left(\begin{array}{cc}
                    0&  e^{-{\rm i}\varphi}\\
                     e^{{\rm i}\varphi} &0
                     \end{array}\right),\quad
\alpha_\varphi=\frac{r^2}{\nu^2}\,\alpha^\varphi.
\end{equation}
Decomposing function $\psi_E(\textbf{x})$ as
\begin{equation}\label{1.16}
\psi_E(\textbf{x}) = \sum_{n \in \mathbb{Z}}
                   \left(\begin{array}{c}
                   f_n(r,E )e^{ {\rm i} n\varphi} \\
                   g_n(r,E )e^{ {\rm i} (n+1)\varphi}
                    \end{array}\right)
\end{equation}
($\mathbb{Z}$   is the set of integer numbers), we present the Dirac
equation as a system of two first-order differential equations for
radial functions:
\begin{equation}\label{1.17}
 \left\{
 \begin{array}{c}
\left\{-\partial_r + r^{-1} [\nu (n-n_{\rm c})-G]\right\} f_n(r,E) =(E+m) g_n(r,E) \\
\left\{\partial_r + r^{-1} [\nu (n-n_{\rm c})+1-G]\right\} g_n(r,E)
=(E-m) f_n(r,E)
\end{array}
\right\},
\end{equation}
where
\begin{equation}\label{1.18}
n_{\rm c}=\left[\!\left| \frac{\tilde e \Phi}{2\pi}
\right|\!\right],\quad F = \left\{\!\!\left| \frac{\tilde
e \Phi}{2\pi}  \right|\!\!\right\},\quad G =\nu \left(F  - \frac12 \right)+ \frac12,
\end{equation}
$\left[\!\left| u \right|\!\right]$ is the integer part of quantity $u$ (i.e., the integer that
is less than or equal to $u$), and $ \left\{\!\!\left|  u \right|\!\!\right\} = u - \left[\!\left| u \right|\!\right]$ is the
fractional part of quantity $u$, $ 0\leq  \left\{\!\!\left| u \right|\!\!\right\}<1 $.

 Using \eqref{1.15} and \eqref{1.16}, one gets $j_r =0$, and the
 only component of the induced vacuum current,
\begin{equation}\label{1.19}
j_\varphi(r) = - \frac r\nu \sum\hspace{-1.4em}\int \sum_{n \in
\mathbb{Z}} {\rm sgn} (E) f_n(r,E) g_n(r,E),
\end{equation}
is independent of the angular variable. The induced vacuum magnetic
field strength is directed along the vortex axis,
\begin{equation}\label{1.20}
B_{\rm I}(r) = e \nu \int\limits_r^\infty \frac{dr'}{r'} \,
j_\varphi(r'),
\end{equation}
with total flux
\begin{equation}\label{1.21}
\Phi_{\rm I} = \frac{2\pi}\nu \int\limits_{r_0}^\infty dr\, r B_{\rm
I}(r),
\end{equation}
where it is assumed without a loss of generality that the vortex
core has the form of a tube of radius $r_0$.

\section{Solution to the Dirac equation}\setcounter{equation}{0}

The solution to the system of equations, \eqref{1.17}, is given in
terms of cylindrical functions. Let us define
\begin{multline}\label{2.1}
\left(
\begin{array}{c}
f_n^{(\wedge)} \\ g_n^{(\wedge )}
\end{array}
\right)= \frac12 \sqrt{\frac{\nu}{\pi} } \\
\times \left(
\begin{array}{c}
\sqrt{1+m/E} \left[\sin(\mu^{(\wedge)}_{\nu l +1 -G}) J_{\nu l-G}(kr) + \cos(\mu^{(\wedge)}_{\nu l +1 -G}) Y_{\nu l-G}(kr)\right] \\
{\rm sgn}(E) \sqrt{1- m/E} \left[\sin(\mu^{(\wedge)}_{\nu l +1 -G})
J_{\nu l+1-G}(kr) + \cos(\mu^{(\wedge)}_{\nu l +1 -G}) Y_{\nu
l+1-G}(kr)\right]
\end{array}
\right),
\end{multline}
where $l=n-n_{\rm c}$, and
\begin{multline}\label{2.2}
\left(
\begin{array}{c}
f_n^{(\vee)} \\ g_n^{(\vee )}
\end{array}
\right)=  \frac12 \sqrt{\frac{\nu}{\pi} } \\
\times \left(
\begin{array}{c}
\sqrt{1+m/E} \left[\sin(\mu^{(\vee)}_{\nu l' +G}) J_{\nu l'+G}(kr) + \cos(\mu^{(\vee)}_{\nu l' +G}) Y_{\nu l'+G}(kr)\right] \\
-{\rm sgn}(E) \sqrt{1- m/E} \left[\sin(\mu^{(\vee)}_{\nu l' +G})
J_{\nu l'-1+G}(kr) + \cos(\mu^{(\vee)}_{\nu l' +G}) Y_{\nu
l'-1+G}(kr)\right]
\end{array}
\right),
\end{multline}
where $l'=-n+n_{\rm c}$; here $J_\rho(u)$ and  $Y_\rho(u)$ are the Bessel
and Neumann functions of order $\rho$, $k = \sqrt{E^2 - m^2}$.

In the case of $\nu > 1$ and $0 < F <\frac12 \left( 1-\frac1\nu \right)$
$\quad$ $\left( \frac12 ( 1-\nu ) < G <0 \right)$, the complete set
of solutions to \eqref{1.17} is given by
\begin{equation}\label{2.3}
\left.
\left( \begin{array}{c} f_n \\ g_n
\end{array}
\right)\right|_{n \geq n_{\rm c}} = \left(
\begin{array}{c}
f_n^{(\wedge)} \\ g_n^{(\wedge )}
\end{array}
\right), \quad
\left. \left( \begin{array}{c} f_n \\ g_n
\end{array}
\right)\right|_{n \leq n_{\rm c}-1} = \left(
\begin{array}{c}
f_n^{(\vee)} \\ g_n^{(\vee )}
\end{array}
\right).
\end{equation}
In the case of $\nu >1 $ and $\frac12 (1+\frac1\nu) < F <1$
$\quad$ $\left( 1<G < \frac12(1+\nu)\right)$, the complete set of
solutions to \eqref{1.17} is given by
\begin{equation}\label{2.4}
\left. \left( \begin{array}{c} f_n \\ g_n
\end{array}
\right)\right|_{n \geq  n_{\rm c}+1} = \left(
\begin{array}{c}
f_n^{(\wedge)} \\ g_n^{(\wedge )}
\end{array}
\right), \quad \left. \left( \begin{array}{c} f_n \\ g_n
\end{array}
\right)\right|_{n \leq n_{\rm c}} = \left(
\begin{array}{c}
f_n^{(\vee)} \\ g_n^{(\vee )}
\end{array}
\right).
\end{equation}
One can note that both upper and lower components of each mode
consist of two terms: one (given by the Bessel function) is
vanishing and another one (given by the Neumann function) is
diverging in the limit of $r\rightarrow 0$.

In the case of $\nu \geq 1$ and $\frac12 (1-\frac1\nu) < F < \frac12 (1+\frac1\nu)$ $\quad$ $(0<G<1)$,
there is a peculiar mode corresponding to $n=n_{\rm c}$. This mode can be composed either
from the pair of columns
$$ \left(
\begin{array}{c}
\sqrt{1+m/E} \, J_{-G}(kr)  \\
{\rm sgn}(E) \sqrt{1- m/E}\, J_{1-G}(kr)
\end{array}\right)\, {\rm and}\,
 \left(
\begin{array}{c}
\sqrt{1+m/E} \, Y_{-G}(kr)  \\
{\rm sgn}(E) \sqrt{1- m/E}\, Y_{1-G}(kr)
\end{array}
\right),
$$
or from the pair of columns
$$ \left(
\begin{array}{c}
\sqrt{1+m/E} \, J_{G}(kr)  \\
-{\rm sgn}(E) \sqrt{1- m/E}\, J_{-1+G}(kr)
\end{array}
\right) \,{\rm and}\, \left(
\begin{array}{c}
\sqrt{1+m/E} \, Y_{G}(kr)  \\
-{\rm sgn}(E) \sqrt{1- m/E}\, Y_{-1+G}(kr)
\end{array}
\right);$$
both terms in the first variant have divergent upper
components, whereas both terms in the second variant have divergent
lower components. Instead of these variants we choose the following
form:
\begin{multline}\label{2.5}
\left(
\begin{array}{c}
f_{n_{\rm c}} \\ g_{n_{\rm c}}
\end{array}
\right)= \frac12 \sqrt{\frac{\nu}{\pi} }
\frac1{\sqrt{1+\sin(2\mu_{1-G})\cos(G\pi) } } \\ \times \left(
\begin{array}{c}
\sqrt{1+m/E} \left[\sin(\mu_{1 -G}) J_{-G}(kr) + \cos(\mu_{1 -G}) J_{G}(kr)\right] \\
{\rm sgn}(E) \sqrt{1- m/E} \left[\sin(\mu_{1 -G}) J_{1-G}(kr) -
\cos(\mu_{1 -G}) J_{-1+G}(kr)\right]
\end{array}
\right).
\end{multline}
Modes
\begin{equation}\label{2.6}
\left. \left( \begin{array}{c} f_n \\ g_n
\end{array}
\right)\right|_{n \geq  n_{\rm c}+1} = \left(
\begin{array}{c}
f_n^{(\wedge)} \\ g_n^{(\wedge )}
\end{array}
\right), \quad \left. \left( \begin{array}{c} f_n \\ g_n
\end{array}
\right)\right|_{n \leq n_{\rm c}-1} = \left(
\begin{array}{c}
f_n^{(\vee)} \\ g_n^{(\vee )}
\end{array}
\right)
\end{equation}
together with mode \eqref{2.5} comprise the set of all solutions with $|E|>m$ in this case.

In the case of $\frac12 \leq \nu< 1$ and $\frac12
\left(\frac1\nu-1\right) < F < \frac12 \left(3-\frac1\nu\right)$
$\quad$ $(1-\nu < G <\nu)$, the set of all solutions with $|E|>m$ is
also given by \eqref{2.5} and \eqref{2.6}. In the case of $\frac12 \leq
\nu < 1$ and $0 < F < \frac12 \left(\frac1\nu-1\right)$ $\quad$
$\left( \frac12(1-\nu)< G< 1-\nu \right)$, there is an additional
peculiar mode:
\begin{multline}\label{2.7}
\left(
\begin{array}{c}
f_{n_{\rm c}-1} \\ g_{n_{\rm c}-1}
\end{array}
\right)= \frac12 \sqrt{\frac{\nu}{\pi} }
\frac1{\sqrt{1+\sin(2\mu_{1-\nu-G}) } }\\ \times \left(
\begin{array}{c}
\sqrt{1+m/E} \left[\sin(\mu_{1-\nu -G}) J_{-\nu-G}(kr) + \cos(\mu_{1-\nu -G}) J_{\nu+G}(kr)\right] \\
{\rm sgn}(E) \sqrt{1- m/E} \left[\sin(\mu_{1-\nu -G})
J_{1-\nu-G}(kr) - \cos(\mu_{1-\nu -G}) J_{-1+\nu+G}(kr)\right]
\end{array}
\right).
\end{multline}
Modes
\begin{equation}\label{2.8}
\left. \left( \begin{array}{c} f_n \\ g_n
\end{array}
\right)\right|_{n \geq  n_{\rm c}+1} = \left(
\begin{array}{c}
f_n^{(\wedge)} \\ g_n^{(\wedge )}
\end{array}
\right), \quad \left. \left( \begin{array}{c} f_n \\ g_n
\end{array}
\right)\right|_{n \leq n_{\rm c}-2} = \left(
\begin{array}{c}
f_n^{(\vee)} \\ g_n^{(\vee )}
\end{array}
\right)
\end{equation}
together with modes \eqref{2.5} and \eqref{2.7} comprise the set of
all solutions with $|E|>m$ in this case. An additional peculiar mode
also appears in the case of $\frac12 \leq \nu < 1$ and  $\frac12
\left(3\! -\!\frac1\nu\right) < F < 1$ $\quad$ $\left( \nu< G < \frac12
(1+\nu) \right)$:
\begin{multline}\label{2.9}
\left(
\begin{array}{c}
f_{n_{\rm c}+1} \\ g_{n_{\rm c}+1}
\end{array}
\right)= \frac12 \sqrt{\frac{\nu}{\pi} }
\frac1{\sqrt{1+\sin(2\mu_{1+\nu-G}) } } \\ \times \left(
\begin{array}{c}
\sqrt{1+m/E} \left[\sin(\mu_{1+\nu -G}) J_{\nu-G}(kr) + \cos(\mu_{1+\nu -G}) J_{-\nu+G}(kr)\right] \\
{\rm sgn}(E) \sqrt{1- m/E} \left[\sin(\mu_{1+\nu -G})
J_{1+\nu-G}(kr) - \cos(\mu_{1+\nu -G}) J_{-1-\nu+G}(kr)\right]
\end{array}
\right).
\end{multline}
Modes
\begin{equation}\label{2.10}
\left. \left( \begin{array}{c} f_n \\ g_n
\end{array}
\right)\right|_{n \geq  n_{\rm c}+2} = \left(
\begin{array}{c}
f_n^{(\wedge)} \\ g_n^{(\wedge )}
\end{array}
\right), \quad \left. \left( \begin{array}{c} f_n \\ g_n
\end{array}
\right)\right|_{n \leq n_{\rm c}-1} = \left(
\begin{array}{c}
f_n^{(\vee)} \\ g_n^{(\vee )}
\end{array}
\right)
\end{equation}
together with modes \eqref{2.5} and \eqref{2.9} comprise the set of all solutions with $|E|>m$ in this case. In the case of
$0< \nu < \frac12$ there are two and more peculiar modes.

Certainly, the limit of $r\rightarrow 0$ is of no sense for vortices
of nonzero transverse size. However, it is instructive to discuss an
infinitely  thin vortex, and we shall touch upon this subject in the
rest of the Section. Most of the modes in the $r_0 = 0$ case are obtained by
putting $\mu^{( \wedge )}_\rho = \mu^{( \vee )}_\rho =\pi/2$ in
\eqref{2.3}, \eqref{2.4}, \eqref{2.6}, \eqref{2.8}, and
\eqref{2.10}; these modes are regular at $r\rightarrow 0$.
However, peculiar modes \eqref{2.5}, \eqref{2.7}, and \eqref{2.9} cannot be
made regular at $r \rightarrow 0$; they are irregular but square
integrable. The latter circumstance requires a quest for a
self-adjoint extension, and the Weyl-von Neumann theory of
deficiency indices (see \cite{Neu,Ree}) has to be employed. In the case of
$\nu \geq 1$  and   $\frac12 (1-\frac1\nu) < F < \frac12 (1+\frac1\nu)$, as well as in the case of
$\frac12 \leq \nu < 1$ and $\frac12 \left(\frac1\nu-1\right) < F <
\frac12 \left(3-\frac1\nu\right)$, when there is one irregular mode,
the deficiency index is (1,1), and the one-parametric family of
self-adjoint extensions can be introduced with the use of condition
\begin{equation}\label{2.11}
    \lim_{r\rightarrow0}
    (mr)^G\cos\left(\frac{\Theta}{2}+\frac\pi4\right) f_{n_{\rm
    c}}=-\lim_{r\rightarrow 0} (mr)^{1-G}\sin\left(\frac{\Theta}{2}+\frac\pi4\right) g_{n_{\rm
    c}},
\end{equation}
where $\Theta$ is the self-adjoint extension parameter \cite{Jac,Ger}. In view of
relations
\begin{equation}\label{2.12}
\int\limits_0^\infty dr\,r J_\rho(kr)J_\rho(k'r) =
\frac{\delta(k-k')}{\sqrt{kk'} },\quad \rho> -1
\end{equation}
and
\begin{equation}\label{2.13}
\int\limits_0^\infty dr\,r J_\rho(kr)J_{-\rho}(k'r) =\cos(\rho\pi)
\frac{\delta(k-k')}{\sqrt{kk'} },\quad -1<\rho<1,
\end{equation}
the modes are orthonormalized as the modes corresponding to the
continuous spectrum:
\begin{equation}\label{2.14}
\int d^2x\, \sqrt{g}\,[f_n(r,E) f_n(r,E')+g_n(r,E)
g_n(r,E')]=\frac12[1+{\rm sgn}(EE')]\frac{\delta(k-k')}{\sqrt{kk'}
}.
\end{equation}
In addition, there is a bound state at $\cos\Theta<0$ with energy
$E_{BS}$ in the gap between the continuums, $-m<E_{BS}<m$. Its
mode is
\begin{equation}\label{2.15}
\left( \begin{array}{c} f_{n_{\rm c}}^{(BS)} \vphantom{\int\limits_0^0}\\
g_{n_{\rm c}}^{(BS)}
\end{array}
\right)=\frac1\pi
\sqrt{\frac{\nu(m^2-E_{BS}^2)\sin(G\pi)}{1+(2G-1)E_{BS}/m} } \left(
\begin{array}{c} \sqrt{1+E_{BS}/m} \,K_G(r\sqrt{m^2-E_{BS}^2}) \\  \sqrt{1-E_{BS}/m} \,K_{1-G}(r\sqrt{m^2-E_{BS}^2})
\end{array}
\right),
\end{equation}
and the value of its energy is determined from relation
\begin{equation}\label{2.16}
\frac{(1+E_{BS}/m)^{1-G}}{(1-E_{BS}/m)^{G} }=-2^{1-2G}
\frac{\Gamma(1-G)}{\Gamma(G)
}\tan\left(\frac{\Theta}{2}+\frac\pi4\right),
\end{equation}
$\Gamma(u)$ is the Euler gamma function, and $K_\rho(u)$ is the
Macdonald function of order $\rho$. The induced vacuum current and
other vacuum polarization effects were comprehensively and exhaustively studied for
$\nu=1$ in \cite{Si6,Si7,SiR,Sit9,Si9} and for carbon nanocones
in \cite{SiV7,SiV1,SiV2,Si18}.

In the case of $\frac12 \leq \nu<1$ and  $0 < F < \frac12
\left(\frac1\nu-1\right)$, or
 $\frac12 \leq \nu<1$ and $\frac12\left(3-\frac1\nu\right) < F < 1$, and other cases, when there
are two irregular square integrable modes (of the kind given by the
pair of \eqref{2.5} and \eqref{2.7}, or \eqref{2.5} and
\eqref{2.9}), the deficiency index is (2,2), and there are four
self-adjoint extension parameters. These cases remain unstudied yet.

\section{Self-adjointness and choice of boundary conditions}\setcounter{equation}{0}

The Dirac hamiltonian operator in the background of the ANO vortex
of nonzero radius $r_0$ is self-adjoint, if condition
\begin{equation}\label{3.1}
\left.\tilde\psi^\dag\alpha^r \psi\right|_{r=r_0}=0
\end{equation}
is valid, see \eqref{1.10} and \eqref{1.11}, and sets of functions
$\psi$ and $\tilde\psi$ coincide. Ergo, the quest is for a
boundary condition in the form
\begin{equation}\label{3.2}
\left. \psi\right|_{r=r_0}=\left. K\psi \right|_{r=r_0}, \quad
\left. \tilde\psi\right|_{r=r_0}=\left. K\tilde \psi
\right|_{r=r_0},
\end{equation}
where $K$ is a matrix (element of the Clifford algebra) which without a loss of generality can be chosen to be Hermitian and has to obey conditions
\begin{equation}\label{3.4}
[K,\alpha^r]_+=0
\end{equation}
and
\begin{equation}\label{3.3}
K^2=I.
\end{equation}
One can simply go through four linearly independent elements of the
Clifford algebra in $2+1$-dimensional spacetime which is a section orthogonal
to the ANO vortex axis, and find
\begin{equation}\label{3.7}
K= c_1 \beta + c_2 {\rm i} \beta \alpha^r
\end{equation}
with real coefficients satisfying
\begin{equation}\label{3.8}
c_1^2+c_2^2 = 1.
\end{equation}
Using obvious
parametrization
$$ c_1 = \sin\theta,\quad c_2= \cos\theta,
$$
we finally obtain
\begin{equation}\label{3.9}
K = {\rm i} \beta \alpha^r {\rm e}^{-{\rm i}\theta \alpha^r}.
\end{equation}
Thus, boundary condition \eqref{3.2} with $K$ given by \eqref{3.9}
is the most general boundary condition ensuring the self-adjointness
of the Dirac hamiltonian operator in the background of the ANO
vortex of nonzero radius $r_0$ in transverse section $z={\rm const}$, and
parameter $\theta$ can be interpreted as the self-adjoint extension
parameter. Value $\theta=0$ corresponds to the MIT bag boundary condition,
which was proposed long ago as the condition ensuring the confinement of the
matter field \cite{Joh}. However, it should be comprehended  that a condition with an arbitrary value of $\theta$ ensures the confinement equally as well as that with $\theta=0$.

Imposing boundary condition \eqref{3.2} with matrix $K$
\eqref{3.9} on the solution to the Dirac equation,
$\psi_E(\textbf{x})$ \eqref{1.16}, we obtain the condition for the
modes:
\begin{equation}\label{3.10}
\cos\left(\frac{\theta}{2}+\frac{\pi}4 \right)
f_n(r_0,E)=-\sin\left(\frac{\theta}{2}+\frac{\pi}4
\right)g_n(r_0,E),
\end{equation}
which allows us to determine their coefficients:
\begin{align}
& \tan(\mu^{(\wedge)}_\rho  ) = \frac{\cos\left(\frac{\theta}{2}+\frac{\pi}4 \right) k Y_{\rho-1}(kr_0)- \sin\left(\frac{\theta}{2}+\frac{\pi}4 \right) (m-E) Y_{\rho}(kr_0)}
{-\cos\left(\frac{\theta}{2}+\frac{\pi}4 \right) k J_{\rho-1}(kr_0)+ \sin\left(\frac{\theta}{2}+\frac{\pi}4 \right) (m-E) J_{\rho}(kr_0)},  \label{3.11}\\
& \tan(\mu^{(\vee)}_\rho  ) =
\frac{\cos\left(\frac{\theta}{2}+\frac{\pi}4 \right) (m+E)
Y_{\rho}(kr_0)- \sin\left(\frac{\theta}{2}+\frac{\pi}4 \right)k
Y_{\rho-1}(kr_0)}
{-\cos\left(\frac{\theta}{2}+\frac{\pi}4 \right) (m+E) J_{\rho}(kr_0)+ \sin\left(\frac{\theta}{2}+\frac{\pi}4 \right)k J_{\rho-1}(kr_0)},  \label{3.12}\\
& \tan(\mu_\rho  ) = \frac{\cos\left(\frac{\theta}{2}+\frac{\pi}4
\right) k J_{1-\rho}(kr_0)+ \sin\left(\frac{\theta}{2}+\frac{\pi}4
\right)(m-E) J_{-\rho}(kr_0)}
{-\cos\left(\frac{\theta}{2}+\frac{\pi}4 \right) k J_{\rho-1}(kr_0)+
\sin\left(\frac{\theta}{2}+\frac{\pi}4 \right)(m-E) J_{\rho}(kr_0)}.
\label{3.13}
\end{align}
Because of condition \eqref{3.10}, in addition to the continuous
spectrum, there is a bound state at $\cos\theta<0$ for $n=n_{\rm c}$
$\quad$ ($\nu \geq 1$ and $\frac12(1-\frac1\nu) < F < \frac12 (1+\frac1\nu)$, or
$\frac12 \leq \nu<1$), as well as for $n=n_{\rm c}-1$ $\quad$ ($\frac12 \leq \nu <1$ and  $0 < F < \frac12\left(\frac1\nu-1\right)$), or
$n=n_{\rm c}+1$ $\quad$ ($\frac12 \leq \nu <1$ and  $\frac12(3-\frac1\nu) < F < 1$).
The bound state modes are
\begin{multline}\label{3.14}
\left( \begin{array}{c} f_{n_{\rm c}}^{(BS)} \vphantom{\int\limits_0^0}\\
g_{n_{\rm c}}^{(BS)}
\end{array}
\right) = \sqrt{\frac{\nu\kappa m}{2\pi r_0}}\Biggl\{ m K_G(\kappa
r_0)
K_{1-G}(\kappa r_0)  \Biggr. \\
\Biggl.+E_{BS}\Biggl[\kappa r_0 K^2_{1-G}(\kappa r_0)-\kappa r_0
K^2_G(\kappa r_0)+ (2G-1)K_G(\kappa r_0) K_{1-G}(\kappa r_0) \Biggr]
\Biggr\}^{-1/2} \\ \times \left(
\begin{array}{c} \sqrt{1+E_{BS}/m} \,K_G(\kappa r_0) \\  \sqrt{1-E_{BS}/m} \,K_{1-G}(\kappa r_0)
\end{array}
\right),
\end{multline}
\begin{multline}\label{3.15}
\left( \begin{array}{c} f_{n_{\rm c}\mp1}^{(BS)} \vphantom{\int\limits_0^0}\\
g_{n_{\rm c}\mp1}^{(BS)}
\end{array}
\right) = \sqrt{\frac{\nu\kappa m}{2\pi r_0}}\,\Biggl\{ m K_{G\pm
\nu}(\kappa r_0)
K_{1-G\mp \nu}(\kappa r_0)  \Biggr. \\
\Biggl.+E_{BS}\Biggl[\kappa r_0 K^2_{1-G \mp \nu}(\kappa r_0)-\kappa
r_0 K^2_{G \pm \nu}(\kappa r_0)+ (2G\pm2\nu-1)K_{G\pm
\nu}(\kappa r_0) K_{1-G\mp \nu}(\kappa r_0) \Biggr] \Biggr\}^{-1/2} \\
\times \left(
\begin{array}{c} \sqrt{1+E_{BS}/m} \,K_{G\pm \nu}(\kappa r_0) \\  \sqrt{1-E_{BS}/m} \,K_{1-G\mp \nu}(\kappa r_0)
\end{array}
\right),
\end{multline}
where $\kappa = \sqrt{m^2-E^2_{BS} }$. The bound state energy for
$n=n_{\rm c}$ is determined from relation
\begin{equation}\label{3.16}
\sqrt{ \frac{1+E_{BS}/m}{1-E_{BS}/m }}=- \frac{K_{1-G}(\kappa
r_0)}{K_{G}(\kappa r_0) }
\tan\left(\frac{\theta}{2}+\frac\pi4\right);
\end{equation}
by changing $G$ to $G\pm \nu$ in \eqref{3.16}, one obtains the
relation for $n=n_{\rm c}\mp 1$.

Comparing the case of a vortex of nonzero transverse size with that
of an infinitely thin one, we conclude that in the first case the
total hamiltonian is extended with the use of the only one
self-adjoint extension parameter, whereas in the second case
several partial hamiltonians are extended, and the number of
self-adjoint extension parameters can be zero (no need for
extension, the operator is essentially self-adjoint), one, four,
etc. The values of the self-adjoint extension parameters in the
second case can be fixed from the first case by limiting
procedure $r_0\rightarrow 0$ \cite{Alf}. The nonpeculiar modes
($\rho>1$) in this limit become regular and independent of $\theta$,
since, as was already noted,
$$
\lim_{r_0\rightarrow 0}\mu^{(\wedge)}_\rho = \lim_{r_0 \rightarrow
0}\mu^{(\vee)}_\rho = \frac\pi2.
$$
The peculiar modes $(0<\rho<1)$ in this limit become irregular and
square integrable, and
\begin{equation}\label{3.17}
\lim_{r_0\rightarrow 0}\mu_\rho=\left\{
\begin{array}{l}
\frac\pi2, \quad \frac12<\rho \quad \left(\theta \neq \pm\frac\pi2\right), \quad 0<\rho \quad (\theta=\frac\pi2), \\
\vphantom{\int\limits_0^0}
{\rm sgn}(E)\arctan\left[\sqrt{
\frac{1-m/E}{1+m/E}}
\tan\left(\frac\theta2+\frac\pi4\right) \right], \quad \rho=\frac12,\\
0,\quad \rho<\frac12 \quad \left(\theta\neq \pm
\frac\pi2\right),\quad \rho<1 \quad \left(\theta= -
\frac\pi2\right).
\end{array}
\right.
\end{equation}
Namely in this way, the condition of minimal irregularity
\cite{Si6,Si7} is obtained, which in the case of the deficiency
index  equal to (1,1) (i.e., only one peculiar mode) takes form
\begin{equation}\label{3.18}
\Theta=\left\{
\begin{array}{l}
\frac\pi2, \quad \frac12\left(1-\frac1\nu\right) < F < \frac12 \quad
(\nu \geq 1),\quad \frac12\left(\frac1\nu-1\right) < F < \frac12 \quad \left(\frac12 < \nu<1\right),\\
\vphantom{\int\limits_0^0} \theta, \quad F = \frac12 \quad \left(\nu \geq\frac12\right),\\
-\frac\pi2,\quad \frac12 < F < \frac12\left(1+\frac1\nu\right)
\quad (\nu \geq 1),\quad \frac12 < F < \frac12\left(3-\frac1\nu\right)
\quad \left(\frac12 < \nu<1\right).
\end{array}
\right.
\end{equation}

\section{Induced vacuum current}\setcounter{equation}{0}

We start with the case of $\nu \geq 1$ and $\frac12\left(1-\frac1\nu
\right) < F < \frac12\left(1+\frac1\nu\right)$, or $\frac12
\leq \nu <1$ and $\frac12\left(\frac1\nu-1 \right) < F <
\frac12\left(3-\frac1\nu\right)$, when there is  one peculiar mode.
Inserting the contribution of the appropriate modes (see
\eqref{2.1}, \eqref{2.2}, \eqref{2.5}, \eqref{2.6}, and \eqref{3.14})
to \eqref{1.19}, we obtain
\begin{equation}\label{4.0}
j_\varphi(r) = j_\varphi^{(1)}(r)+j_\varphi^{(2)}(r)+j_\varphi^{(3)}(r),  \end{equation}
where
\begin{equation}\label{4.1}
j_\varphi^{(1)}(r)=-\frac{r}{2\pi}\int\limits_0^\infty \frac{dk\,
k^2}{\sqrt{k^2+m^2}  }\sum_{l=1}^\infty\left[ J_{\nu
l+1-G}(kr)J_{\nu l-G}(kr)- J_{\nu l+G}(kr)J_{\nu l-1+G}(kr)\right],
\end{equation}
\begin{multline}\label{4.2}
j_\varphi^{(2)}(r)\!=\!-\frac{r}{4\pi}\!\int\limits_0^\infty \!\!
\frac{dk\, k^2}{\sqrt{k^2+m^2}  } \\
\times \sum_{{\rm
sgn}(E)}\sum_{l=1}^\infty \!\!\left\{\vphantom{\frac12}
\!\!\cos^2(\mu^{(\wedge)}_{\nu l+1-G})\left[Y_{\nu l+1-G}(kr)Y_{\nu
l-G}(kr)\!-\! J_{\nu l+1-G}(kr)J_{\nu l-G}(kr) \right]\right.\\
+\frac12\sin(2\mu^{(\wedge)}_{\nu l+1-G}) \left[J_{\nu
l+1-G}(kr)Y_{\nu l-G}(kr)+ Y_{\nu l+1-G}(kr)J_{\nu l-G}(kr)
\right]-\\
-\cos^2(\mu^{(\vee)}_{\nu l+G})\left[Y_{\nu l+G}(kr)Y_{\nu
l-1+G}(kr)- J_{\nu l+G}(kr)J_{\nu l-1+G}(kr) \right]-\\
\left.-\frac12\sin(2\mu^{(\vee)}_{\nu l+G}) \left[J_{\nu
l+G}(kr)Y_{\nu l-1+G}(kr)+ Y_{\nu l+G}(kr)J_{\nu l-1+G}(kr)
\right]\right\},
\end{multline}
\begin{multline}\label{4.3}
j_\varphi^{(3)}(r)\!=\!-\frac{r}{4\pi}\!\int\limits_0^\infty
\!\! \frac{dk\, k^2}{\sqrt{k^2+m^2}  }\!\!\sum_{{\rm
sgn}(E)}\left[\tan(\mu_{1-G})+2\cos(G\pi)+\cot(\mu_{1-G})\right]^{-1} \\
\times \left[\tan(\mu_{1-G})J_{-G}(kr)J_{1-G}(kr)+J_{G}(kr)J_{1-G}(kr) \right. \\
-\left. J_{-G}(kr)J_{-1+G}(kr)-\cot(\mu_{1-G})J_{G}(kr)J_{-1+G}(kr)\right]\\
+\frac{\kappa^2}{4\pi r_0}\,\frac{[1-{\rm sgn}(\cos\theta)]\,{\rm
sgn}\!\!\left[\tan\left(\frac\theta4+\frac\pi4\right)
+\frac{K_G(\kappa r_0)}{K_{1-G}(\kappa r_0)} \right] K_G(\kappa r)
K_{1-G}(\kappa r) }{m K_G(\kappa r_0)K_{1-G}(\kappa r_0)
\!+\!E_{BS}\left\{\kappa r_0 [K_{1-G}^2(\kappa
r_0)\!-\!K_{G}^2(\kappa r_0)]\!+\!(2G\!-\!1)K_G(\kappa
r_0)K_{1-G}(\kappa r_0) \right\} }.
\end{multline}

In Appendix A, the summation in \eqref{4.1} is performed, yielding
\begin{multline}\label{4.4}
j_\varphi^{(1)}(r)\!=\!-\frac{r}{2\pi^2}\!\int\limits_m^\infty \!\!
\frac{dq\, q^2}{\sqrt{q^2-m^2}
}\left[I_{1-G}(qr)K_G(qr)-I_G(qr)K_{1-G}(qr) \right] \\
-\frac{m}{(2\pi)^2}\left\{ \int\limits_0^\infty \!\!
\frac{du}{\cosh(u/2)}\left[1+\frac1{2mr \cosh(u/2)}\right]{\rm
e}^{-2mr
\cosh(u/2)}\right. \\
\times \frac{\sin(G\pi)\sinh(\nu u)
\sinh\left[\left(G-\frac12\right)u\right]- \cos(G\pi)\sin(\nu \pi)
\cosh\left[\left(G-\frac12\right)u\right]}{\cosh(\nu u)-\cos(\nu
\pi)} \\
- \frac{2\pi}{\nu} \sum_{p=1}^{\left[\!\left| {\nu}/2
\right|\!\right]} \left[1 +\frac{1}{2mr\sin(p\pi/\nu)}\right]\,
\exp[-2mr\sin(p\pi/\nu)]
\,\frac{\sin[(2G-1)p\pi/\nu]}{\sin(p\pi/\nu)} \\
\left. - \frac\pi\nu  \left(1 +\frac{1}{2mr}\right){\rm e}^{-2mr}\cos(G\pi) \, \delta_{\nu, \, 2N}\right\},
\end{multline}
where $I_\rho(u)$ is the modified Bessel function of order $\rho$
($p$ and $N$ are the positive integers, $\delta_{\omega, \,
\omega'}$ is the Kronecker symbol, $\delta_{\omega, \, \omega'}=0$
at $\omega' \neq \omega$ and $\delta_{\omega, \, \omega} = 1$),
while \eqref{4.2} is transformed to the following expression:
\begin{multline}\label{4.5}
j_\varphi^{(2)}(r)\!=\!-\frac{r}{\pi^2}\!\int\limits_m^\infty \!\!
\frac{dq\,
q^2}{\sqrt{q^2-m^2}}\sum_{l=1}^\infty\left[C^{(\wedge)}_{\nu
l+1-G}(qr_0) K_{\nu l+1-G}(qr)K_{\nu l-G}(qr)\right. \\-\left.
C^{(\vee)}_{\nu l+G}(qr_0) K_{\nu l+G}(qr)K_{\nu l-1+G}(qr) \right],
\end{multline}
where
\begin{multline}\label{4.6}
C^{(\wedge)}_\rho(v)=\left\{v
I_\rho(v)K_\rho(v)\tan\left(\frac\theta2+\frac\pi4\right)+mr_0\left[I_\rho(v)K_{\rho-1}(v)-I_{\rho-1}(v)K_\rho(v)\right]\right.\\
\left.-
vI_{\rho-1}(v)K_{\rho-1}(v)\cot\left(\frac\theta2+\frac\pi4\right)
\right\} \left[v
K^2_\rho(v)\tan\left(\frac\theta2+\frac\pi4\right)+2mr_0K_\rho(v)K_{\rho-1}(v)\right.\\
\left.+ vK^2_{\rho-1}(v)\cot\left(\frac\theta2+\frac\pi4\right)
\right]^{-1}
\end{multline}
and
\begin{multline}\label{4.7}
C^{(\vee)}_\rho(v)=\left\{v
I_\rho(v)K_\rho(v)\cot\left(\frac\theta2+\frac\pi4\right)+mr_0\left[I_\rho(v)K_{\rho-1}(v)-I_{\rho-1}(v)K_\rho(v)\right]-\right.\\
\left.-
vI_{\rho-1}(v)K_{\rho-1}(v)\tan\left(\frac\theta2+\frac\pi4\right)
\right\} \left[v
K^2_\rho(v)\cot\left(\frac\theta2+\frac\pi4\right)+2mr_0K_\rho(v)K_{\rho-1}(v)+\right.\\
\left.+ vK^2_{\rho-1}(v)\tan\left(\frac\theta2+\frac\pi4\right)
\right]^{-1};
\end{multline}
note that $C^{(\wedge)}_{\nu l+1-G}(v) \leftrightarrow C^{(\vee)}_{\nu l+G}(v)$ under simultaneous
changes $F \rightarrow 1-F$ and $\theta \rightarrow -\theta$.

In Appendix B, $j^{(3)}_\varphi$ \eqref{4.4} is transformed to
the following expression:
\begin{multline}\label{4.8}
j^{(3)}_\varphi(r)\!=\!\frac{r}{2\pi^2}\!\int\limits_m^\infty \!\!
\frac{dq\,
q^2}{\sqrt{q^2-m^2}}\left[I_{1-G}(qr)K_{G}(qr)\!-\!I_{G}(qr)K_{1-G}(qr) \right. \\
\left. -2C_{1-G}(qr_0)K_{G}(qr)K_{1-G}(qr)
\right],
\end{multline}
where
\begin{multline}\label{4.9}
C_{1-G}(v)\!=\!\left\{v
\left[I_{1-G}(v)+\frac{\sin(G\pi)}{\pi}K_{1-G}(v)\right]\!K_{1-G}(v)\tan\left(\frac\theta2+\frac\pi4\right)+\!mr_0
\left[I_{1-G}(v)K_G(v)\right.\right. \\
\left.\left.-I_{G}(v)K_{1-G}(v)\right]-
v\left[I_{G}(v)+\frac{\sin(G\pi)}{\pi}K_{G}(v)\right]
K_{G}(v)\cot\left(\frac\theta2+\frac\pi4\right) \right\} \\
\times\left[v
K^2_{1-G}(v)\tan\left(\frac\theta2+\frac\pi4\right)+2mr_0K_G(v)K_{1-G}(v)+
vK^2_G(v)\cot\left(\frac\theta2+\frac\pi4\right) \right]^{-1};
\end{multline}
note that $C_{1-G}(v)$ changes sign under simultaneous
changes $F \rightarrow 1-F$ and $\theta \rightarrow -\theta$.

Summing \eqref{4.4}, \eqref{4.5} and \eqref{4.8}, we obtain the
final form for the induced vacuum current and express it in terms of
$F$ instead of $G$ (see \eqref{1.18}),
\begin{multline}\label{4.10}
j_\varphi(r)=-\frac{m}{(2\pi)^2}\left\{ \int\limits_0^\infty
\frac{du}{\cosh(u/2)}\left[1+\frac1{2mr \cosh(u/2)}\right]{\rm e}^{-2mr
\cosh(u/2)}\right. \\
\times \frac{\cos\left[\nu\left(F-\frac12\right)\pi\right]\sinh(\nu u)
\sinh\left[\nu\left(F-\frac12\right)u\right]+ \sin\left[\nu\left(F-\frac12\right)\pi\right]\sin(\nu \pi)
\cosh\left[\nu\left(F-\frac12\right)u\right]}{\cosh(\nu u)-\cos(\nu
\pi)}\\
- \frac{2\pi}{\nu}\sum_{p=1}^{\left[\!\left| {\nu}/2 \right|\!\right]} \left[1 +\frac{1}{2mr\sin(p\pi/\nu)}\right]\,
\exp[-2mr\sin(p\pi/\nu)] \,\frac{\sin[(2F-1)p\pi]}{\sin(p\pi/\nu)}\\
\left. + \frac{\pi}{2N} \left(-1\right)^{N}\sin\left(2NF \pi \right) \left(1 +\frac{1}{2mr}\right){\rm e}^{-2mr} \, \delta_{\nu, \, 2N}\right\} \\
-\frac{r}{\pi^2}\!\int\limits_m^\infty \!\! \frac{dq\,
q^2}{\sqrt{q^2-m^2}}\left[C_{\frac12-\nu\left(F-\frac12\right)}(qr_0) K_{\frac12-\nu\left(F-\frac12\right)}(qr)
K_{\frac12+\nu\left(F-\frac12\right)}(qr)+\Sigma(qr,qr_0) \right],
\end{multline}
where
\begin{multline}\label{4.11}
\Sigma(w,v)=\sum_{l=1}^\infty \left[C^{(\wedge)}_{\nu \left(l-F+\frac12\right)+\frac12}(v)
K_{\nu \left(l-F+\frac12\right)+\frac12} (w)
K_{\nu \left(l-F+\frac12\right)-\frac12}(w) \right. \\
\left. - C^{(\vee)}_{\nu \left(l+F-\frac12\right)+\frac12}(v) K_{\nu
\left(l+F-\frac12\right)+\frac12}(w)K_{\nu \left(l+F-\frac12\right)-\frac12}(w)\right].
\end{multline}

The analysis in Appendix A is sufficient to consider cases when
there are no peculiar modes. In the case of $\nu>1$ and $0 < F <
\frac12\left(1-\frac1\nu\right)$ $\quad$ $\left(\frac12(1-\nu) <G <
0\right)$, we obtain
\begin{multline}\label{4.12}
j_\varphi(r)=-\frac{m}{(2\pi)^2}\left\{ \int\limits_0^\infty
\frac{du}{\cosh(u/2)}\left[1+\frac1{2mr \cosh(u/2)}\right]{\rm
e}^{-2mr
\cosh(u/2)}\right. \\
\times \frac{\cos\left[\nu\left(F-\frac12\right)\pi\right] \cosh\left[\nu\left(F+\frac12 \right)u\right]
- \cos[\nu\left(F+\frac12 \right)\pi)]
\cosh\left[\nu\left(F-\frac12\right)u\right]}{\cosh(\nu
u)-\cos(\nu \pi)} \\
- \frac{2\pi}{\nu}\sum_{p=1}^{\left[\!\left| {\nu}/2 \right|\!\right]} \left[1 +\frac{1}{2mr\sin(p\pi/\nu)}\right]\,
\exp[-2mr\sin(p\pi/\nu)] \,\frac{\sin[(2F-1)p\pi]}{\sin(p\pi/\nu)}\\
\left. + \frac{\pi}{2N} \left(-1\right)^{N}\sin\left(2NF \pi \right) \left(1 +\frac{1}{2mr}\right){\rm e}^{-2mr} \, \delta_{\nu, \, 2N}\right\} \\
-\frac{r}{\pi^2}\!\int\limits_m^\infty \!\! \frac{dq\,
q^2}{\sqrt{q^2-m^2}}\left[C^{(\wedge)}_{\frac12-\nu\left(F-\frac12\right)}(qr_0) K_{\frac12-\nu\left(F-\frac12\right)}(qr)
K_{\frac12+\nu\left(F-\frac12\right)}(qr)+\Sigma(qr,qr_0) \right].
\end{multline}
In the case of $\nu>1$ and $\frac12\left(1+\frac1\nu \right) < F < 1$
$\quad$ $\left(1<G < \frac12(1+\nu) \right)$, we obtain
\begin{multline}\label{4.13}
j_\varphi(r)=\frac{m}{(2\pi)^2}\left\{ \int\limits_0^\infty
\frac{du}{\cosh(u/2)}\left[1+\frac1{2mr \cosh(u/2)}\right]{\rm
e}^{-2mr
\cosh(u/2)}\right. \\
\times \frac{\cos\left[\nu\left(F-\frac12\right)\pi\right]
\cosh\left[\nu\left(F-\frac32 \right)u\right] -
\cos[\nu\left(F-\frac32 \right)\pi)]
\cosh\left[\nu\left(F-\frac12\right)u\right]}{\cosh(\nu u)-\cos(\nu
\pi)}\\
 + \frac{2\pi}{\nu}\sum_{p=1}^{\left[\!\left| {\nu}/2
\right|\!\right]} \left[1 +\frac{1}{2mr\sin(p\pi/\nu)}\right]\,
\exp[-2mr\sin(p\pi/\nu)] \,\frac{\sin[(2F-1)p\pi]}{\sin(p\pi/\nu)}\\
\left. - \frac{\pi}{2N} \left(-1\right)^{N}\sin\left(2NF \pi \right) \left(1 +\frac{1}{2mr}\right){\rm e}^{-2mr} \, \delta_{\nu, \, 2N}\right\} \\
+\frac{r}{\pi^2}\!\int\limits_m^\infty \!\! \frac{dq\,
q^2}{\sqrt{q^2-m^2}}\left[C^{(\vee)}_{\frac12+\nu\left(F-\frac12\right)}(qr_0) K_{\frac12+\nu\left(F-\frac12\right)}(qr)
K_{\frac12-\nu\left(F-\frac12\right)}(qr)-\Sigma(qr,qr_0) \right].
\end{multline}
Note that both \eqref{4.12} and \eqref{4.13} consist of two parts:
one (with a factor of $m/(2\pi)^2$) is independent of $r_0$, and
another one (with a factor of $r/\pi^2$) is vanishing in the limit
of $r_0\rightarrow 0$,
\begin{equation}\label{4.14}
j_\varphi(r)= j_\varphi^{(a)}(r) + j_\varphi^{(b)}(r),\qquad
j_\varphi^{(a)}(r)=\lim_{r_0\rightarrow 0}j_\varphi(r).
\end{equation}
It is instructive to present result \eqref{4.10} in the same way; evidently, $j_\varphi^{(a)}(r)$ then coincides with the current
that is obtained by imposing the condition of minimal irregularity in the case of an infinitely thin vortex \cite{Si6,Si7},
see \eqref{2.11} and \eqref{3.18}. We obtain for the decomposition of \eqref{4.10} according to \eqref{4.14}:
\begin{multline}\label{4.15}
\left.j_\varphi^{(a)}(r)\right|_{F\neq 1/2, \,\theta\neq \pm
\pi/2}\!=\!\frac{m}{(2\pi)^2}\left\{ {\rm
sgn}\left(F-\frac12\right)\int\limits_0^\infty
\frac{du}{\cosh(u/2)}\left[1+\frac1{2mr \cosh(u/2)}\right]{\rm
e}^{-2mr
\cosh(u/2)}\right. \\
\times \frac{\cos\left[\nu\left(F-\frac12\right)\pi)\right]
\cosh\left[\nu\left(\left|F-\frac12\right|-1\right)u\right]-
\cos\left[\nu\left(\left|F-\frac12\right|-1\right)\pi\right]\cosh\left[\nu\left(F-\frac12\right)u\right]}{\cosh(\nu
u)-\cos(\nu \pi)} \\
+ \frac{2\pi}{\nu}\sum_{p=1}^{\left[\!\left| {\nu}/2
\right|\!\right]} \left[1 +\frac{1}{2mr\sin(p\pi/\nu)}\right]\,
\exp[-2mr\sin(p\pi/\nu)]
\,\frac{\sin[(2F-1)p\pi]}{\sin(p\pi/\nu)}\\
\left. - \frac{\pi}{2N} \left(-1\right)^{N}\sin\left(2NF \pi \right) \left(1 +\frac{1}{2mr}\right){\rm e}^{-2mr} \, \delta_{\nu, \, 2N}\right\},
\end{multline}
\begin{multline}\label{4.16}
\left.j_\varphi^{(a)}(r)\right|_{F\neq 1/2, \,\theta= \pm
\pi/2}\!=\mp\frac{m}{(2\pi)^2}\left\{ \int\limits_0^\infty
\frac{du}{\cosh(u/2)}\left[1+\frac1{2mr \cosh(u/2)}\right]{\rm
e}^{-2mr
\cosh(u/2)}\right. \\
\times \frac{\cos\left[\nu\left(F-\frac12\right)\pi\right] \cosh\left[\nu\left(F-\frac12 \pm 1\right)u\right]
- \cos\left[\nu\left(F-\frac12 \pm 1\right)\pi\right]\cosh\left[\nu\left(F-\frac12\right)u\right]}{\cosh(\nu
u)-\cos(\nu \pi)} \\
 \mp \frac{2\pi}{\nu}\sum_{p=1}^{\left[\!\left| {\nu}/2
\right|\!\right]} \left[1 +\frac{1}{2mr\sin(p\pi/\nu)}\right]\,
\exp[-2mr\sin(p\pi/\nu)]
\,\frac{\sin[(2F-1)p\pi]}{\sin(p\pi/\nu)}\\
\left. \pm \frac{\pi}{2N} \left(-1\right)^{N}\sin\left(2NF \pi \right) \left(1 +\frac{1}{2mr}\right){\rm e}^{-2mr} \, \delta_{\nu, \, 2N}\right\},
\end{multline}
\begin{equation}\label{4.17}
\left.j_\varphi^{(a)}(r)\right|_{F= 1/2}\!= -
\frac{\sin\theta}{2\pi^2}\!\int\limits_m^\infty \!\! \frac{dq\,
q^2}{\sqrt{q^2-m^2}} \,\, \frac{ {\rm e}^{-2qr}}{q+m\cos\theta},
\end{equation}
\begin{multline}\label{4.18}
\left.j_\varphi^{(b)}(r)\right|_{F < 1/2, \,\theta\neq -\pi/2}\!=
-\frac{r}{\pi^2}\!\int\limits_m^\infty \!\! \frac{dq\,
q^2}{\sqrt{q^2-m^2}} \left[C^{(\wedge)}_{\frac12-\nu\left(F-\frac12\right)}(qr_0) K_{\frac12-\nu\left(F-\frac12\right)}(qr)
K_{\frac12+\nu\left(F-\frac12\right)}(qr) \right. \\
\left. +\Sigma(qr,qr_0) \right],
\end{multline}
\begin{multline}\label{4.19}
\left.j_\varphi^{(b)}(r)\right|_{F > 1/2, \,\theta\neq \pi/2}\!=
\frac{r}{\pi^2}\!\int\limits_m^\infty \!\! \frac{dq\,
q^2}{\sqrt{q^2-m^2}} \left[C^{(\vee)}_{\frac12+\nu\left(F-\frac12\right)}(qr_0) K_{\frac12+\nu\left(F-\frac12\right)}(qr)
K_{\frac12-\nu\left(F-\frac12\right)}(qr) \right. \\
\left. -\Sigma(qr,qr_0) \right],
\end{multline}
\begin{multline}\label{4.20}
\left.j_\varphi^{(b)}(r)\right|_{F\neq 1/2, \,\theta= \pm \pi/2}\!=
\mp \frac{r}{\pi^2}\!\int\limits_m^\infty \!\! \frac{dq\,
q^2}{\sqrt{q^2-m^2}} \\
\times \left\{\frac{I_{\frac12\pm
\nu\left(\frac12-F\right) }(qr_0)}{K_{\frac12\pm \nu\left(\frac12-F\right)}(qr_0)}
K_{\frac12+\nu\left(F-\frac12\right)}(qr) K_{\frac12-\nu\left(F-\frac12\right)}(qr) \right. \\
+ \sum_{l=1}^\infty\left[ \frac{I_{\nu \left(l-F+\frac12\right) \pm \frac12}(qr_0)}
{K_{\nu \left(l-F+\frac12\right) \pm \frac12}(qr_0)} K_{\nu
\left(l-F+\frac12\right) + \frac12}(qr) K_{\nu \left(l-F+\frac12\right) - \frac12}(qr) \right. \\
\left. \left. + \frac{I_{\nu \left(l+F-\frac12\right) \mp \frac12}(qr_0)}
{K_{\nu \left(l+F-\frac12\right) \mp \frac12}(qr_0)} K_{\nu
\left(l+F-\frac12\right) + \frac12}(qr) K_{\nu \left(l+F-\frac12\right) - \frac12}(qr) \right] \right\},
\end{multline}
and
\begin{multline}\label{4.21}
\left.j_\varphi^{(b)}(r)\right|_{F= 1/2}\!= -
\frac{\sin\theta}{2\pi^2}\!\int\limits_m^\infty \!\! \frac{dq\,
q^2}{\sqrt{q^2-m^2}}\left[
\frac{{\rm e}^{-2qr}\left({\rm e}^{2qr_0}-1\right)}{q+m\cos\theta} \right.\\
\left.  +4r\sum_{l=1}^\infty {\tilde C}_{\nu l+\frac12}(qr_0)K_{\nu
l+\frac12}(qr)K_{\nu l-\frac12}(qr)\right],
\end{multline}
where
\begin{multline}\label{4.22}
{\tilde C}_{\nu l+\frac12}(v)=\left\{ 2v K_{\nu
l+\frac12}(v)K_{\nu l-\frac12}(v) + mr_0\cos\theta\left[K_{\nu
l+\frac12}^2(v)+K_{\nu l-\frac12}^2(v)\right] \right\} \\
\times \left\{ v\cos\theta\left[K_{\nu
l+\frac12}^2(v)+K_{\nu l-\frac12}^2(v)\right]\left[
v\cos\theta\left(K_{\nu l+\frac12}^2(v)+K_{\nu
l-\frac12}^2(v)\right)
+4mr_0 K_{\nu l+\frac12}(v)K_{\nu
l-\frac12}(v) \right] \right. \\
\left. +4(v^2\sin^2\theta+m^2r_0^2
\cos^2\theta)K^2_{\nu l+\frac12}(v)K^2_{\nu
l-\frac12}(v) \right\}^{-1},
\end{multline}
and the use is made of relations
\begin{equation}\label{4.231}
- \frac{1}{\pi}\cos\left[\nu\left(F-\frac12\right)\pi\right] + C_{\frac12-\nu\left(F-\frac12\right)}(v)=
C^{(\wedge)}_{\frac12-\nu\left(F-\frac12\right)}
\end{equation}
and
\begin{equation}\label{4.232}
\frac{1}{\pi}\cos\left[\nu\left(F-\frac12\right)\pi\right] + C_{\frac12-\nu\left(F-\frac12\right)}(v)=
- C^{(\vee)}_{\frac12+\nu\left(F-\frac12\right)}(v).
\end{equation}

It should be noted that the $r_0$-independent part of the current in cases $\frac12\left(1-\frac1{\nu}\right)<F<\frac12$
and $\frac12<F<\frac12\left(1+\frac1{\nu}\right)$ $\quad$ ($\nu \geq 1$), or $\frac12\left(\frac1\nu-1 \right) < F <\frac12$ and
$\frac12<F < \frac12\left(3-\frac1\nu \right)$ $\quad$ ($\frac12 \leq \nu<1$), is independent of $\theta$ if  $\theta \neq \pm \pi/2$,
see \eqref{4.15}, whereas it depends on  $\theta$ if  $\theta=\pm \pi/2$, see \eqref{4.16}. The latter is distinct from cases of
the absence of peculiar modes, when the $r_0$-independent part of the current is always independent of $\theta$, see the first
four lines in \eqref{4.12} and \eqref{4.13}. Note also that limits $F\rightarrow 1/2$ and $r_0\rightarrow 0$ in general do not commute. Indeed, we obtain a discontinuity at $F=1/2$, if limit $r_0\rightarrow 0$ is taken first,
\begin{equation}\label{4.233}
\left.\lim_{F\rightarrow (1/2)_{\pm}}
\lim_{r_0\rightarrow 0}j_\varphi(r)\right|_{\theta \neq \pm \pi/2}\!=
\left.\lim_{F\rightarrow (1/2)_{\pm}}
j_\varphi^{(a)}(r)\right|_{\theta \neq \pm \pi/2}\!=\pm \frac{m}{2\pi^{2}} K_1(2mr),
\end{equation}
where the use is made of relation
\begin{equation}\label{4.a28}
\frac{m}{2}\int\limits_0^\infty
\frac{du}{\cosh(u/2)}\left[1+\frac1{2mr \cosh(u/2)}\right]{\rm
e}^{-2mr \cosh(u/2)}= \int\limits_m^\infty \!\! \frac{dq\,
q}{\sqrt{q^2-m^2}} \,\, {\rm e}^{-2qr} = m  K_1(2mr).
\end{equation}
When the order of limits is reversed, then $\left.j_\varphi^{(b)}(r)\right|_{\theta \neq \pm \pi/2}$ contributes, because of relation
\begin{multline}\label{4.234}
\left.\lim_{F\rightarrow (1/2)_{\pm}}
j_\varphi^{(b)}(r)\right|_{\theta \neq \pm \pi/2}\!=\mp
\frac{m}{2\pi^{2}} K_1(2mr)  -
\frac{\sin\theta}{2\pi^2}\!\int\limits_m^\infty \!\! \frac{dq\,
q^2}{\sqrt{q^2-m^2}} \,\, \frac{ {\rm e}^{2q(r_0-r)}}{q+m\cos\theta}\\
 + m\,O\left[mr_0 \left(\frac{r_0}{r}\right)^{2\nu-1}\right],
\end{multline}
which follows from particular cases of \eqref{4.231} and \eqref{4.232},
\begin{equation*}
C^{(\wedge)}_{1/2}(qr_0)=-\frac1{\pi}+\frac{q \sin\theta}{q+m \cos\theta}
\,e^{2qr_0}
\end{equation*}
and
\begin{equation*}
 C^{(\vee)}_{1/2}(qr_0)=-\frac1{\pi}-\frac{q \sin\theta}{q+m \cos\theta}
\,e^{2qr_0}.
\end{equation*}
Adding $\left.\lim_{F\rightarrow (1/2)_{\pm}} j_\varphi^{(a)}(r)\right|_{\theta\neq \pm \pi/2}$ to
\eqref{4.234} and taking limit $r_0\rightarrow 0$, we get
\begin{equation}\label{4.a28}
\lim_{r_0\rightarrow0}\, \left.\lim_{F\rightarrow (1/2)_{\pm}}
j_\varphi(r)\right|_{\theta\neq \pm \pi/2} = \left.
j_\varphi^{(a)}(r)\right|_{F=1/2,\,\theta\neq \pm \pi/2}.
\end{equation}
The limits do commute in special cases only:
\begin{multline}\label{4.a29}
\lim_{F\rightarrow 1/2}\, \lim_{r_0\rightarrow0} \left.
j_\varphi(r)\right|_{\theta= \pm \pi/2} = \lim_{r_0\rightarrow0}\,
\lim_{F\rightarrow 1/2} \left. j_\varphi(r)\right|_{\theta= \pm
\pi/2}\\
= \left. j_\varphi^{(a)}(r)\right|_{F=1/2,\,\theta= \pm
\pi/2}=\mp \frac{m}{2\pi^2} K_1(2mr);
\end{multline}
the discontinuity at $F=1/2$ is
absent in these cases.

We can summarize our results for the current at $F \neq 1/2$ in
cases when there is one peculiar mode: (i)
$\left.j_\varphi (r)\right|_{F < 1/2, \,\theta\neq -\pi/2}$ is given
by the right-hand side of \eqref{4.12} at $\theta \neq -\pi/2$,
whereas $\left.j_\varphi (r)\right|_{F < 1/2, \,\theta=-\pi/2}$ is
given by the right-hand side of \eqref{4.13} at $\theta=-\pi/2$, and
(ii) $\left.j_\varphi (r)\right|_{F > 1/2, \,\theta\neq \pi/2}$ is
given by the right-hand side of \eqref{4.13} at $\theta\neq \pi/2$,
whereas $\left.j_\varphi (r)\right|_{F > 1/2, \,\theta=\pi/2}$ is
given by the right-hand side of \eqref{4.12} at $\theta=\pi/2$. Note
also relation
\begin{equation}\label{4.24}
\left.j_\varphi(r)\right|_{F, \,\theta} = - \left.j_\varphi(r)\right|_{1-F, \,-\theta},
\end{equation}
which holds in all cases considered in the present Section.

In the case of $\nu=1$ expression \eqref{4.10} takes form
\begin{multline}\label{4.25}
\left.j_\varphi(r)\right|_{\nu=1}\!=-\frac{r}{\pi^2}\!\int\limits_m^\infty
\!\! \frac{dq\, q^2}{\sqrt{q^2-m^2}} \left\{
\frac{\sin(F\pi)}{\pi}\,
qr\left[K_{1-F}^2(qr)-K_{F}^2(qr)\right]\right.
\\ \left.+
\left[(2F-1)\frac{\sin(F\pi)}{\pi}+C_{1-F}(qr_0)\right]K_{F}(qr)K_{1-F}(qr)+
\left.\Sigma(qr,qr_0)\right|_{\nu=1} \right\},
\end{multline}
where the use is made of \eqref{a17} and relation
\begin{multline}\label{4.26}
\!\!\!\int\limits_m^\infty \!\! \frac{dq\, q^3}{\sqrt{q^2-m^2}}\!
\left[K_{1-F}^2(qr)\!-\!K_{F}^2(qr)\right]=\frac{\pi
m}{4r^2}\!\int\limits_0^\infty\!\!
\frac{du}{\cosh(u/2)}\!\left[1+\frac1{2mr \cosh(u/2)}\right]\!{\rm
e}^{-2mr
\cosh(u/2)}\\
\times
\left\{\tanh(u/2)\sinh\left[\left(F-\frac12\right)u\right]-(2F-1)\cosh\left[\left(F-\frac12\right)u\right]
\right\}.
\end{multline}
Decomposing \eqref{4.25} according to \eqref{4.14}, we get
\begin{multline}\label{4.27}
\left.j_\varphi^{(a)}(r)\right|_{\nu=1,F\neq 1/2, \,\theta\neq \pm
\pi/2}=-\frac{r}{\pi^3}\sin(F\pi)\!\int\limits_m^\infty \!\!
\frac{dq\, q^2}{\sqrt{q^2-m^2}} \left\{
qr\left[K_{1-F}^2(qr)-K_{F}^2(qr)\right]\right.
\\ \left.+
{\rm sgn}\left(F-\frac12\right) (|2F-1|-1)
K_{F}(qr)K_{1-F}(qr)\right\},
\end{multline}
\begin{multline}\label{4.28}
\left.j_\varphi^{(a)}(r)\right|_{\nu=1,F\neq 1/2, \,\theta= \pm
\pi/2}=-\frac{r}{\pi^3}\sin(F\pi)\!\int\limits_m^\infty \!\!
\frac{dq\, q^2}{\sqrt{q^2-m^2}} \left\{
qr\left[K_{1-F}^2(qr)-K_{F}^2(qr)\right]\right.
\\ \left.+
 (2F-1 \pm 1)
K_{F}(qr)K_{1-F}(qr)\right\},
\end{multline}
$\left.j_\varphi^{(a)}(r)\right|_{\nu=1,F= 1/2}$ is given by
\eqref{4.17}, while $\left.j_\varphi^{(b)}(r)\right|_{\nu=1,F < 1/2,
\,\theta\neq -\pi/2}$, $\left.j_\varphi^{(b)}(r)\right|_{\nu=1,F >
1/2, \,\theta\neq \pi/2}$,
$\left.j_\varphi^{(b)}(r)\right|_{\nu=1,F\neq 1/2, \,\theta= \pm
\pi/2}$, and $\left.j_\varphi^{(b)}(r)\right|_{\nu=1,F= 1/2}$ are
obtained by putting $\nu=1$ in \eqref{4.18}, \eqref{4.19}, \eqref{4.20}, and \eqref{4.21}, respectively.

\section{Induced vacuum magnetic field and its flux}\setcounter{equation}{0}

In the case of $\nu \geq 1$ and $\frac12\left(1-\frac1\nu \right) < F <
\frac12\left(1+\frac1\nu \right)$, or $\frac12 \leq \nu < 1$ and
$\frac12\left(\frac1\nu -1\right) < F < \frac12\left(3-\frac1\nu
\right)$, we obtain the following expression for $B_{\rm I}$
\eqref{1.20}:
\begin{multline}\label{5.1}
B_{\rm I}(r)=-\frac{\nu e}{2(2\pi)^2}\frac1r\left\{ \int\limits_0^\infty
\frac{du}{\cosh^2(u/2)}\,{\rm e}^{-2mr
\cosh(u/2)}\right. \\
\times \frac{\cos\left[\nu\left(F-\frac12\right)\pi\right]\sinh(\nu u)
\sinh\left[\nu\left(F-\frac12\right)u\right]+ \sin\left[\nu\left(F-\frac12\right)\pi\right]\sin(\nu \pi)
\cosh\left[\nu\left(F-\frac12\right)u\right]}{\cosh(\nu u)-\cos(\nu
\pi)}\\
\left.- \frac{2\pi}{\nu}\sum_{p=1}^{\left[\!\left| {\nu}/2 \right|\!\right]}
\exp[-2mr\sin(p\pi/\nu)] \,\frac{\sin[(2F-1)p\pi]}{\sin^2(p\pi/\nu)} + \frac{\pi}{2N} \left(-1\right)^{N}\sin\left(2NF \pi \right) {\rm e}^{-2mr}
\, \delta_{\nu, \, 2N}\right\} \\
-\frac{\nu e}{\pi^2}\!\int\limits_r^\infty dr'\!\int\limits_m^\infty
\!\! \frac{dq\, q^2}{\sqrt{q^2-m^2}}\left[C_{\frac12-\nu\left(F-\frac12\right)}(qr_0) K_{\frac12-\nu\left(F-\frac12\right)}(qr')
K_{\frac12+\nu\left(F-\frac12\right)}(qr')+\Sigma(qr',qr_0) \right],
\end{multline}
where $\Sigma(w,v)$ is given by \eqref{4.11}, while
$C^{(\wedge)}_\rho(v)$, $C^{(\vee)}_\rho(v)$ and $C_{1-G}(v)$ are
given by \eqref{4.6}, \eqref{4.7} and \eqref{4.9}, respectively.
Expression \eqref{5.1} can be decomposed as
\begin{equation}\label{5.2}
B_{\rm I}(r)=B^{(a)}_{\rm I}(r)+B^{(b)}_{\rm I}(r),\qquad
B^{(a)}_{\rm I}(r)=\lim_{r_0\rightarrow 0}B_{\rm I}(r),
\end{equation}
where
\begin{multline}\label{5.3}
\left. B^{(a)}_{\rm I}(r)\right|_{F<1/2, \,\theta\neq
-\pi/2}=\left. B^{(a)}_{\rm I}(r)\right|_{F>1/2, \,\theta =
\pi/2}= -\frac{\nu e}{2(2\pi)^2}\frac1r\left\{
\int\limits_0^\infty \frac{du}{\cosh^2(u/2)}\,{\rm e}^{-2mr
\cosh(u/2)}\right. \\
\times \frac{\cos\left[\nu\left(F-\frac12\right)\pi\right] \cosh\left[\nu\left(F+\frac12 \right)u\right]
- \cos[\nu\left(F+\frac12 \right)\pi)]
\cosh\left[\nu\left(F-\frac12\right)u\right]}{\cosh(\nu u)-\cos(\nu
\pi)}\\
\left.- \frac{2\pi}{\nu}\sum_{p=1}^{\left[\!\left| {\nu}/2
\right|\!\right]} \exp[-2mr\sin(p\pi/\nu)]
\,\frac{\sin[(2F-1)p\pi]}{\sin^2(p\pi/\nu)} + \frac{\pi}{2N} \left(-1\right)^{N}\sin\left(2NF \pi \right) {\rm e}^{-2mr}
\, \delta_{\nu, \, 2N}\right\},
\end{multline}
\begin{multline}\label{5.4}
\left. B^{(a)}_{\rm I}(r)\right|_{F>1/2, \,\theta\neq
\pi/2}=\left. B^{(a)}_{\rm I}(r)\right|_{F<1/2, \,\theta =-
\pi/2}= \frac{\nu e}{2(2\pi)^2}\frac1r\left\{
\int\limits_0^\infty \frac{du}{\cosh^2(u/2)}\,{\rm e}^{-2mr
\cosh(u/2)}\right. \\
\times \frac{\cos\left[\nu\left(F-\frac12\right)\pi\right] \cosh\left[\nu\left(F-\frac32 \right)u\right]
- \cos[\nu\left(F-\frac32 \right)\pi)]
\cosh\left[\nu\left(F-\frac12\right)u\right]}{\cosh(\nu u)-\cos(\nu
\pi)}\\
\left.+ \frac{2\pi}{\nu}\sum_{p=1}^{\left[\!\left| {\nu}/2
\right|\!\right]} \exp[-2mr\sin(p\pi/\nu)]
\,\frac{\sin[(2F-1)p\pi]}{\sin^2(p\pi/\nu)} - \frac{\pi}{2N} \left(-1\right)^{N}\sin\left(2NF \pi \right) {\rm e}^{-2mr}
\, \delta_{\nu, \, 2N}\right\},
\end{multline}
\begin{equation}\label{5.5}
\left. B^{(a)}_{\rm I}(r)\right|_{F=1/2}=-\frac{\nu e
\sin\theta}{2\pi^2} \int\limits_m^\infty \!\! \frac{dq\,
q^2}{\sqrt{q^2-m^2}}\,\frac{\Gamma(0,2qr)}{q+m\cos\theta},
\end{equation}
and
$$\Gamma(z,u)=\int\limits_u^\infty dy\,y^{z-1}{\rm e}^{-y} $$
is the incomplete gamma function. The $r_0$-dependent part of
$B_{\rm I}(r)$ is given by
\begin{multline}\label{5.6}
\left.B^{(b)}_{\rm I}(r)\right|_{F < 1/2, \,\theta\neq -\pi/2}\!= -
\frac{\nu e}{\pi^2}\int\limits_r^\infty dr'\!\int\limits_m^\infty
\!\! \frac{dq\, q^2}{\sqrt{q^2-m^2}} \\
\times \left[C^{(\wedge)}_{\frac12-\nu\left(F-\frac12\right)}(qr_0)
K_{\frac12-\nu\left(F-\frac12\right)}(qr') K_{\frac12+\nu\left(F-\frac12\right)}(qr') + \Sigma(qr',qr_0) \right],
\end{multline}
\begin{multline}\label{5.7}
\left.B^{(b)}_{\rm I}(r)\right|_{F > 1/2, \,\theta\neq \pi/2}\!=
\frac{\nu e}{\pi^2}\int\limits_r^\infty dr'\!\int\limits_m^\infty
\!\! \frac{dq\, q^2}{\sqrt{q^2-m^2}} \\
\times \left[C^{(\vee)}_{\frac12+\nu\left(F-\frac12\right)}(qr_0)
K_{\frac12+\nu\left(F-\frac12\right)}(qr') K_{\frac12-\nu\left(F-\frac12\right)}(qr') - \Sigma(qr',qr_0) \right],
\end{multline}
\begin{multline}\label{5.8}
\left. B^{(b)}_{\rm I}(r)\right|_{F\neq 1/2, \,\theta= \pm \pi/2}\!=
\mp \frac{\nu e}{\pi^2}\int\limits_r^\infty
dr'\!\int\limits_m^\infty \!\! \frac{dq\, q^2}{\sqrt{q^2-m^2}} \\
\times \left\{\frac{I_{\frac12 \pm \nu \left(\frac12-F\right)
}(qr_0)}{K_{\frac12 \pm \nu \left(\frac12-F\right) }(qr_0)} K_{\frac12+\nu\left(F-\frac12\right)}(qr')
K_{\frac12-\nu\left(F-\frac12\right)}(qr') \right. \\
+  \sum_{l=1}^\infty\left[ \frac{I_{\nu
\left(l-F+\frac12\right) \pm \frac12}(qr_0)}
{K_{\nu \left(l-F+\frac12\right) \pm \frac12}(qr_0)} K_{\nu \left(l-F+\frac12\right) + \frac12}(qr')
K_{\nu \left(l-F+\frac12\right) - \frac12}(qr') \right. \\
\left. \left. + \frac{I_{\nu \left(l+F-\frac12\right) \mp \frac12}(qr_0)}{K_{\nu \left(l+F-\frac12\right) \mp
\frac12}(qr_0)} K_{\nu \left(l+F-\frac12\right) + \frac12}(qr') K_{\nu \left(l+F-\frac12\right) - \frac12}(qr') \right]
\right\},
\end{multline}
\begin{multline}\label{5.9}
\left. B^{(b)}_{\rm I}(r)\right|_{F=1/2}=-\frac{\nu e
\sin\theta}{2\pi^2} \int\limits_m^\infty \!\! \frac{dq\,
q^2}{\sqrt{q^2-m^2}}\left[\frac{ {\rm
e}^{2qr_0}-1}{q+m\cos\theta}\,\Gamma(0,2qr)\right.
\\ \left.+ 4\int\limits_r^\infty dr' \sum_{l=1}^\infty {\tilde C}_{\nu
l+\frac12}(qr_0)K_{\nu l+\frac12}(qr')K_{\nu l-\frac12}(qr')\right],
\end{multline}
and ${\tilde C}_{\nu l+\frac12}(v) $ is given by \eqref{4.22}.

In the case of $\nu> 1$ and $0 < F < \frac12\left(1-\frac1\nu \right)$, the
induced vacuum magnetic field is given by \eqref{5.2} with
$B^{(a)}_{\rm I}$ given by the right-hand side of \eqref{5.3} and $B^{(b)}_{\rm I}$ given
by the right-hand side of \eqref{5.6}.
In the case of $\nu> 1$ and $\frac12\left(1+\frac1\nu \right) < F < 1$,
the induced vacuum magnetic field is given by \eqref{5.2} with
$B^{(a)}_{\rm I}$ given by the right-hand side of \eqref{5.4} and
$B^{(b)}_{\rm I}$ given by the right-hand side of \eqref{5.7}.

Turning to the total flux of the induced vacuum magnetic field, see
\eqref{1.21}, we present it as
\begin{equation}\label{5.10}
\Phi_{\rm I}=\Phi_{\rm I}^{(a)} + \Phi_{\rm I}^{(b)},
\end{equation}
where
\begin{equation}\label{5.11}
\Phi_{\rm I}^{(a)} = \frac{2\pi}{\nu}\int\limits_{r_0}^\infty dr\, r
B_{\rm I}^{(a)}(r),\qquad \Phi_{\rm I}^{(b)} =
\frac{2\pi}{\nu}\int\limits_{r_0}^\infty dr\, r B_{\rm I}^{(b)}(r).
\end{equation}
We obtain in the case of $\nu \geq 1$ and $\frac12\left(1-\frac1\nu \right) < F <
\frac12\left(1+\frac1\nu \right)$, or $\frac12 \leq \nu < 1$ and
$\frac12\left(\frac1\nu -1\right) < F < \frac12\left(3-\frac1\nu
\right)$:
\begin{multline}\label{5.12}
\left. \Phi^{(a)}_{\rm I}\right|_{F<1/2, \,\theta\neq
-\pi/2}=\left. \Phi^{(a)}_{\rm I}(r)\right|_{F>1/2, \,\theta =
\pi/2}= -\frac{e}{8\pi}\frac1m\left\{
\int\limits_0^\infty \frac{du}{\cosh^3(u/2)}\,{\rm e}^{-2mr_0
\cosh(u/2)}\right. \\
\times \frac{\cos\left[\nu\left(F-\frac12\right)\pi\right] \cosh\left[\nu\left(F+\frac12 \right)u\right]
- \cos[\nu\left(F+\frac12 \right)\pi)]
\cosh\left[\nu\left(F-\frac12\right)u\right]}{\cosh(\nu u)-\cos(\nu
\pi)}\\
\left.- \frac{2\pi}{\nu}\sum_{p=1}^{\left[\!\left| {\nu}/2
\right|\!\right]} \exp[-2mr_0\sin(p\pi/\nu)]
\,\frac{\sin[(2F-1)p\pi]}{\sin^3(p\pi/\nu)} + \frac{\pi}{2N} \left(-1\right)^{N}\sin\left(2NF \pi \right) {\rm e}^{-2mr_0}
\, \delta_{\nu, \, 2N}\right\},
\end{multline}
\begin{multline}\label{5.13}
\left. \Phi^{(a)}_{\rm I}\right|_{F>1/2, \,\theta\neq
\pi/2}=\left. \Phi^{(a)}_{\rm I}(r)\right|_{F<1/2, \,\theta =-
\pi/2}= \frac{e}{8\pi}\frac1m\left\{
\int\limits_0^\infty \frac{du}{\cosh^3(u/2)}\,{\rm e}^{-2mr_0
\cosh(u/2)}\right. \\
\times \frac{\cos\left[\nu\left(F-\frac12\right)\pi\right] \cosh\left[\nu\left(F-\frac32 \right)u\right]
- \cos[\nu\left(F-\frac32 \right)\pi)]
\cosh\left[\nu\left(F-\frac12\right)u\right]}{\cosh(\nu u)-\cos(\nu
\pi)}\\
\left.+ \frac{2\pi}{\nu}\sum_{p=1}^{\left[\!\left| {\nu}/2
\right|\!\right]} \exp[-2mr_0\sin(p\pi/\nu)]
\,\frac{\sin[(2F-1)p\pi]}{\sin^3(p\pi/\nu)} - \frac{\pi}{2N} \left(-1\right)^{N}\sin\left(2NF \pi \right) {\rm e}^{-2mr_0}
\, \delta_{\nu, \, 2N}\right\},
\end{multline}
and
\begin{equation}\label{5.14}
\left. \Phi^{(a)}_{\rm I}\right|_{F=1/2} = -\frac{ e
\sin\theta}{8\pi} \int\limits_m^\infty \!\! \frac{dq}{\sqrt{q^2-m^2}}\,\frac{1}{q+m\cos\theta} \left[
\Gamma(2,2qr_0) - 4q^2 r_0^2\,\Gamma(0,2qr_0)\right].
\end{equation}
In the case of $\nu> 1$ and $0 < F < \frac12\left(1-\frac1\nu \right)$,
$\Phi^{(a)}_{\rm I}$ is given by the right-hand side of \eqref{5.12}.
In the case of $\nu> 1$ and $\frac12\left(1+\frac1\nu \right) < F < 1$,
$\Phi^{(a)}_{\rm I}$ is given by the right-hand side of \eqref{5.13}.

As follows from \eqref{5.12}-\eqref{5.14}, $\Phi^{(a)}_{\rm I}$ is damped exponentially at $r_0\rightarrow \infty$.
In the case of the opposite limit, i.e., at $r_0\rightarrow 0$, all integrations in $\Phi_{\rm I}^{(a)} $ can be
explicitly performed. It is straightforward to obtain
\begin{equation}\label{5.14a}
\left.\lim_{r_0\rightarrow 0} \Phi_{\rm I}^{(a)}\right|_{F= 1/2}= -\frac{e}{4\pi
m}\arctan\left(\tan\frac\theta2\right).
\end{equation}
The analysis at $F \neq 1/2$ requires more efforts. Let us consider first the case of $\nu>1$ and
present $j^{(a)}_{\varphi}$ defined in \eqref{4.14} as
\begin{multline}\label{5.15}
\hspace{-1em}\left.j^{(a)}_{\varphi}(r)\right|_{\frac12\left(1-\frac1{\nu}\right)<F<\frac12, \,
\theta \neq -\frac{\pi}2}\!=\!
\left.j^{(a)}_{\varphi}(r)\right|_{\frac12<F<\frac12\left(1+\frac1{\nu}\right), \, \theta =
\frac{\pi}2}\!=\!\left.j^{(a)}_{\varphi}(r)\right|_{0 < F <\frac12\left(1-\frac1{\nu}\right)}\\
=\!j^{(1,1)}_\varphi(r)+j^{(1,2)}_\varphi(r)-2j^{(1,3)}_\varphi(r)
=\frac{m}{8\pi}\frac1{2\pi {\rm
i}}\int\limits_{C_0}dz\left[1+\frac1{2mr
\sqrt{-\sinh^2(z/2)}}\right]\\
\times\exp\left[-2mr\sqrt{-\sinh^2(z/2)}\right]\frac{\sinh(\nu Fz)
}{\sinh(z/2)\sinh(\nu z/2)}
\end{multline}
and
\begin{multline}\label{5.16}
\left.j^{(a)}_{\varphi}(r)\right|_{\frac12<F<\frac12\left(1+\frac1{\nu}\right), \, \theta \neq
\frac{\pi}2}= \left.j^{(a)}_{\varphi}(r)\right|_{\frac12\left(1-\frac1{\nu}\right)<F< \frac12, \,
\theta =-
\frac{\pi}2}=\left.j^{(a)}_{\varphi}(r)\right|_{\frac12\left(1+\frac1{\nu}\right)<F < 1}\\
=j^{(1,1)}_\varphi(r)+j^{(1,2)}_\varphi(r)
=-\frac{m}{8\pi}\frac1{2\pi {\rm i}}\int\limits_{
C_0}dz\left[1+\frac1{2mr \sqrt{-\sinh^2(z/2)}}\right]\\
\times\exp\left[-2mr\sqrt{-\sinh^2(z/2)}\right]
\frac{\sinh\left[\nu\left(1-F\right)z\right]
}{\sinh(z/2)\sinh(\nu z/2)},
\end{multline}
where $j^{(1,1)}_\varphi$, $j^{(1,2)}_\varphi$, and
$j^{(1,3)}_\varphi$ are given by \eqref{a13}, \eqref{a15}, and
\eqref{a18} in Appendix A, contour $C_0$ in the complex $z$ plane is
depicted in Fig. 8. Consequently, we get
\begin{multline}\label{5.17}
\left.B^{(a)}_{\rm I}(r)\right|_{\frac12\left(1-\frac1{\nu}\right)<F<\frac12, \,
\theta \neq -\frac{\pi}2}=
\left.B^{(a)}_{\rm I}(r)\right|_{\frac12<F<\frac12\left(1+\frac1{\nu}\right), \, \theta =
\frac{\pi}2}=\left.B^{(a)}_{\rm I}(r)\right|_{0 < F <\frac12\left(1-\frac1{\nu}\right)}\\
=\frac{\nu e}{16\pi r}\frac1{2\pi {\rm i}}\int\limits_{C_0}dz\,
\frac{\exp\left[-2mr\sqrt{-\sinh^2(z/2)}\right]}{\sqrt{-\sinh^2(z/2)}}\,
\frac{\sinh(\nu Fz)
}{\sinh(z/2)\sinh(\nu z/2)},
\end{multline}
\begin{multline}\label{5.18}
\left.B^{(a)}_{\rm I}(r)\right|_{\frac12<F<\frac12\left(1+\frac1{\nu}\right), \, \theta \neq
\frac{\pi}2}= \left.B^{(a)}_{\rm I}(r)\right|_{\frac12\left(1-\frac1{\nu}\right)<F< \frac12, \,
\theta =-
\frac{\pi}2}=\left.B^{(a)}_{\rm I}(r)\right|_{\frac12\left(1+\frac1{\nu}\right)<F < 1}\\
=-\frac{\nu e}{16\pi r}\frac1{2\pi {\rm i}}\int\limits_{ C_0}dz\,
\frac{\exp\left[-2mr\sqrt{-\sinh^2(z/2)}\right]}{\sqrt{-\sinh^2(z/2)}}\,
\frac{\sinh\left[\nu\left(1-F\right)z\right]
}{\sinh(z/2)\sinh(\nu z/2)},
\end{multline}
\begin{multline}\label{5.19}
\left. \lim_{r_0\rightarrow 0}\Phi^{(a)}_{\rm I}\right|_{\frac12\left(1-\frac1{\nu}\right)<F<\frac12, \,
\theta \neq -\frac{\pi}2}=
\left.\lim_{r_0\rightarrow 0}\Phi^{(a)}_{\rm I}\right|_{\frac12<F<\frac12\left(1+\frac1{\nu}\right), \, \theta =
\frac{\pi}2}=\left.\lim_{r_0\rightarrow 0}\Phi^{(a)}_{\rm I}\right|_{0 < F <\frac12\left(1-\frac1{\nu}\right)}\\=- \frac{e}{16m}\,\frac1{2\pi {\rm
i}}\int\limits_{C_0}
dz\,\frac{\sinh(\nu Fz)}{\sinh^3(z/2)\sinh(\nu
z/2)},
\end{multline}
and
\begin{multline}\label{5.20}
\left. \lim_{r_0\rightarrow 0}\Phi^{(a)}_{\rm I}\right|_{\frac12<F<\frac12\left(1+\frac1{\nu}\right), \, \theta \neq
\frac{\pi}2}= \left.\lim_{r_0\rightarrow 0}\Phi^{(a)}_{\rm I}\right|_{\frac12\left(1-\frac1{\nu}\right)<F< \frac12, \,
\theta =-
\frac{\pi}2}=\left.\lim_{r_0\rightarrow 0}\Phi^{(a)}_{\rm I}\right|_{\frac12\left(1+\frac1{\nu}\right)<F < 1}\\
=\frac{e}{16m}\,\frac1{2\pi {\rm
i}}\int\limits_{C_0}
dz\,\frac{\sinh\left[\nu\left(1-F\right)z\right]}{\sinh^3(z/2)\sinh(\nu
z/2)}.
\end{multline}
Only a simple pole at $z=0$ of the integrands contributes to the integrals in
\eqref{5.19} and \eqref{5.20}. Calculation of its residue yields
\begin{multline}\label{5.21}
\left. \lim_{r_0\rightarrow 0} \Phi^{(a)}_{\rm I}\right|_{ \frac12\left(1-\frac1{\nu}\right)<F<\frac12, \,
\theta \neq -\frac{\pi}2}=
\left. \lim_{r_0\rightarrow 0} \Phi^{(a)}_{\rm I}\right|_{\frac12<F<\frac12\left(1+\frac1{\nu}\right), \, \theta =
\frac{\pi}2}=\left. \lim_{r_0\rightarrow 0} \Phi^{(a)}_{\rm I}\right|_{0 < F <\frac12\left(1-\frac1{\nu}\right)}\\
= -\frac{e}{6m} F \left[\frac14
\left(\nu^2 + 3\right) -\nu^2 F^2 \right]
\end{multline}
and
\begin{multline}\label{5.22}
\left. \lim_{r_0\rightarrow 0} \Phi^{(a)}_{\rm I}\right|_{\frac12<F<\frac12\left(1+\frac1{\nu}\right), \, \theta \neq
\frac{\pi}2}= \left. \lim_{r_0\rightarrow 0} \Phi^{(a)}_{\rm I}\right|_{\frac12\left(1-\frac1{\nu}\right)<F< \frac12, \,
\theta =-
\frac{\pi}2}=\left. \lim_{r_0\rightarrow 0} \Phi^{(a)}_{\rm I}\right|_{\frac12\left(1+\frac1{\nu}\right)<F < 1}\\
= \frac{e}{6m} \left(1-F\right)\left[\frac14
\left(\nu^2 + 3\right) -\nu^2 \left(1-F\right)^2 \right].
\end{multline}
Considering the case of $\frac12<\nu \leq 1$ and
$\frac12\left(\frac1\nu-1 \right) < F < \frac12\left(3-\frac1\nu
\right)$ at $F \neq 1/2$, we obtain that $$\left.
\lim_{r_0\rightarrow 0} \Phi^{(a)}_{\rm
I}\right|_{\frac12\left(\frac1\nu-1 \right) < F <\frac12,\, \theta
\neq -\frac\pi2} = \left. \lim_{r_0\rightarrow 0} \Phi^{(a)}_{\rm
I}\right|_{ \frac12<F < \frac12\left(3-\frac1\nu \right),\,
\theta=\frac\pi2}$$ and $$\left. \lim_{r_0\rightarrow 0}
\Phi^{(a)}_{\rm I}\right|_{\frac12<F < \frac12\left(3-\frac1\nu
\right),\, \theta \neq \frac\pi2} = \left. \lim_{r_0\rightarrow 0}
\Phi^{(a)}_{\rm I}\right|_{ \frac12\left(\frac1\nu-1 \right) < F
<\frac12,\, \theta = -\frac\pi2}$$ are given by the right-hand sides
of \eqref{5.21} and \eqref{5.22}, respectively. Note that $\left.
\lim_{r_0\rightarrow 0} \Phi^{(a)}_{\rm I}\right|_{\theta \neq \pm
\pi/2}$ is discontinuous at $F=1/2$ and its limiting values are
independent of $\nu$:
\begin{equation}\label{5.231}
\left.\lim_{F\rightarrow (1/2)_{+}}
\lim_{r_0\rightarrow 0} \Phi^{(a)}_{\rm I}\right|_{\theta \neq \pm \pi/2}\!=\left.-\lim_{F\rightarrow (1/2)_{-}} \lim_{r_0\rightarrow 0} \Phi^{(a)}_{\rm I}\right|_{\theta \neq \pm \pi/2}\!=\frac{e}{16 m};
\end{equation}
this is clearly a consequence of \eqref{4.233}.

As for the remaining part of the total flux, $\Phi_{\rm I}^{(b)}$, it
can be presented as
\begin{equation}\label{5.23}
\Phi_{\rm I}^{(b)} = e \pi \int\limits_{r_0}^\infty \frac{dr}{r} \,
j^{(b)}_\varphi(r) \, (r^2-r_0^2),
\end{equation}
provided that the following condition holds:
\begin{equation}\label{5.24}
\lim_{r \rightarrow r_0}j^{(b)}_\varphi(r) \, (r-r_0)^2 = 0;
\end{equation}
otherwise, the total flux diverges.

\begin{figure}[h]
\begin{center}
\includegraphics[width=95mm]{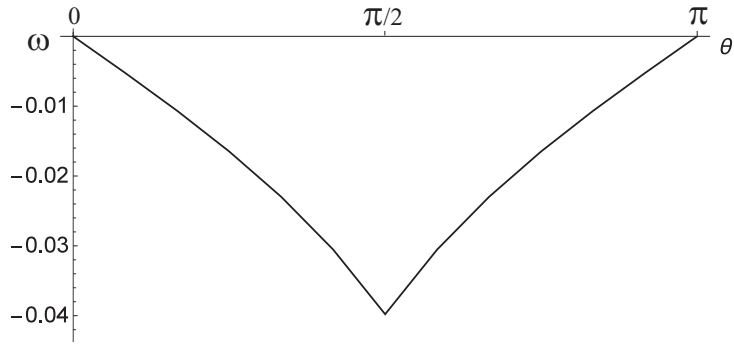}
\end{center}
\caption{${\displaystyle \omega(\theta)=\lim_{r \rightarrow r_0} \nu r
j_\varphi (r) \left(\frac{r-r_0}{r_0}\right)^2 }$ is the same at
$\nu=3/4, 1, 2, 3, 5, 10$, $m r_0= 10^{-5}, 10^{-3}, 10^{-2},10^{-1},1 $, and
different values of $F$.}\label{Triangle}
\end{figure}

By performing a numerical analysis, we find that quantity
${\displaystyle \lim_{r \rightarrow r_0} \nu r j_\varphi (r)
\left(\frac{r-r_0}{r_0}\right)^2 }$ depends on $\theta$, actually being independent of other parameters ($F$, $\nu$ and $m r_0$);
see Fig. 1. As follows from this analysis, relation \eqref{5.24} is
fulfilled in cases $\theta=0$ and $\theta=\pi$ only. The case of
$F=1/2$ needs a special comment, since, due to relation \eqref{4.24},
the current in this case is odd function of $\theta$. Whereas the
current and, consequently, the induced magnetic field with its flux
vanish at $\theta=0$, they can be nonvanishing if discontinuous at $\theta=\pi$. Indeed, periodicity in $\theta$ with period $2\pi$,
\begin{equation}\label{5.23a}
\left.j_\varphi(r)\right|_{F=1/2,\,\theta=\pi_\pm} =\left.j_\varphi(r)\right|_{F=1/2,\,\theta=-\pi_\mp}
\end{equation}
together with oddness in $\theta$,
\begin{equation}\label{5.23b}
\left.j_\varphi(r)\right|_{F=1/2,\,\theta=\pi_\pm} = -\left.j_\varphi(r)\right|_{F=1/2,\,\theta=-\pi_\pm}
\end{equation}
results in
\begin{equation}\label{5.23c}
\left.j_\varphi(r)\right|_{F=1/2,\,\theta=\pi_\pm} = - \left.j_\varphi(r)\right|_{F=1/2,\,\theta=\pi_\mp}.
\end{equation}
Namely this is obtained from the appropriate formulas at $F=1/2$ and $\theta=\pi_\pm$:
\begin{equation}\label{5.24a}
\left.j_\varphi(r)\right|_{F=1/2,\,\theta=\pi_\pm}=\pm\frac{m}{2\pi}\,e^{2m(r_0-r)}.
\end{equation}
As a consequence, we obtain
\begin{equation}\label{5.24b}
\left.B_{\rm I}(r)\right|_{F=1/2,\,\theta=\pi_\pm}=\pm\frac{e\nu
m}{2\pi}\,e^{2mr_0}\Gamma(0,2mr)
\end{equation}
and
\begin{equation}\label{5.24c}
\left.\Phi_{\rm
I}\right|_{F=1/2,\,\theta=\pi_\pm}=\pm\frac{e}{8m}\,e^{2mr_0}\left[\Gamma(2,2mr_0)-4m^2r_0^2\Gamma(0,2mr_0)\right];
\end{equation}
in particular, cf. \eqref{5.14a},
\begin{equation}\label{5.24d}
\lim_{r_0\rightarrow0}\left.\Phi_{\rm
I}\right|_{F=1/2,\,\theta=\pi_\pm}=\pm\frac{e}{8m}.
\end{equation}
In the case of $F\neq 1/2$ continuity in $\theta$ is maintained, and the induced vacuum current in this case takes form
\begin{multline}\label{5.25}
\left.j_\varphi(r)\right|_{\theta=\frac{\pi}{2} \mp \frac{\pi}{2}}\!=\!\frac{m}{(2\pi)^2}\Biggl\{ {\rm
sgn}\left(F-\frac12\right)\int\limits_0^\infty
\frac{du}{\cosh(u/2)}\left[1+\frac1{2mr \cosh(u/2)}\right]{\rm
e}^{-2mr
\cosh(u/2)}\Biggr. \\
\times \frac{\cos\left[\nu\left(F-\frac12\right)\pi)\right]
\cosh\left[\nu\left(\left|F-\frac12\right|-1\right)u\right]-
\cos\left[\nu\left(\left|F-\frac12\right|-1\right)\pi\right]\cosh\left[\nu\left(F-\frac12\right)u\right]}{\cosh(\nu
u)-\cos(\nu \pi)} \\
+ \frac{2\pi}{\nu}\sum_{p=1}^{\left[\!\left| {\nu}/2
\right|\!\right]} \left[1 +\frac{1}{2mr\sin(p\pi/\nu)}\right]\,
\exp[-2mr\sin(p\pi/\nu)]
\,\frac{\sin[(2F-1)p\pi]}{\sin(p\pi/\nu)}\\
\Biggl. - \frac{\pi}{2N} \left(-1\right)^{N}\sin\left(2NF \pi \right)
\left(1 +\frac{1}{2mr}\right){\rm e}^{-2mr} \, \delta_{\nu, \, 2N}\Biggr\}\\
+\frac{r}{\pi^2}\!\int\limits_m^\infty \!\! \frac{dq\,
q^2}{\sqrt{q^2-m^2}}\Biggl\{\frac12\Biggl[C^{(\pm)}_{\frac12+\nu\left(F-\frac12\right)}(qr_0)-C^{(\pm)}_{\frac12-\nu\left(F-\frac12\right)}(qr_0)\Biggr.\Biggr. \\
\Biggl. +{\rm
sgn}\left(F-\frac12\right)\Biggl(C^{(\pm)}_{\frac12+\nu\left(F-\frac12\right)}(qr_0)+C^{(\pm)}_{\frac12-\nu\left(F-\frac12\right)}(qr_0)\Biggr)\Biggr]K_{\frac12+\nu\left(F-\frac12\right)}(qr)
K_{\frac12-\nu\left(F-\frac12\right)}(qr) \\
 + \sum_{l=1}^\infty \Biggl[C^{(\pm)}_{\nu \left(l+F-\frac12\right)+\frac12}(qr_0) K_{\nu
\left(l+F-\frac12\right)+\frac12}(qr)K_{\nu \left(l+F-\frac12\right)-\frac12}(qr)\Biggr. \\
\Biggl.\Biggl. - C^{(\pm)}_{\nu \left(l-F+\frac12\right)+\frac12}(qr_0)
K_{\nu \left(l-F+\frac12\right)+\frac12} (qr)
K_{\nu \left(l-F+\frac12\right)-\frac12}(qr)\Biggr]\Biggr\},
\end{multline}
where\newpage

\phantom{hhh}

\vspace{-3.5 em}

\begin{multline}\label{5.26}
C^{(\pm)}_\rho(v)=\left\{v
I_\rho(v)K_\rho(v) \pm mr_0\left[I_\rho(v)K_{\rho-1}(v)-I_{\rho-1}(v)K_\rho(v)\right] -
vI_{\rho-1}(v)K_{\rho-1}(v)
\right\} \\
\times \left[v
K^2_\rho(v)\pm 2mr_0K_\rho(v)K_{\rho-1}(v) + vK^2_{\rho-1}(v)\right]^{-1}.
\end{multline}
Consequently, we obtain the following expressions for the induced
vacuum magnetic field,
\begin{multline}\label{5.27}
\left. B_{\rm I}(r)\right|_{\theta=\frac{\pi}{2} \mp \frac{\pi}{2}} = \frac{\nu e}{2(2\pi)^2}\frac1r\left\{ {\rm
sgn}\left(F-\frac12\right)
\int\limits_0^\infty \frac{du}{\cosh^2(u/2)}\,{\rm e}^{-2mr
\cosh(u/2)}\right. \\
\times \frac{\cos\left[\nu\left(F-\frac12\right)\pi)\right]
\cosh\left[\nu\left(\left|F-\frac12\right|-1\right)u\right]-
\cos\left[\nu\left(\left|F-\frac12\right|-1\right)\pi\right]\cosh\left[\nu\left(F-\frac12\right)u\right]}{\cosh(\nu u)-\cos(\nu
\pi)}\\
\left.+ \frac{2\pi}{\nu}\sum_{p=1}^{\left[\!\left| {\nu}/2
\right|\!\right]} \exp[-2mr\sin(p\pi/\nu)]
\,\frac{\sin[(2F-1)p\pi]}{\sin^2(p\pi/\nu)} - \frac{\pi}{2N} \left(-1\right)^{N}\sin\left(2NF \pi \right) {\rm e}^{-2mr}
\, \delta_{\nu, \, 2N}\right\}\\
+\frac{\nu e}{\pi^2}\int\limits_r^\infty dr'\!\int\limits_m^\infty \!\! \frac{dq\,
q^2}{\sqrt{q^2-m^2}}\Biggl\{\frac12\Biggl[C^{(\pm)}_{\frac12+\nu\left(F-\frac12\right)}(qr_0)-C^{(\pm)}_{\frac12-\nu\left(F-\frac12\right)}(qr_0)\Biggr.\Biggr. \\
\Biggl. +\,{\rm
sgn}\left(F-\frac12\right)\Biggl(C^{(\pm)}_{\frac12+\nu\left(F-\frac12\right)}(qr_0)+C^{(\pm)}_{\frac12-\nu\left(F-\frac12\right)}(qr_0)\Biggr)\Biggr]K_{\frac12+\nu\left(F-\frac12\right)}(qr')
K_{\frac12-\nu\left(F-\frac12\right)}(qr') \\
 + \sum_{l=1}^\infty \Biggl[C^{(\pm)}_{\nu \left(l+F-\frac12\right)+\frac12}(qr_0) K_{\nu
\left(l+F-\frac12\right)+\frac12}(qr')K_{\nu \left(l+F-\frac12\right)-\frac12}(qr')\Biggr. \\
\Biggl.\Biggl. - C^{(\pm)}_{\nu \left(l-F+\frac12\right)+\frac12}(qr_0)
K_{\nu \left(l-F+\frac12\right)+\frac12} (qr')
K_{\nu \left(l-F+\frac12\right)-\frac12}(qr')\Biggr]\Biggr\}, \quad F \neq 1/2,
\end{multline}
and its flux,
\begin{multline}\label{5.28}
\left. \Phi_{\rm I}\right|_{\theta = \frac{\pi}{2} \mp
\frac{\pi}{2}} = \frac{e}{8\pi}\frac1m\left\{ {\rm
sgn}\left(F-\frac12\right) \int\limits_0^\infty
\frac{du}{\cosh^3(u/2)}\,{\rm e}^{-2mr_0
\cosh(u/2)}\right. \\
\times \frac{\cos\left[\nu\left(F-\frac12\right)\pi)\right]
\cosh\left[\nu\left(\left|F-\frac12\right|-1\right)u\right]-
\cos\left[\nu\left(\left|F-\frac12\right|-1\right)\pi\right]\cosh\left[\nu\left(F-\frac12\right)u\right]}{\cosh(\nu u)-\cos(\nu
\pi)}\\
\left.+ \frac{2\pi}{\nu}\sum_{p=1}^{\left[\!\left| {\nu}/2
\right|\!\right]} \exp[-2mr_0\sin(p\pi/\nu)]
\,\frac{\sin[(2F-1)p\pi]}{\sin^3(p\pi/\nu)} - \frac{\pi}{2N}
\left(-1\right)^{N}\sin\left(2NF \pi \right) {\rm e}^{-2mr_0} \,
\delta_{\nu, \, 2N}\right\}\\
 + \frac{e}{\pi}\int\limits_{r_0}^\infty
dr\,(r^2-r_0^2)\!\int\limits_m^\infty \!\! \frac{dq\,
q^2}{\sqrt{q^2-m^2}}\Biggl\{\frac12\Biggl[C^{(\pm)}_{\frac12+\nu\left(F-\frac12\right)}(qr_0)-C^{(\pm)}_{\frac12-\nu\left(F-\frac12\right)}(qr_0)\Biggr.\Biggr. \\
\Biggl. +\,{\rm
sgn}\left(F-\frac12\right)\Biggl(C^{(\pm)}_{\frac12+\nu\left(F-\frac12\right)}(qr_0)+C^{(\pm)}_{\frac12-\nu\left(F-\frac12\right)}(qr_0)\Biggr)\Biggr]K_{\frac12+\nu\left(F-\frac12\right)}(qr)
K_{\frac12-\nu\left(F-\frac12\right)}(qr) \\
 + \sum_{l=1}^\infty \Biggl[C^{(\pm)}_{\nu \left(l+F-\frac12\right)+\frac12}(qr_0) K_{\nu
\left(l+F-\frac12\right)+\frac12}(qr)K_{\nu \left(l+F-\frac12\right)-\frac12}(qr)\Biggr. \\
\Biggl.\Biggl. - C^{(\pm)}_{\nu \left(l-F+\frac12\right)+\frac12}(qr_0)
K_{\nu \left(l-F+\frac12\right)+\frac12} (qr)
K_{\nu \left(l-F+\frac12\right)-\frac12}(qr)\Biggr]\Biggr\}, \quad F \neq 1/2.
\end{multline}

With the use of relations (see \cite{Prud})
\begin{equation*}
\int\limits_{v}^\infty dw \,w
K^2_\rho(w) = \frac{v^2}{2}\left[\frac{d}{d v} K_{\rho}(v) \right]^2 -
\frac{v^2 + {\rho}^2}{2} K_\rho^2(v),
\end{equation*}
\begin{equation*}
\int\limits_{v}^\infty \frac{dw}{w} \,
K_\rho(w) K_{\rho\,'}(w) =  \frac{v}{{\rho}^2 - {\rho\,'}^2}
\left[ K_{\rho}(v) \frac{d}{d v} K_{\rho\,'}(v) -
K_{\rho\,'}(v) \frac{d}{d v} K_{\rho}(v)\right]
\end{equation*}
and the Schl$\ddot{a}$fli contour integral representation for the
Macdonald function,
\footnote{There are no poles in this case, and contour $C_0$ can be straightened to two horizontal lines at $z=\pm {\rm i}\pi$.}
\begin{equation*}
K_\rho(w) = \frac1{4{\rm i} \sin(\rho \pi)} \int\limits_{C_0}
dz\, {\rm e}^{w \cosh z+\rho z },
\end{equation*}
the  integration over $r$ can be performed. As a result, we obtain the following representation for the induced vacuum magnetic flux
\begin{multline}\label{5.29}
\left. \Phi_{\rm I}\right|_{\theta = \frac{\pi}{2} \mp
\frac{\pi}{2}} = \frac{e}{8\pi}\frac1m\left\{ {\rm
sgn}\left(F-\frac12\right) \int\limits_0^\infty
\frac{du}{\cosh^3(u/2)}\,{\rm e}^{-2mr_0
\cosh(u/2)}\right. \\
\times \frac{\cos\left[\nu\left(F-\frac12\right)\pi)\right]
\cosh\left[\nu\left(\left|F-\frac12\right|-1\right)u\right]-
\cos\left[\nu\left(\left|F-\frac12\right|-1\right)\pi\right]\cosh\left[\nu\left(F-\frac12\right)u\right]}{\cosh(\nu u)-\cos(\nu
\pi)}\\
\left.+ \frac{2\pi}{\nu}\sum_{p=1}^{\left[\!\left| {\nu}/2
\right|\!\right]} \exp[-2mr_0\sin(p\pi/\nu)]
\,\frac{\sin[(2F-1)p\pi]}{\sin^3(p\pi/\nu)} - \frac{\pi}{2N}
\left(-1\right)^{N}\sin\left(2NF \pi \right) {\rm e}^{-2mr_0} \,
\delta_{\nu, \, 2N}\right\} \\ + \frac{e}{2\pi} r_0 \int\limits_{m
r_0}^\infty
 \frac{dv\, v}{\sqrt{v^2-m^2 r_0^2}}\Biggl\{\frac12\Biggl[C^{(\pm)}_{\frac12+\nu\left(F-\frac12\right)}(v)-C^{(\pm)}_{\frac12-\nu\left(F-\frac12\right)}(v)\Biggr.\Biggr. \\
\Biggl. +\,{\rm
sgn}\left(F-\frac12\right)\Biggl(C^{(\pm)}_{\frac12+\nu\left(F-\frac12\right)}(v)+C^{(\pm)}_{\frac12-\nu\left(F-\frac12\right)}(v)\Biggr)\Biggr]
D_{\frac12+\nu\left|F-\frac12\right|}(v)\\
\Biggl. + \sum_{l=1}^\infty \Biggl[C^{(\pm)}_{\nu \left(l+F-\frac12\right)+\frac12}(v) D_{\nu
\left(l+F-\frac12\right)+\frac12}(v)
 - C^{(\pm)}_{\nu \left(l-F+\frac12\right)+\frac12}(v)
D_{\nu \left(l-F+\frac12\right)+\frac12}(v) \Biggr]\Biggr\}, \quad F \neq 1/2,
\end{multline}
where
\begin{equation}\label{5.30}
D_{\rho}(v) = \rho K_\rho^2(v)-(\rho-1)K_{\rho+1}(v)K_{\rho-1}(v)
+ v \left[ K_{\rho}(v)\frac{d}{d \rho} K_{\rho-1}(v) -
K_{\rho-1}(v)\frac{d}{d \rho} K_{\rho}(v) \right];
\end{equation}
in particular,
\begin{multline}\label{5.31}
\left. \lim_{r_0\rightarrow 0} \Phi_{\rm I}\right|_{\theta =
\frac{\pi}{2} \mp \frac{\pi}{2}} = -\frac{e}{6m} \left[F -\frac12 -
\frac12{\rm sgn}\left(F-\frac12\right)\right] \Biggl\{\frac34 -\nu^2
\left[\frac14 - \left|F -\frac12\right| - F(1-F)\right] \Biggr\},\\
F \neq 1/2.
\end{multline}

As follows from \eqref{4.24}, $\left.j_\varphi(r)\right|_{F\neq
\frac12, \, \theta= \frac{\pi}{2} \mp \frac{\pi}{2}}$ and,
consequently, $\left.\Phi_{\rm I}\right|_{F\neq \frac12, \, \theta =
\frac{\pi}{2} \mp \frac{\pi}{2}}$ changes signs under $F \rightarrow
1-F$. To be more precise, the dimensionless induced vacuum magnetic
flux, $e^{-1}m\left.\Phi_{\rm I}\right|_{F\neq \frac12, \, \theta =
\frac{\pi}{2} \mp \frac{\pi}{2}}$, is positive at $F>1/2$ and
negative at $F<1/2$; its absolute value increases with the increase
of $\nu$. Whereas $\left.\Phi_{\rm I}\right|_{F = \frac12, \, \theta
= 0}$ vanishes, $e^{-1}m\left.\Phi_{\rm I}\right|_{F = \frac12, \,
\theta = \pi_+}$ is positive and $e^{-1}m\left.\Phi_{\rm
I}\right|_{F = \frac12, \, \theta = \pi_-}$ is negative, being of
the same absolute value that is independent of $\nu$, see
\eqref{5.24c}. Continuity of the results in $\theta$ is broken at
$\theta = \pi$ and $F = 1/2$ only.

\begin{figure}[t]
\begin{center}
\includegraphics[width=80mm]{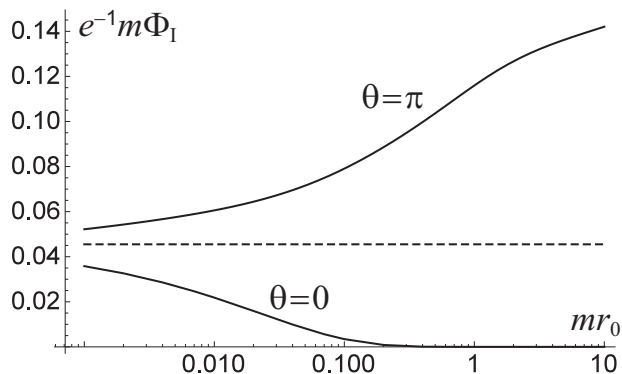}
\end{center}
\caption{The dimensionless induced flux, $e^{-1}m\left.\Phi_{\rm
I}\right|_{\theta = \frac{\pi}{2} \mp \frac{\pi}{2}}$, as a function
of $m r_0$ at $\nu=1$ and $F=0.7$ (solid lines); the dashed line
corresponds to the case of $m r_0 = 0$.}\label{fig2}
\end{figure}

A more detailed analysis of the behavior of the induced vacuum
magnetic flux can be obtained with the use of numerical
computations. Taking $F=0.7$ and $\nu=1$, we calculate the
dimensionless flux, $e^{-1}m\left.\Phi_{\rm I}\right|_{\theta =
\frac{\pi}{2} \mp \frac{\pi}{2}}$, as a function of $m r_0$, see
Fig. 2. In the case of $\theta = 0$, this function decreases with the
increase of $m r_0$, becoming vanishingly small ($\lesssim 10^{-7}$)
at $m r_0 \geq 1$. On the contrary, in the case of $\theta = \pi$,
this function increases at no allowance with the increase of $m
r_0$.

The dimensionless flux in the case of $\theta = 0$ at several values
of $\nu$, as well as of $m r_0$, is presented as a function of $F$
in Fig. 3. As $m r_0$ increases, the absolute value of this function
decreases as compared to the value at $m r_0 = 0$, becoming
negligible in the vicinity of $F=1/2$. However, the vicinity is
shrinked as $\nu$ increases (this is also demonstrated by Fig. 4),
and the flux at $m r_0 \geq 1$ can equal its value at $m r_0 = 0$
for sufficiently large values of $\nu$, unless $F=1/2$.

The dimensionless flux in the case of $\theta = \pi$ at several
values of $\nu$, as well as of $m r_0$, is presented as a function
of $F$ in Fig. 5. In the case of $\frac12 < \nu \leq 2$ the absolute
value of this function increases with the increase of $m r_0$, see
Figs. 5a and 5b. However, in the case of $\nu > 2$, the increase
takes place in the vicinity of $F=1/2$ and, otherwise, there is a
decrease, see Figs. 5c and 5d (this is also demonstrated by
Fig. 6a). Note that the flux at large values of $m r_0$ fails to
depend on $\nu$ (lines corresponding to different values of $\nu$
merge together) at least in the case of $\frac12 <\nu \leq 4$, see
Fig. 6b. \pagebreak

\begin{figure}[t]
\includegraphics[width=160mm]{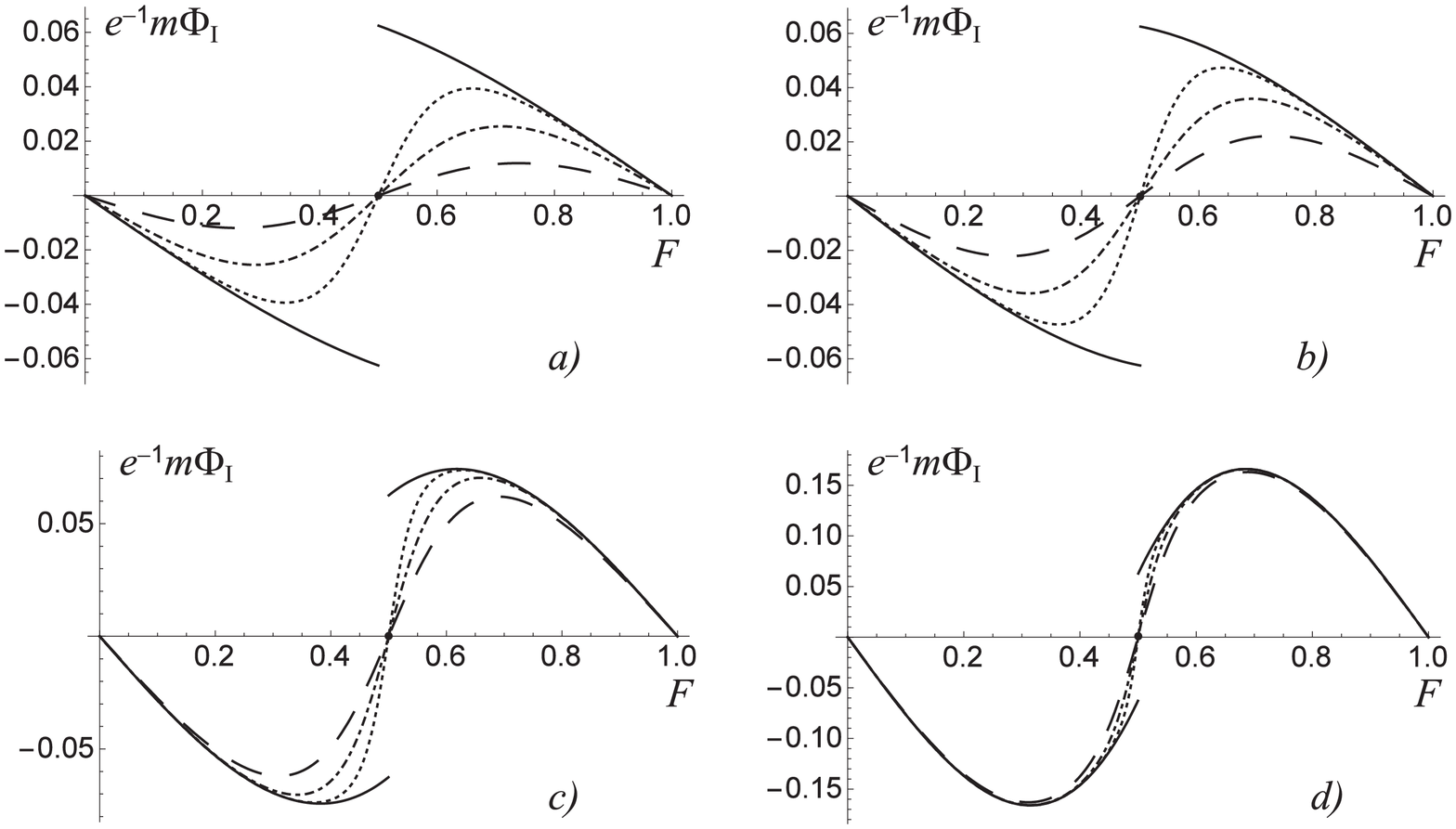}
\caption{The dimensionless induced flux at $\theta=0$ as a function
of $F$ in the cases of $m r_0 = 0$ (solid line), $m r_0 = 10^{-5}$
(dotted line), $m r_0 = 10^{-3}$ (dash-dotted line), and $m r_0 =
10^{-2}$ (dashed line):
a) $\nu=3/4$, b) $\nu=1$, c) $\nu=2$, d) $\nu=4$. The point at $F=1/2$ corresponds to the case of $m r_0 = 0$.}\label{fig3}
\end{figure}

\phantom{hvhvh}

\phantom{hfhfh}

\begin{figure}[h]
\begin{center}
\includegraphics[width=80mm]{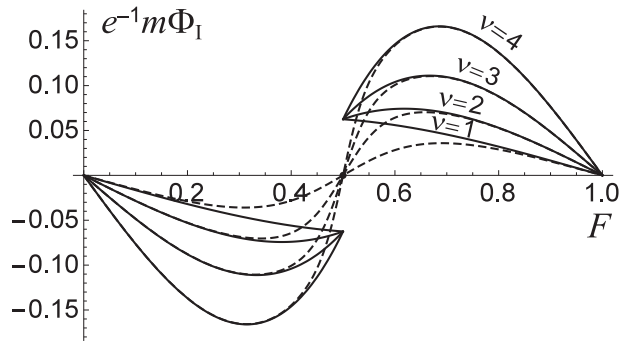}
\end{center}
\caption{The dimensionless induced flux at $\theta=0$ as a function
of $F$ in the cases of $m r_0 = 0$ (solid lines) and $m r_0 =
10^{-3}$ (dashed lines).}\label{fig4}
\end{figure}

\pagebreak

\begin{figure}[h]
\includegraphics[width=160mm]{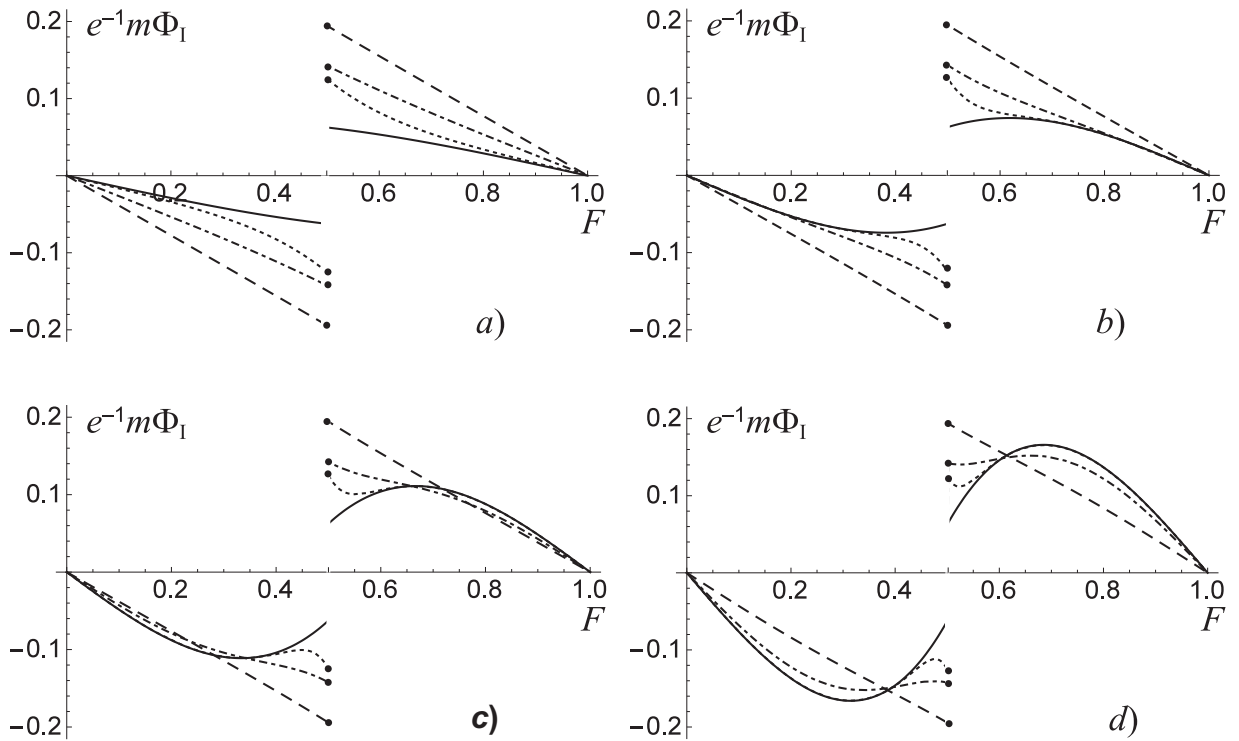}
\caption{The dimensionless induced flux at $\theta=\pi$ as a
function of $F$ in the cases of $m r_0 = 0$ (solid line), $m r_0 =
10^{-3}$ (dotted line), $m r_0 = 10^{-1}$ (dash-dotted line), and $m
r_0 = 1$ (dashed line): a) $\nu=1$, b) $\nu=2$,
c) $\nu=3$, d) $\nu=4$. The points at $F=1/2$
correspond to $\theta=\pi_+$ (positive values) and to $\theta=\pi_-$
(negative values), with the absolute values increasing with the
increase of $m r_0$. The point in the case of $m r_0 = 10^{-3}$ (or
less) actually coincides with the point in the case of $m r_0 = 0$;
moreover, in cases $m r_0 \gtrsim  10^{-3}$, the points at $F=1/2$
coincide with the end points of the appropriate curves at $F \neq
1/2$.} \label{fig5}
\end{figure}

\begin{figure}[t]
\begin{center}
\includegraphics[width=160mm]{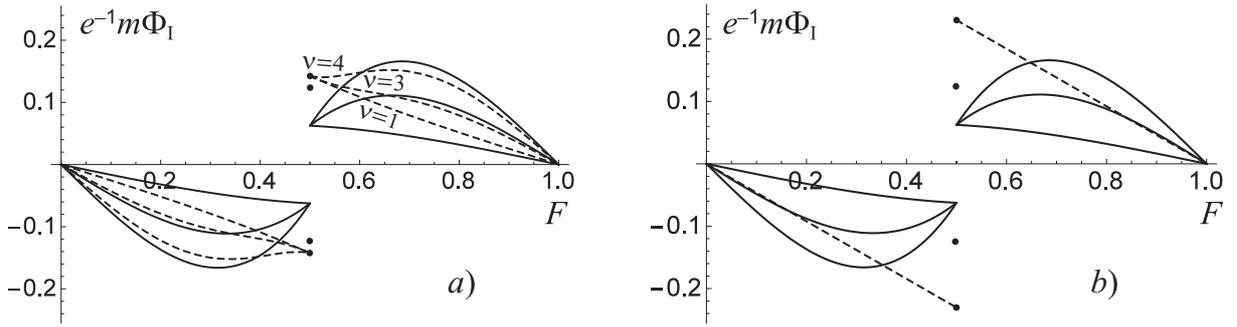}
\end{center} \caption{The dimensionless induced flux at $\theta=\pi$ as a
function of $F$: a) $m r_0 = 0$ (solid lines) and
 $m r_0 = 10^{-1}$ (dashed lines), b) $m r_0 = 0$ (solid lines) and $m r_0
= 5$ (dashed line).}\label{fig6}
\end{figure}

\section{Summary and discussion}
\setcounter{equation}{0}
\renewcommand{\theequation}{\arabic{section}.\arabic{equation}}

In the present paper, we have studied the impact of boundary
conditions at the edge of the ANO vortex core on the vacuum
polarization effects in quantum relativistic spinor matter in
two-dimensional space. The most general boundary condition that is
compatible with the self-adjointness of the energy operator in
first-quantized theory, Dirac hamiltonian \eqref{1.14}, is (see
\eqref{3.2} and \eqref{3.9})
\begin{equation}\label{7.1}
    (I-{\rm i} \beta\alpha^r\, e^{-{\rm i} \theta
    \alpha^r})\left.\psi\right|_{r=r_0}=0,
\end{equation}

\noindent where $\theta$ is the self-adjoint extension parameter.
This condition is also the most general one ensuring the
impenetrability of the vortex core edge, i.e., the confinement of the
matter field to the region out of  the vortex core. We find that a
current circulating in the vacuum around the vortex is given by
expression \eqref{4.10} in the case of $\nu \geq 1$ and
$\frac12\left(1-\frac1\nu \right) < F <
\frac12\left(1+\frac1\nu\right)$, or $\frac12 \leq \nu <1$ and
$\frac12\left(\frac1\nu-1 \right) < F <
\frac12\left(3-\frac1\nu\right)$; it is given by expression
\eqref{4.12} in the case of $\nu>1$ and $0 < F <
\frac12\left(1-\frac1\nu \right)$ and by expression \eqref{4.13} in
the case of $\nu>1$ and $\frac{1}{2} \left( 1+\frac{1}{\nu} \right)
< F < 1$. At large distances from the vortex, $r \rightarrow
\infty$, the current decreases exponentially as
$$\left\{
\begin{array}{l}
\sqrt{m/r}\,\, {\rm exp}(-2mr), \quad \frac12 \leq \nu < 2, \quad F \neq \frac12  \\
\vphantom{\int\limits_0^0}
m \, \frac{\sin[(2F-1)\pi]}{\sin(\pi/\nu)}\,\, {\rm exp}[-2mr\sin(\pi/\nu)], \quad \nu\geq 2, \quad F \neq \frac12\\
\sqrt{m/r}\,\, {\rm exp}(-2mr),\quad \nu \geq \frac12, \quad F = \frac12
\end{array}
\right\},
$$
whereas it decreases as $1/r$ in the case of massless spinor matter; see Appendix D.

As a manifestation of the Aharonov-Bohm effect, the current is
periodic in the value of the vortex flux, $\Phi$; i.e., it depends on
$F$ and not on $n_c$, see \eqref{1.18}; moreover, it changes sign
under simultaneous changes $F\rightarrow 1-F$ and
$\theta\rightarrow-\theta$, see \eqref{4.24}. One can introduce the
charge conjugation transformation with the vortex flux changing its
sign,
\begin{equation}\label{7.2}
C: \quad \Phi\rightarrow -\Phi, \quad E\rightarrow-E,\quad
\Psi\rightarrow \sigma^1\Psi^*.
\end{equation}
The boundary condition for a conjugated wave function differs from
\eqref{7.1}:
\begin{equation}\label{7.3}
    (I-{\rm i}\beta \alpha^r\, e^{{\rm i} \theta
    \alpha^r})\left.\sigma^1\psi^*\right|_{r=r_0}=0.
\end{equation}
By requiring invariance of the boundary condition under such a
charge conjugation, one restricts the values of the self-adjoint
extension parameter to $\theta=\frac{\pi}2 \mp \frac{\pi}2$. In the
latter case the current changes sign under change $F\rightarrow
1-F$; i.e., it is odd, as well as periodic in the value of the vortex
flux. Consequently, it vanishes at $\theta=0$ and $F=1/2$, while it
is discontinuous in $\theta$ at $\theta=\pi$ and $F=1/2$, see \eqref{5.24a}.

It is appropriate here to discuss the dependence on the transverse
size of the vortex core and the limiting procedure as this size
tends to zero, $r_0\rightarrow 0$. For this task it is instructive
to decompose the current into $r_0$-independent, $j_\varphi^{(a)}$,
and $r_0$-dependent, $j_\varphi^{(b)}$, pieces, see \eqref{4.14} --
\eqref{4.21}; the $r_0$-dependent piece vanishes at
$r_0\rightarrow0$. It should be noted that, in the case of the
infinitely thin $(r_0=0)$ vortex, the Dirac hamiltonian is essentially self-adjoint
for $\nu>1$ and either $0<F<
\frac12\left(1-\frac1\nu\right)$ or
$\frac12\left(1+\frac1\nu\right)<F<1$; otherwise, there emerges the
self-adjoint extension problem with one, or four, or more
parameters. One self-adjoint extension parameter, $\Theta$, appears
for
 $\nu \geq 1$ and
$\frac12\left(1-\frac1\nu \right) < F <
\frac12\left(1+\frac1\nu\right)$, or for $\frac12 \leq \nu <1$ and
$\frac12\left(\frac1\nu-1 \right) < F <
\frac12\left(3-\frac1\nu\right)$. The results for $\nu \geq 1$ and
$0 < F < 1$, as well as for $\frac12 \leq \nu <1$ and
$\frac12\left(\frac1\nu-1 \right) < F <
\frac12\left(3-\frac1\nu\right)$, are comprehensively presented in Appendix C. The value of $\Theta$ can be fixed by limiting
procedure $r_0\rightarrow 0$ starting from the $r_0>0$ case. In this
way, the condition of minimal irregularity in the $r_0=0$ case is
obtained in the form of \eqref{3.18}. If this condition  is
supplemented with the requirements of invariance under charge
conjugation \eqref{7.2} and continuity in $\theta$, then it takes
the form of \eqref{3.18} with $\theta=0$ at $F=1/2$; namely in the
latter form it was first proposed in \cite{Si6,Si7}.

As a consequence of the Maxwell equation, the magnetic field strength is also induced in the vacuum, pointing along the vortex axis; the relevant expressions in the case of the most general boundary condition, \eqref{7.1}, are given by \eqref{5.1}-\eqref{5.9}. This allows us to consider the total magnetic flux which is induced in the vacuum. As follows from our numerical analysis, the latter is finite at $\theta=\frac{\pi}2 \mp \frac{\pi}2$ only, but otherwise, it is divergent. Thus, the physical condition that the induced vacuum magnetic flux be finite corresponds to the requirement of invariance under charge conjugation \eqref{7.2}. The flux for boundary conditions maintaining the charge conjugation invariance is given at $F \neq 1/2$ by expression \eqref{5.29}, it vanishes at $\theta=0$ and $F=1/2$, and it is nonvanishing and discontinuous in 
$\theta$ at $\theta=\pi$ and $F=1/2$, see \eqref{5.24c}. The flux in the case of $r_0=0$ is discontinuous at $F=1/2$; moreover, its absolute value at $\theta=\pi$ and $F=1/2$, see  \eqref{5.24d}, is twice as large as its absolute value at $\theta=\pi$ in the limit $F\rightarrow 1/2$, see \eqref{5.231}. The case of an infinitely thin vortex is an idealization that is inappropriate to physics reality, since, as has already been noted in Introduction, the transverse size of the vortex, $r_0$, is of the order of the correlation length. In the case of $r_0>0$, the differences in behavior of the flux at $\theta=0$ and at $\theta=\pi$ are comprehensively illustrated by Figs. 2--6. Whereas the flux at $\theta=0$ decreases in its absolute value with the increase of $r_0$, the flux at $\theta=\pi$ in general increases at no allowance in its absolute value with the increase of $r_0$ (although there is a moderate decrease in vicinities of $F=0$ and $F=1$ at $\nu > 2$). Such a behavior of the flux, as that at $\theta=\pi$, can hardly be regarded as physical. Quantity $r_0^{-1}$ is identified with the energy scale of spontaneous symmetry breaking, i.e., the mass of the corresponding Higgs particle. It looks rather unlikely that a topological defect influences the surrounding quantum matter with the matter particle mass, $m$, exceeding the Higgs particle mass, $m_{\rm cond}$; the more unlikely is the unrestricted growth of this influence with the increase of quotient $m/m_{\rm cond}$. The influence of a topological defect on the surrounding quantum matter at $m_{\rm cond} > m$ looks much more physically plausible. Namely this situation is realized in the case of quantum scalar matter obeying the Dirichlet boundary condition at the vortex edge, see \cite{Gor1,Gor2,Gor3,Gor4}.

We conclude that, although we have solved the problem analytically with the use of the most general set of boundary conditions ensuring the impenetrability of the vortex core, the analysis of the behavior of the induced vacuum magnetic flux allows us to restrict the variety of admissible boundary conditions. The requirement of the flux to be finite, which is equivalent to the requirement of invariance under charge conjugation \eqref{7.2}, restricts the values of the self-adjoint extension parameter to $\theta=\frac{\pi}2 \mp \frac{\pi}2$. The further requirement of physical plausibility of the finite flux behavior, which is equivalent to the formal requirement of continuity in the dependence on $\theta$, yields $\theta=0$. As long as the transverse size of the vortex is taken into account, the induced vacuum effects at $\theta=0$ are continuous in $F$ and vanishing at $F=1/2$. At this point we would like to emphasize the crucial distinction between the cases of massive and massless quantum spinor matter. The latter case requires an introduction of the maximal size of the system, $r_{\rm max}$. We discover that, for $r_{\rm max}\gg r_0$ (in conformance to the reality), the induced vacuum effects for both $\theta=0$ and $\theta=\pi$ are physically plausible; moreover, they coincide, being independent of the transverse size of the vortex, see \eqref{d15} and \eqref{d16} in Appendix D. Because of this distinction, the results in the massless case are discontinuous at $F=1/2$ with a jump which is independent of $\nu$ (see \eqref{d19}). Note in this respect that the current and the magnetic field that are induced in the vacuum of quantum scalar matter under the Dirichlet boundary condition are continuous in $F$ and vanishing at $F=1/2$ even in the $r_0=0$ case, see \cite{SiB1,SiB2,SiV9}. In contrast to this, the emergence of a peculiar mode in the solution to the Dirac equation (see Section 3) leads to a discontinuity of the  results at $F=1/2$. For massive quantum spinor matter, the discontinuity is present in the $r_0=0$ case, but is eliminated by a choice of the physically motivated boundary condition in the $r_0>0$ case. The discontinuity nonetheless stays on for massless quantum spinor matter. It would be interesting to perform a similar study for other characteristics of the vacuum, for instance, for the induced vacuum energy-momentum tensor.

\section*{Acknowledgments}
The work of Yu.A.S. was supported by the National Academy of
Sciences of Ukraine (Project No.01172U000237), by the Program of
Fundamental Research of the Department of Physics and Astronomy of
the National Academy of Sciences of Ukraine (Project No.0117U000240),
and by the SEENET-MTP -- ICTP(Abdus Salam International Centre for Theoretical Physics) project NT-03 ``Cosmology - Classical
and Quantum Challenges''.


\setcounter{equation}{0}
\renewcommand{\theequation}{A.\arabic{equation}}
\section*{Appendix A: contribution of nonpeculiar modes to the current}

\newcounter{p}  
\renewcommand{\thep}{\arabic{p}}

Using relations (see, e.g., \cite{Abra})
\begin{align}
& J_\rho({\rm i} z) = {\rm e}^{{\rm i}\rho \pi/2}I_\rho(z),\qquad
Y_\rho({\rm i} z) = {\rm i}{\rm e}^{{\rm i}\rho \pi/2}I_\rho(z)-\frac2\pi
{\rm e}^{-{\rm i}\rho\pi/2}K_\rho(z),\qquad -\pi< \arg z\leq \pi/2,\nonumber \\
& I_\rho(-z)={\rm e}^{{\rm i} \rho \pi}I_\rho(z),\qquad K_\rho(- z)
= {\rm e}^{-{\rm i}\rho \pi}K_\rho(z)-{\rm i}\pi I_\rho(z),\qquad
-\pi< \arg z < 0,\nonumber
\end{align}
one can obtain
\begin{multline}\label{a1}
J_\rho(kr)J_{\rho-1}(kr)=\frac1{2\pi}\left[I_{\rho}(-{\rm i} kr)
K_{\rho-1}(-{\rm i} kr)- I_{\rho-1}(-{\rm i} kr) K_{\rho}(-{\rm
i}kr)\right. \\
\left.+I_{\rho}({\rm i} kr) K_{\rho-1}({\rm i} kr) -I_{\rho-1}({\rm
i} kr) K_{\rho}({\rm i} kr)\right],
\end{multline}
\begin{multline}\label{a2}
J_\rho(kr)Y_{\rho-1}(kr)+Y_\rho(kr)J_{\rho-1}(kr)\\
=-\frac{2}{\pi^2}\left[ {\rm e}^{-{\rm i}\rho\pi }K_{\rho}(-{\rm i}
kr) K_{\rho-1}(-{\rm i} kr)+ {\rm e}^{{\rm i}\rho\pi } K_{\rho}({\rm
i} kr) K_{\rho-1}({\rm i}kr)\right],
\end{multline}
\begin{multline}\label{a3}
Y_\rho(kr)Y_{\rho-1}(kr)-J_\rho(kr)J_{\rho-1}(kr)\\
=\frac{2{\rm i}}{\pi^2}\left[ {\rm e}^{-{\rm i}\rho\pi
}K_{\rho}(-{\rm i} kr) K_{\rho-1}(-{\rm i} kr)- {\rm e}^{{\rm
i}\rho\pi } K_{\rho}({\rm i} kr) K_{\rho-1}({\rm i}kr)\right].
\end{multline}
With the help of these relations, $j^{(1)}_\varphi$ \eqref{4.1} and
$j^{(2)}_\varphi$ \eqref{4.2} are presented as
\begin{multline}\label{a4}
j^{(1)}_\varphi(r)\!=\!-\frac{r}{(2\pi)^2}\!\!\int\limits_{-\infty}^\infty\!\!\frac{dk
\,k^2}{\sqrt{k^2+m^2}  }\sum_{l=1}^\infty \left[I_{\nu l+1-G}(-{\rm
i} kr) K_{\nu l-G}(-{\rm i} kr)\! -\! I_{\nu l-G}(-{\rm i} kr)
K_{\nu
l+1-G}(-{\rm i} kr)    \right.\\
-\left. I_{\nu l+G}(-{\rm i} kr) K_{\nu l-1+G}(-{\rm i} kr)+I_{\nu
l-1+G}(-{\rm i} kr) K_{\nu l+G}(-{\rm i} kr) \right]
\end{multline}
and
\begin{equation}\label{a5}
j^{(2)}_\varphi(r)=\sum_{{\rm sgn}(E)}\sum_{l=1}^\infty
\left(\lambda^{(\wedge)}_{\nu l+1-G}(r) - \lambda^{(\vee)}_{\nu
l+G}(r)\right),
\end{equation}
where
\begin{multline}\label{a6}
\lambda^{(\wedge/\vee)}_{\rho}(r)=-\frac{r}{2\pi^2}\int\limits_{-\infty}^\infty\!\!\frac{dk
\,k^2}{\sqrt{k^2+m^2}
}\left[\cos^2(\mu^{(\wedge/\vee)}_{\rho})\frac{(-{\rm i}
k)^{2\rho-1} }{(\sqrt{k^2}\,)^{2\rho-1}  } - \frac12
\sin(2\mu^{(\wedge/\vee)}_{\rho}) \frac{(-{\rm i} k)^{2\rho}
}{(\sqrt{k^2}\,)^{2\rho}  }\right]\\ \times K_{\rho}(-{\rm i} kr)
K_{\rho-1}(-{\rm i} kr)
\end{multline}
and it is implied that $\mu^{(\wedge)}_{\rho}$ and
$\mu^{(\vee)}_{\rho}$, determined by relations \eqref{3.11} and
\eqref{3.12}, depend on $\sqrt{k^2}$ instead of $k$. The integral
over real $k$ can be transformed into the integral over a contour
circumventing anticlockwise the positive imaginary semiaxis in the
complex $k$ plane. It is evident in the case of $j^{(1)}_\varphi$
that the latter contour is reduced to a contour circumventing a part
of the positive imaginary semiaxis, see Fig. 7. As a result, we get
\begin{multline}\label{a7}
j^{(1)}_\varphi(r)=\frac{r}{2\pi^2}\int\limits_{m}^\infty\frac{dq
\,q^2}{\sqrt{q^2-m^2}  }\sum_{l=1}^\infty \left[I_{\nu l+1-G}(qr)
K_{\nu l-G}(qr)\! -\! I_{\nu l-G}(qr) K_{\nu
l+1-G}(qr)    \right.\\
-\left. I_{\nu l+G}(qr) K_{\nu l-1+G}(qr)+I_{\nu l-1+G}(qr) K_{\nu
l+G}(qr) \right],
\end{multline}

\begin{figure}[t]
\begin{center}
\includegraphics[width=80mm]{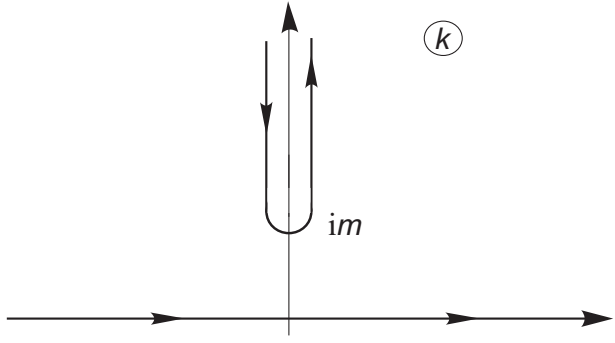}
\end{center} \caption{The integral over real $k$
in \eqref{a4} is transformed into the integral over a contour in the
complex $k$ plane.}\label{cont1}
\end{figure}

\noindent which can be decomposed as
\begin{multline}\label{a8}
j^{(1)}_\varphi(r)=j^{(1,1)}_\varphi(r)+j^{(1,2)}_\varphi(r)-\frac{r}{\pi^2}\int\limits_{m}^\infty
\frac{dq \,q^2}{\sqrt{q^2-m^2}
}\left\{\frac{\sin(G\pi)}{\pi}K_{G}(qr)
K_{1-G}(qr)   \right.\\
+\left.\frac12 \left[ I_{1-G}(qr) K_{G}(qr)-I_{G}(qr)
K_{1-G}(qr)\right] \right\},
\end{multline}
where
\begin{multline}\label{a9}
j^{(1,1)}_\varphi(r)=\frac{1}{2\pi^2}\int\limits_{m}^\infty\frac{dq
\,q^2}{\sqrt{q^2-m^2} }\,\frac{d}{dq} \sum_{n \in \mathbb{Z}}
I_{|\nu n-G|}(qr) K_{\nu n-G}(qr)\\
=\frac{1}{(2\pi)^2}\int\limits_{m}^\infty\frac{dq
\,q^2}{\sqrt{q^2-m^2} }\,\frac{d}{dq}\int\limits_0^\infty
\frac{dy}{y}\,\exp\left(-\frac{q^2r^2}{2y}-y\right) \sum_{n \in
\mathbb{Z}} I_{|\nu n-G|}(y)
\end{multline}
and
\begin{multline}\label{a10}
j^{(1,2)}_\varphi(r)=-\frac{1}{\pi^2}\int\limits_{m}^\infty\frac{dq
\,q}{\sqrt{q^2-m^2} } \sum_{n \in \mathbb{Z}} (\nu n-G)
I_{|\nu n-G|}(qr) K_{\nu n-G}(qr)\\
=-\frac{1}{2\pi^2} \int\limits_{m}^\infty \frac{dq
\,q}{\sqrt{q^2-m^2} } \int\limits_0^\infty
\frac{dy}{y}\,\exp\left(-\frac{q^2r^2}{2y}-y\right) \sum_{n \in
\mathbb{Z}} (\nu n-G)I_{|\nu n-G|}(y).
\end{multline}
Using the Schl\"{a}fli contour integral representation,
\begin{equation*}
I_{\rho}(y)=\frac1{2\pi {\rm i}}\int\limits_{C_+} dz\, {\rm e}^{y \cosh
z-\rho z}=-\frac1{2\pi {\rm i}} \int\limits_{C_-} dz\, {\rm e}^{y \cosh
z+\rho z}
\end{equation*}
we obtain
\begin{multline}\label{a11}
\sum_{n \in \mathbb{Z}} I_{|\nu n-G|}(y)= \frac1{2\pi {\rm i}}\left[
\,\, \int\limits_{C_+} dz\, {\rm e}^{y \cosh z-Gz} \,
\frac{1}{1-{\rm e}^{-\nu z}} - \int\limits_{C_-} dz\, {\rm e}^{y
\cosh
z-Gz} \, \frac{{\rm e}^{\nu z}}{1-{\rm e}^{\nu z}} \right] \\
=\frac1{4\pi {\rm i}}\int\limits_{C_0} dz\, {\rm e}^{y \cosh z}
\frac{\cosh\left[\left(G-\frac{\nu}2\right)z \right]}{\sinh(\nu
z/2)}+\frac{{\rm e}^y }{\nu},
\end{multline}
where contours $C_+$, $C_-$, and $C_0$ in the complex $z$ plane are
shown in Fig. 8. The vertical segments of contours $C_+$, $C_-$ and
$C_0$ are infinitesimally close to the imaginary axis, not
coinciding with it in order to avoid simple poles of the integrand
at $z=0$ and $z=\pm 2{\rm i}p\pi/\nu$ ($p$ is the positive integer,
$1\leq p \leq \left[\!\left| {\nu}/2 \right|\!\right]$). Contour
$C_0$ circumvents poles out of the origin (existing at $\nu \geq
2$), whereas the contribution of the pole at the origin (existing at
$\nu \neq 0 $) is explicitly separated in \eqref{a11}. Substituting
\eqref{a11} into \eqref{a9}, we change integration variable
$y\rightarrow v= y (q r)^{-2}$ and take a derivative,
\begin{multline}\label{a12}
j^{(1,1)}_\varphi(r)\!=\!\frac{r^2}{2\pi^2}\!\!\int\limits_{m}^\infty\!\!\frac{dq
\,q^3}{\sqrt{q^2\!-\!m^2} }\int\limits_0^\infty dv\,\frac1{2\pi
{\rm i}}\int\limits_{C_0}\! dz \, \exp\left[-\frac1{2v}+ 2v q^2 r^2
\sinh^2(z/2)\right] \sinh^2(z/2)\, \\
\times \frac{\cosh\left[\left(G\!-\! \frac\nu2\right)z
\right]}{\sinh(\nu z/2)}.
\end{multline}

\begin{figure}[t]
\begin{center}
\includegraphics[width=160mm]{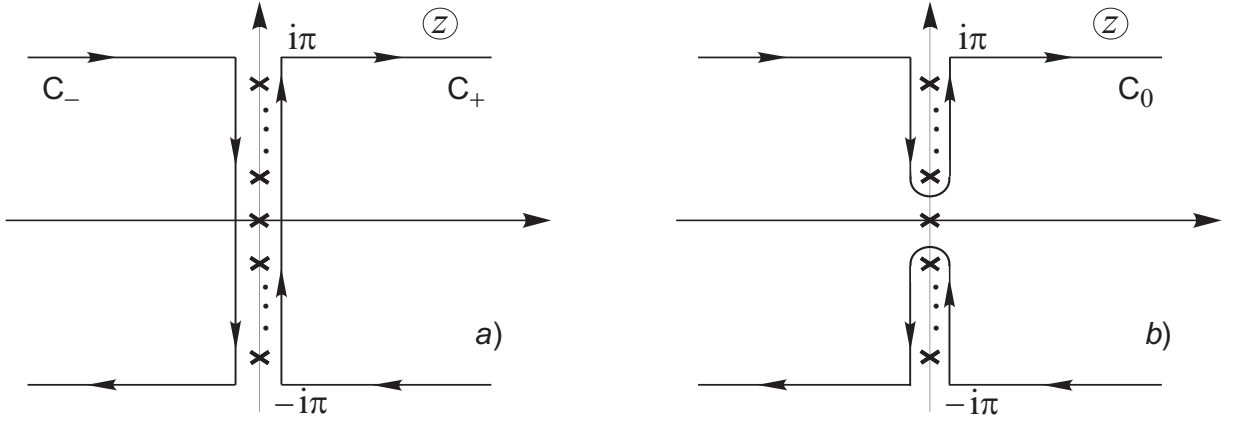}
\end{center} \caption{Complex $z$ plane with
simple poles indicated by crosses: a) contours $C_+$ and $C_-$, b) contour $C_0$.}\label{cont2}
\end{figure}

\noindent Integrating over $q$ and $v$, we get
\begin{multline}\label{a13}
j^{(1,1)}_\varphi(r)\!=\!-\frac{m}{8\pi}\frac1{2\pi {\rm
i}}\int\limits_{C_0} dz\,
\left[1+\frac1{2mr\sqrt{-\sinh^2(z/2)}}\right]\,\exp\left[-2mr\sqrt{-\sinh^2(z/2)}\right] \\
\times\frac{\cosh\left[\left(G\!-\! \frac\nu2\right)z
\right]}{\sinh(\nu z/2)} =\frac{m}{(2\pi)^2}\left\{
\int\limits_0^\infty du\,\left[1+
\frac1{2mr\cosh(u/2)} \right]\,{\rm e}^{-2mr\cosh(u/2)}\right. \\
\times\frac{\sin(G\pi)\cosh\left[\left(G-\nu\right)u\right]-\sin[(G-\nu)\pi]\cosh(G
u) }{\cosh(\nu u)-\cos(\nu\pi)} \\
- \frac{2\pi}{\nu}\sum_{p=1}^{\left[\!\left| {\nu}/2
\right|\!\right]} \left[1 +\frac{1}{2mr\sin(p\pi/\nu)}\right]\, {\rm
e}^{-2mr\sin(p\pi/\nu)}\,\cos(2Gp\pi/\nu)\\
\left. + \frac\pi\nu  \left(1 +\frac{1}{2mr}\right){\rm e}^{-2mr}\cos(G\pi) \, \delta_{\nu, \, 2N}\right\},
\end{multline}
where the finite sum over integer $p$ and the last term with the Kronecker $\delta$ symbol ($N$ is the positive integer) are
due to a contribution of simple poles on the imaginary axis out of the origin.

In a similar way we calculate the sum in \eqref{a10}:
\begin{multline}\label{a14}
\sum_{n \in \mathbb{Z}} (\nu n-G) I_{|\nu n-G|}(y)=\frac{y}2\left\{
\sum_{l=1}^\infty\left[I_{\nu l-1-G}(y)-I_{\nu l+1-G}(y)\right]\right. \\
\left. +\sum_{l=0}^\infty\left[I_{\nu l+1+G}(y)-I_{\nu
l-1+G}(y)\right]\right\}
=\frac{y}{4\pi{\rm i} }\int\limits_{C_0} dz\, {\rm e}^{y\cosh z
}\,\sinh(z)\,\,\frac{\sinh\left[\left(G\!-\! \frac\nu2\right)z
\right]}{\sinh(\nu z/2)}.
\end{multline}
Substituting \eqref{a14} into \eqref{a10} and integrating over $q$
and $y$, we get
\begin{multline}\label{a15}
j^{(1,2)}_\varphi(r)\!=\!\frac{m}{8\pi}\frac1{2\pi {\rm
i}}\int\limits_{C_0} dz\,
\left[1+\frac1{2mr\sqrt{-\sinh^2(z/2)}}\right]\,\exp\left[-2mr\sqrt{-\sinh^2(z/2)}\right]\\
\times\frac{\coth(z/2)\sinh\left[\left(G\!-\! \frac\nu2\right)z
\right]}{\sinh(\nu z/2)} =-\frac{m}{(2\pi)^2}\left\{
\int\limits_0^\infty du\,\left[1+
\frac1{2mr\cosh(u/2)} \right]\,{\rm e}^{-2mr\cosh(u/2)}\right. \\
\times\tanh(u/2)\frac{\sin(G\pi)\sinh\left[\left(G-\nu\right)u\right]-\sin[(G-\nu)\pi]\sinh(G
u) }{\cosh(\nu u)-\cos(\nu\pi)} \\
\left.- \frac{2\pi}{\nu}\sum_{p=1}^{\left[\!\left| {\nu}/2
\right|\!\right]} \left[1 +\frac{1}{2mr\sin(p\pi/\nu)}\right]\, {\rm
e}^{-2mr\sin(p\pi/\nu)}\,\cot(p\pi/\nu)\sin(2Gp\pi/\nu)\right\}.
\end{multline}
As a result, we obtain the following expression for
$j^{(1)}_\varphi$ \eqref{a7}:
\begin{multline}\label{a16}
j^{(1)}_\varphi(r)\!=\!\frac{m}{(2\pi)^2}\left\{ \int\limits_0^\infty
\frac{du}{\cosh(u/2)}\,\left[1+
\frac1{2mr\cosh(u/2)} \right]\,{\rm e}^{-2mr\cosh(u/2)}\right. \\
\times\frac{\sin(G\pi)\cosh\left[\left(G-\nu-\frac12\right)u\right]-\sin[(G-\nu)\pi]\cosh\left[\left(G-\frac12\right)
u\right] }{\cosh(\nu u)-\cos(\nu\pi)}\\
+ \frac{2\pi}{\nu}\sum_{p=1}^{\left[\!\left| {\nu}/2 \right|\!\right]} \left[1 +\frac{1}{2mr\sin(p\pi/\nu)}\right]\,
\exp[-2mr\sin(p\pi/\nu)] \,\frac{\sin[(2G-1)p\pi/\nu]}{\sin(p\pi/\nu)}\\
\left. + \frac\pi\nu  \left(1 +\frac{1}{2mr}\right){\rm e}^{-2mr}\cos(G\pi) \, \delta_{\nu, \, 2N}\right\} \\
-\frac{r}{\pi^2}\!\!\int\limits_{m}^\infty\!\!\frac{dq
\,q^2}{\sqrt{q^2-m^2}  }\left\{\frac{\sin(G\pi)}{\pi}K_{G}(qr)
K_{1-G}(qr) +\frac12 \left[ I_{1-G}(qr) K_{G}(qr)-I_{G}(qr)
K_{1-G}(qr)\right] \right\}.
\end{multline}
Note that $j^{(1)}_\varphi$, see \eqref{4.1} or \eqref{a7}, changes
sign under substitution $G\rightarrow 1-G$. In view of this one gets
\begin{multline}\label{a17}
\!\int\limits_{m}^\infty\!\!\frac{dq \,q^2}{\sqrt{q^2-m^2}
}K_{G}(qr) K_{1-G}(qr)\!=\!\frac{\pi m}{4r}\!\!\int\limits_0^\infty
\!\!\frac{du}{\cosh(u/2)}\,\,
\cosh\left[\left(G\!-\!\frac12\right)u\right]\\
\times \left[1\!\!+ \frac1{2mr\cosh(u/2)} \right]{\rm
e}^{-2mr\cosh(u/2)}.
\end{multline}
Defining
\begin{multline}\label{a18}
j^{(1,3)}_\varphi(r)\!=\!-\frac{m}{8\pi}\frac1{2\pi {\rm
i}}\int\limits_{C_0} dz\,
\left[1+\frac1{2mr\sqrt{-\sinh^2(z/2)}}\right]\,\exp\left[-2mr\sqrt{-\sinh^2(z/2)}\right] \\
\times\frac{\cosh\left[\left(G\!-\! \frac12\right)z
\right]}{\sinh(z/2)} =\frac{m\sin(G\pi)}{(2\pi)^2}
\int\limits_0^\infty \!\!\frac{du}{\cosh(u/2)}\,\,
\cosh\left[\left(G\!-\!\frac12\right)u\right]\\
\times \left[1\!\!+ \frac1{2mr\cosh(u/2)} \right]{\rm
e}^{-2mr\cosh(u/2)},
\end{multline}
we can present \eqref{a16} as
\begin{multline}\label{a18e}
j^{(1)}_\varphi(r)=j^{(1,1)}_\varphi(r)+j^{(1,2)}_\varphi(r)-j^{(1,3)}_\varphi(r) \\
-\frac{r}{2\pi^2}\int\limits_{m}^\infty
\frac{dq \,q^2}{\sqrt{q^2-m^2}
} \left[I_{1-G}(qr) K_{G}(qr)-I_{G}(qr)
K_{1-G}(qr)\right]
\end{multline}
with $j^{(1,1)}_\varphi$, $j^{(1,2)}_\varphi$, and $j^{(1,3)}_\varphi$ given by \eqref{a13}, \eqref{a15}, and \eqref{a18}, respectively.
Using the latter relation, we finally obtain \eqref{4.4}.

Turning now to $j^{(2)}_\varphi$ \eqref{a5}, we obtain by deforming
the integration contour to circumvent the positive imaginary semiaxis
in the complex $k$ plane
\begin{multline}\label{a19}
\lambda^{(\wedge/\vee)}_{\rho}(r)=-\frac{r}{2\pi^3}\left\{\,\,
\int\limits_{0}^m\!\!\frac{dq \,q^2}{\sqrt{m^2-q^2} }\left[{\rm
e}^{{\rm i} \rho\pi}\cos^2(\mu^{(\wedge/\vee)}_{\rho,+})+ {\rm
e}^{-{\rm i}
\rho\pi}\cos^2(\mu^{(\wedge/\vee)}_{\rho,-})\right.\right. \\
\left. - \frac{\rm i}2\,{\rm e}^{{\rm i} \rho\pi}
\sin(2\mu^{(\wedge/\vee)}_{\rho,+}) +\frac{\rm i}2\,{\rm e}^{-{\rm
i} \rho\pi} \sin(2\mu^{(\wedge/\vee)}_{\rho,-})\right] K_{\rho}(qr)
K_{\rho-1}(qr)\\
+\int\limits_{m}^\infty\!\!\frac{dq \,q^2}{\sqrt{q^2-m^2}
}\left[{\rm i}{\rm e}^{{\rm i}
\rho\pi}\cos^2(\mu^{(\wedge/\vee)}_{\rho,+})-{\rm i} {\rm e}^{-{\rm
i}
\rho\pi}\cos^2(\mu^{(\wedge/\vee)}_{\rho,-})\right. \\
\left.\left. + \frac12\,{\rm e}^{{\rm i} \rho\pi}
\sin(2\mu^{(\wedge/\vee)}_{\rho,+}) +\frac12\,{\rm e}^{-{\rm i}
\rho\pi} \sin(2\mu^{(\wedge/\vee)}_{\rho,-})\right] K_{\rho}(qr)
K_{\rho-1}(qr)\vphantom{\int\limits_0^1}\right\},
\end{multline}
where $\mu^{(\wedge/\vee)}_{\rho,+}$ and
$\mu^{(\wedge/\vee)}_{\rho,-}$ are determined by relations
\begin{multline}\label{a20}
\tan(\mu^{(\wedge)}_{\rho,\pm})=\left\{
\cos\left(\frac\theta2+\frac\pi4\right)q\left[\frac2\pi\, {\rm
e}^{\pm{\rm i}\rho\pi}K_{\rho-1}(qr_0) \mp {\rm i} I_{\rho-1}(qr_0)
\right]\right.\\ \left.
+\sin\left(\frac\theta2+\frac\pi4\right)(m-\Delta)\left[\frac2\pi\,
{\rm e}^{\pm{\rm i}\rho\pi}K_{\rho}(qr_0) \pm  {\rm i}
I_{\rho}(qr_0) \right]\right\}\\
\times\left[-\cos\left(\frac\theta2+\frac\pi4\right)q
I_{\rho-1}(qr_0)
+\sin\left(\frac\theta2+\frac\pi4\right)(m-\Delta)I_\rho(qr_0)
\right]^{-1},
\end{multline}
\begin{multline}\label{a21}
\tan(\mu^{(\vee)}_{\rho,\pm})=\left\{
\cos\left(\frac\theta2+\frac\pi4\right)(m+\Delta)\left[\frac2\pi\,
{\rm e}^{\pm{\rm i}\rho\pi}K_{\rho}(qr_0) \pm {\rm i} I_{\rho}(qr_0)
\right]\right.\\ \left.
+\sin\left(\frac\theta2+\frac\pi4\right)q\left[\frac2\pi\, {\rm
e}^{\pm{\rm i}\rho\pi}K_{\rho-1}(qr_0) \mp  {\rm i}
I_{\rho-1}(qr_0) \right]\right\}\\
\times\left[\cos\left(\frac\theta2+\frac\pi4\right)(m+\Delta)
I_{\rho}(qr_0) -\sin\left(\frac\theta2+\frac\pi4\right)q
I_{\rho-1}(qr_0) \right]^{-1}
\end{multline}
and
\begin{equation}\label{a22}
\Delta = \left\{\begin{array}{ll}
            \vphantom{\int\limits_0} {\rm sgn}(E)\sqrt{m^2-q^2},& q<m, \\
            \mp \,{\rm i}\,{\rm sgn}(E)\sqrt{q^2-m^2},& q>m. \\
         \end{array}\right.
\end{equation}
In view of relation
\begin{equation}\label{a23}
\sum_{\pm}{\rm e}^{\pm{\rm
i}\rho\pi}\left[\cos^2(\mu^{(\wedge/\vee)}_{\rho,\pm}) \mp \frac{\rm
i}2 \sin(2\mu^{(\wedge/\vee)}_{\rho,\pm})\right]=0,
\end{equation}
the first integral in \eqref{a19} vanishes, and, as in the case of
$j^{(1)}_\varphi$, only the integral over a contour depicted on
Fig. 7 contributes; namely the latter is given by the second integral
in \eqref{a19}. In view of relation
\begin{equation}\label{a24}
\sum_{{\rm sgn}(E)}\sum_{\pm}{\rm e}^{\pm{\rm i}\rho\pi}\left[\pm
{\rm i}\,\cos^2(\mu^{(\wedge/\vee)}_{\rho,\pm}) + \frac{1}{2}
\sin(2\mu^{(\wedge/\vee)}_{\rho,\pm})\right]=2\pi
C^{(\wedge/\vee)}_{\rho}(qr_0),
\end{equation}
where $C^{(\wedge)}_{\rho}(v)$ and $C^{(\vee)}_{\rho}(v)$ are given
by \eqref{4.6} and \eqref{4.7}, we get
\begin{equation}\label{a25}
\sum_{{\rm sgn}(E)}\lambda^{(\wedge/\vee)}_{\rho}(r) =
-\frac{r}{\pi^2}\!\!\int\limits_{m}^\infty\!\!\frac{dq
\,q^2}{\sqrt{q^2-m^2}  }\, C^{(\wedge/\vee)}_{\rho}(qr_0)
K_{\rho}(qr) K_{\rho-1}(qr).
\end{equation}
As a consequence of \eqref{a5} and \eqref{a25}, we obtain \eqref{4.5}.

\setcounter{equation}{0}
\renewcommand{\theequation}{B.\arabic{equation}}
\section*{Appendix B: contribution of peculiar modes to the current}

Similarly to that in the beginning of Appendix A, one can obtain
\begin{multline}\label{b1}
J_G(kr)J_{1-G}(kr)-J_{-G}(kr)J_{-1+G}(kr)\!=\!-\frac1{2\pi}\left\{
\left[{\rm e}^{{\rm i} G\pi} I_{G}(-{\rm i} kr)\! +\! {\rm e}^{-{\rm
i} G\pi} I_{-G}(-{\rm i} kr)\right] K_{1-G}(-{\rm i} kr)\right.\\
-\left[{\rm e}^{-{\rm i} G\pi} I_{1-G}(-{\rm i} kr)\! +\! {\rm
e}^{{\rm i} G\pi} I_{-1+G}(-{\rm i} kr)\right] K_{G}(-{\rm i} kr)
+\left[{\rm e}^{-{\rm i} G\pi} I_{G}({\rm i} kr)\! +\! {\rm e}^{{\rm
i} G\pi} I_{-G}({\rm i} kr)\right] K_{1-G}({\rm i} kr)\\-\left.
\left[{\rm e}^{{\rm i} G\pi} I_{1-G}({\rm i} kr)\! +\! {\rm
e}^{-{\rm i} G\pi} I_{-1+G}({\rm i} kr)\right] K_{G}({\rm i}
kr)\right\}.
\end{multline}
With the help of \eqref{a1} and \eqref{b1}, $j^{(3)}_\varphi$
\eqref{4.3} is presented as
\begin{multline}\label{b2}
j^{(3)}_\varphi(r)=-\frac{r}{2(2\pi)^2} \int\limits_{-\infty}^\infty
\frac{dk\, k^2}{\sqrt{k^2+m^2}}\sum_{{\rm
sgn}(E)}\left[\tan(\mu_{1-G})+2 \cos(G\pi)+ \cot(\mu_{1-G})
\right]^{-1}\\
\times \left\{\vphantom{\int\limits_{.}^.}
\tan(\mu_{1-G})\left[I_{1-G}(-{\rm i} kr)K_{G}(-{\rm
i} kr)-I_{-G}(-{\rm i} kr)K_{1-G}(-{\rm i} kr)\right]\right.\\
+\left[\frac{(-{\rm i}k)^{2G}}{(\sqrt{k^2})^{2G}}I_{1-G}(-{\rm i}
kr)+ \frac{(\sqrt{k^2})^{2G}}{(-{\rm
i}k)^{2G}}I_{-1+G}(-{\rm i} kr) \right]K_{G}(-{\rm i} kr)\\
-\left[\frac{(\sqrt{k^2})^{2G}}{(-{\rm i}k)^{2G}}I_{G}(-{\rm i} kr)+
\frac{(-{\rm i}k)^{2G}}{(\sqrt{k^2})^{2G}}I_{-G}(-{\rm i} kr)
\right]K_{1-G}(-{\rm i} kr)\\ \left.-
\cot(\mu_{1-G})\left[I_{G}(-{\rm i} kr)K_{1-G}(-{\rm i}
kr)-I_{-1+G}(-{\rm i} kr)K_{G}(-{\rm i} kr)\right]\vphantom{\int\limits_{.}^.}\right\}\\
+\frac{\kappa^2}{4\pi r_0} \, \, \frac{[1-{\rm
sgn}(\cos\theta)]\,{\rm
sgn}\!\!\left[\tan\left(\frac\theta2+\frac\pi4\right)
+\frac{K_G(\kappa r_0)}{K_{1-G}(\kappa r_0)} \right] }{m K_G(\kappa
r_0)K_{1-G}(\kappa r_0) \!+\!E_{BS}\left\{\kappa r_0
[K_{1-G}^2(\kappa r_0)\!-\!K_{G}^2(\kappa
r_0)]\!+\!(2G\!-\!1)K_G(\kappa r_0)K_{1-G}(\kappa r_0) \right\} },
\end{multline}
and it is implied that $\mu_{1-G}$, as determined by \eqref{3.13},
depends on $\sqrt{k^2}$ instead of $k$. The integral over real $k$
can be transformed into the integral over a contour circumventing
anticlockwise the positive imaginary semiaxis in the complex
$k$ plane. The latter contour is reduced to a contour consisting of
two parts: one encircling a simple pole emerging  at  $\cos\theta<0$
and another one circumventing a part of the positive imaginary
semiaxis, see Fig. 9. The contribution of the pole cancels out the
last term in \eqref{b2}, and for the remaining part we get
\begin{multline}\label{b3}
j^{(3)}_\varphi(r)=\frac{r}{(2\pi)^2} \int\limits_{m}^\infty
\frac{dq\, q^2}{\sqrt{q^2-m^2}}\sum_{{\rm sgn}(E)}\sum_\pm
\left[\tan(\mu_{1-G,\pm})+2 \cos(G\pi)+ \cot(\mu_{1-G,\pm})
\right]^{-1}\\
\times \left\{ \left[\tan(\mu_{1-G},\pm)+{\rm e}^{\pm{\rm
i}G\pi}\right]\left[I_{1-G}(qr)K_{G}(qr)-I_{-G}(qr)K_{1-G}(qr)\right]\right.\\
-\left.\left[\cot(\mu_{1-G},\pm)+{\rm e}^{\mp{\rm
i}G\pi}\right]\left[I_{G}(qr)K_{1-G}(qr)-I_{-1+G}(qr)K_{G}(qr)\right]\right\}\\
=\frac{r}{2\pi^2}\int\limits_{m}^\infty \frac{dq\,
q^2}{\sqrt{q^2-m^2}}\left[\vphantom{\sum_{{\rm sgn}(E)}}
I_{1-G}(qr)K_{G}(qr)-I_{G}(qr)K_{1-G}(qr)\right.\\
\left. -\frac{\sin(G\pi)}{2\pi}K_{G}(qr)K_{1-G}(qr)\sum_{{\rm
sgn}(E)}\sum_\pm\frac{\tan(\mu_{1-G,\pm})\pm 2{\rm i}\sin(G\pi)
-\cot(\mu_{1-G,\pm})}{\tan(\mu_{1-G,\pm})+ 2\cos(G\pi)
+\cot(\mu_{1-G,\pm})}\right],
\end{multline}
where
\begin{equation}\label{b4}
\tan(\mu_{1-G,\pm}) = {\rm e}^{\mp{\rm i}G\pi}
\frac{\cos\left(\frac\theta2+\frac\pi4\right)q I_G(qr_0)-
\sin\left(\frac\theta2+\frac\pi4\right)\left(m\pm {\rm
i\,sgn}(E)\sqrt{q^2-m^2}\right)I_{-1+G}(qr_0)}{-\cos\left(\frac\theta2+\frac\pi4\right)q
I_{-G}(qr_0)+ \sin\left(\frac\theta2+\frac\pi4\right)\left(m\pm {\rm
i\,sgn}(E)\sqrt{q^2-m^2}\right)I_{1-G}(qr_0)}.
\end{equation}

\begin{figure}[t]
\begin{center}
\includegraphics[width=80mm]{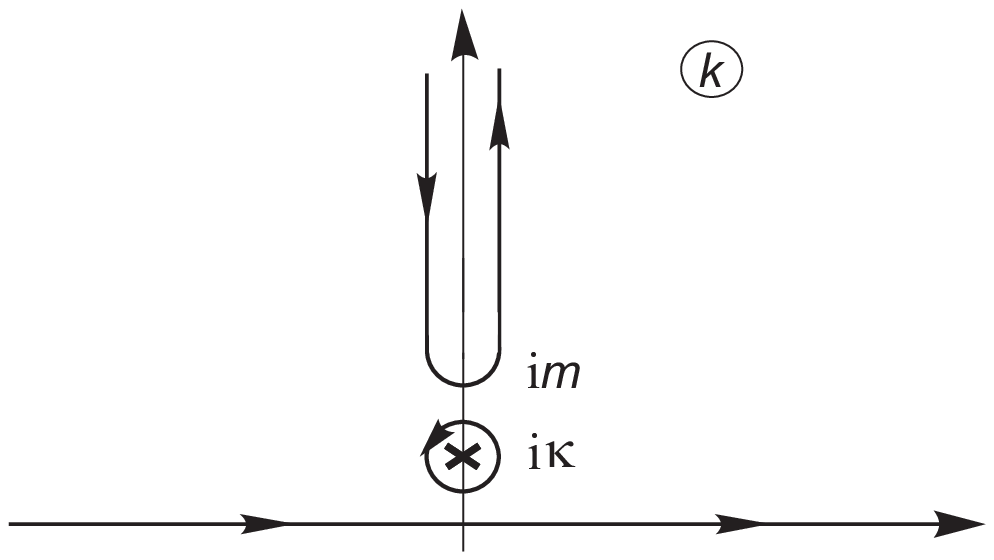}
\end{center} \caption{The integral over real $k$
in \eqref{b2} is transformed into the integral over a contour in the
complex $k$ plane.}\label{cont3}
\end{figure}

\noindent The sum in \eqref{b3} is reduced to the form
\begin{multline}\label{b5}
\sum_{{\rm sgn}(E)}\sum_\pm\frac{\tan(\mu_{1-G,\pm})\pm 2{\rm
i}\sin(G\pi) -\cot(\mu_{1-G,\pm})}{\tan(\mu_{1-G,\pm})+ 2\cos(G\pi)
+\cot(\mu_{1-G,\pm})}\\=\sum_\pm \left[\frac{{\rm e}^{\mp{\rm
i}G\pi}(h_\pm-1)-{\rm e}^{\pm{\rm i}G\pi}(h_\pm^{-1}-1)}{{\rm
e}^{\mp{\rm i}G\pi}(h_\pm+1)+{\rm e}^{\pm{\rm
i}G\pi}(h_\pm^{-1}+1)}+ \frac{{\rm e}^{\mp{\rm i}G\pi}(h_\mp-1)-{\rm
e}^{\pm{\rm i}G\pi}(h_\mp^{-1}-1)}{{\rm e}^{\mp{\rm
i}G\pi}(h_\mp+1)+{\rm e}^{\pm{\rm
i}G\pi}(h_\mp^{-1}+1)}\right]\\=\frac{4(h_+ h_--1)}{h_+h_-+h_+
+h_-+1},
\end{multline}
where
\begin{equation}\label{b6}
h_\pm=\frac{\cos\left(\frac\theta2+\frac\pi4\right)q I_G(qr_0)-
\sin\left(\frac\theta2+\frac\pi4\right)\left(m\pm {\rm
i}\sqrt{q^2-m^2}\right)I_{-1+G}(qr_0)}{-\cos\left(\frac\theta2+\frac\pi4\right)q
I_{-G}(qr_0)+ \sin\left(\frac\theta2+\frac\pi4\right)\left(m\pm {\rm
i}\sqrt{q^2-m^2}\right)I_{1-G}(qr_0)}.
\end{equation}
It is straightforward to get
\begin{equation}\label{b7}
\frac{4(h_+ h_--1)}{h_+h_-+h_+ +h_-+1} =
\frac{4\pi}{\sin(G\pi)}\,C_{1-G}(qr_0),
\end{equation}
where $C_{1-G}(v)$ is given by \eqref{4.9}. Substituting \eqref{b7}
into \eqref{b3}, we obtain \eqref{4.8}.

\setcounter{equation}{0}
\renewcommand{\theequation}{C.\arabic{equation}}
\section*{Appendix C: case of the infinitely thin vortex}

We present here the results for the case of the infinitely thin ANO
vortex (i.e., $r_0=0$).

In the case of $\nu>1$ and $0 < F < \frac12\left(1-\frac1\nu \right)$, partial
hamiltonians with all $n$ are essentially self-adjoint (deficiency
indices equal $(0,0)$), and the modes are given by \eqref{2.3} with
$\mu_\rho^{(\wedge)}=\mu_\rho^{(\vee)}=\pi/2$. We obtain
\begin{multline}\label{c1}
j_\varphi(r)=-\frac{m}{(2\pi)^2}\left\{ \int\limits_0^\infty
\frac{du}{\cosh(u/2)}\left[1+\frac1{2mr \cosh(u/2)}\right]{\rm
e}^{-2mr
\cosh(u/2)}\right. \\
\times \frac{\cos\left[\nu\left(F-\frac12\right)\pi\right] \cosh\left[\nu\left(F+\frac12 \right)u\right]
- \cos[\nu\left(F+\frac12 \right)\pi)]
\cosh\left[\nu\left(F-\frac12\right)u\right]}{\cosh(\nu
u)-\cos(\nu \pi)}\\
- \frac{2\pi}{\nu}\sum_{p=1}^{\left[\!\left| {\nu}/2
\right|\!\right]} \left[1 +\frac{1}{2mr\sin(p\pi/\nu)}\right]\,
\exp[-2mr\sin(p\pi/\nu)]
\,\frac{\sin[(2F-1)p\pi]}{\sin(p\pi/\nu)}\\
\left. + \frac{\pi}{2N} \left(-1\right)^{N}\sin\left(2NF \pi \right) \left(1 +\frac{1}{2mr}\right){\rm e}^{-2mr}  \, \delta_{\nu, \, 2N}\right\},
\end{multline}
\begin{multline}\label{c2}
B_{\rm I}(r)=-\frac{\nu e}{2(2\pi)^2}\frac1r \left\{ \int\limits_0^\infty
\frac{du}{\cosh^2(u/2)}\,{\rm e}^{-2mr
\cosh(u/2)}\right. \\
\times \frac{\cos\left[\nu\left(F-\frac12\right)\pi\right] \cosh\left[\nu\left(F+\frac12 \right)u\right]
- \cos[\nu\left(F+\frac12 \right)\pi)]
\cosh\left[\nu\left(F-\frac12\right)u\right]}{\cosh(\nu
u)-\cos(\nu \pi)}\\
\left.- \frac{2\pi}{\nu}\sum_{p=1}^{\left[\!\left| {\nu}/2
\right|\!\right]} \exp[-2mr\sin(p\pi/\nu)]
\,\frac{\sin[(2F-1)p\pi]}{\sin^2(p\pi/\nu)} + \frac{\pi}{2N} \left(-1\right)^{N}\sin\left(2NF \pi \right) {\rm e}^{-2mr}
\, \delta_{\nu, \, 2N}\right\},
\end{multline}
and
\begin{equation}\label{c3}
\Phi_{\rm I}=-\frac{e}{6m} F \left[\frac14
\left(\nu^2 + 3\right) -\nu^2 F^2\right].
\end{equation}

In the case of $\nu>1$ and $\frac12\left(1+\frac1\nu\right) < F < 1$, partial
hamiltonians with all $n$ are essentially self-adjoint as well, and
the modes are given by \eqref{2.4} with
$\mu_\rho^{(\wedge)}=\mu_\rho^{(\vee)}=\pi/2$. We obtain
\begin{multline}\label{c4}
j_\varphi(r)=\frac{m}{(2\pi)^2}\left\{ \int\limits_0^\infty
\frac{du}{\cosh(u/2)}\left[1+\frac1{2mr \cosh(u/2)}\right]{\rm
e}^{-2mr
\cosh(u/2)}\right. \\
\times \frac{\cos\left[\nu\left(F-\frac12\right)\pi\right] \cosh\left[\nu\left(F-\frac32 \right)u\right]
- \cos[\nu\left(F-\frac32 \right)\pi)]
\cosh\left[\nu\left(F-\frac12\right)u\right]}{\cosh(\nu
u)-\cos(\nu \pi)}\\
+ \frac{2\pi}{\nu}\sum_{p=1}^{\left[\!\left| {\nu}/2
\right|\!\right]} \left[1 +\frac{1}{2mr\sin(p\pi/\nu)}\right]\,
\exp[-2mr\sin(p\pi/\nu)]
\,\frac{\sin[(2F-1)p\pi]}{\sin(p\pi/\nu)}\\
\left. - \frac{\pi}{2N} \left(-1\right)^{N}\sin\left(2NF \pi \right) \left(1 +\frac{1}{2mr}\right){\rm e}^{-2mr}  \, \delta_{\nu, \, 2N}\right\},
\end{multline}
\begin{multline}\label{c5}
B_{\rm I}(r)=\frac{\nu e}{2(2\pi)^2}\frac1r \left\{ \int\limits_0^\infty
\frac{du}{\cosh^2(u/2)}\,{\rm e}^{-2mr
\cosh(u/2)}\right. \\
\times \frac{\cos\left[\nu\left(F-\frac12\right)\pi\right] \cosh\left[\nu\left(F-\frac32 \right)u\right]
- \cos[\nu\left(F-\frac32 \right)\pi)]
\cosh\left[\nu\left(F-\frac12\right)u\right]}{\cosh(\nu
u)-\cos(\nu \pi)}\\
\left.+ \frac{2\pi}{\nu}\sum_{p=1}^{\left[\!\left| {\nu}/2
\right|\!\right]} \exp[-2mr\sin(p\pi/\nu)]
\,\frac{\sin[(2F-1)p\pi]}{\sin^2(p\pi/\nu)} - \frac{\pi}{2N} \left(-1\right)^{N}\sin\left(2NF \pi \right) {\rm e}^{-2mr}
\, \delta_{\nu, \, 2N}\right\},
\end{multline}
and
\begin{equation}\label{c6}
\Phi_{\rm I}=\frac{e}{6m} \left(1-F\right)\left[\frac14
\left(\nu^2 + 3\right) -\nu^2 \left(1-F\right)^2\right].
\end{equation}

In the case of $\nu \geq 1$ and $\frac12\left(1-\frac1\nu\right) < F <
\frac12\left(1+\frac1\nu\right)$ $\quad$ $\left(0<G< 1 \right)$, as
well as in the case of $\frac12 \leq \nu< 1$ and
$\frac12\left(\frac1\nu-1\right) < F < \frac12\left(3-\frac1\nu \right)$
$\quad$ $\left(1-\nu <G<\nu\right)$, the deficiency index for a
partial hamiltonian with $n=n_c$ equals (1,1), and the
one-parametric family of self-adjoint extensions is introduced via
condition \eqref{2.11}. The modes corresponding to the continuous
spectrum $(|E| > m)$ are given by \eqref{2.6} with
$\mu_\rho^{(\wedge)}=\mu_\rho^{(\vee)}=\pi/2 $ and \eqref{2.5} with
$\mu_{1-G}$ determined from relation
\begin{equation}\label{c7}
\tan(\mu_{1-G} )={\rm sgn}(E) \frac{(1-m/E)^G
\Gamma(1-G)}{(1+m/E)^{1-G} \Gamma(G)} \, 2^{1-2G} \,
\tan\left(\frac{\Theta}{2}+\frac\pi4 \right).
\end{equation}
In addition, there is a bound state at $\cos\Theta<0$ with the mode
given by \eqref{2.15} and the energy $(|E_{BS}|<m) $ determined from
relation \eqref{2.16}. We obtain
\begin{multline}\label{c8}
j_\varphi(r)=-\frac{m}{(2\pi)^2}\left\{ \int\limits_0^\infty
\frac{du}{\cosh(u/2)}\left[1+\frac1{2mr \cosh(u/2)}\right]{\rm
e}^{-2mr
\cosh(u/2)}\right. \\
\times \frac{\cos\left[\nu\left(F-\frac12\right)\pi\right]\sinh(\nu u)
\sinh\left[\nu\left(F-\frac12\right)u\right] + \sin\left[\nu\left(F-\frac12\right)\pi\right]\sin(\nu
\pi)\cosh\left[\nu\left(F-\frac12\right)u\right]}{\cosh(\nu u)-\cos(\nu
\pi)} \\
- \frac{2\pi}{\nu}\sum_{p=1}^{\left[\!\left| {\nu}/2 \right|\!\right]} \left[1 +\frac{1}{2mr\sin(p\pi/\nu)}\right]\,
\exp[-2mr\sin(p\pi/\nu)] \,\frac{\sin[(2F-1)p\pi]}{\sin(p\pi/\nu)}\\
\left. + \frac{\pi}{2N} \left(-1\right)^{N}\sin\left(2NF \pi \right) \left(1 +\frac{1}{2mr}\right){\rm e}^{-2mr}  \, \delta_{\nu, \, 2N}\right\} \\
-\frac{r}{\pi^2}\int\limits_m^\infty \frac{dq\, q^2}{\sqrt{q^2-m^2}
}\,\, C\!\!\left(\frac{q}m\right)K_{\frac12+\nu\left(F-\frac12\right)}(qr) K_{\frac12-\nu\left(F-\frac12\right)}(qr),
\end{multline}
\begin{multline}\label{c9}
B_{\rm I}(r)=-\frac{\nu e}{2(2\pi)^2}\frac1r \left\{ \int\limits_0^\infty
\frac{du}{\cosh^2 (u/2)}\,{\rm e}^{-2mr
\cosh(u/2)}\right. \\
\times \frac{\cos\left[\nu\left(F-\frac12\right)\pi\right]\sinh(\nu u)
\sinh\left[\nu\left(F-\frac12\right)u\right] + \sin\left[\nu\left(F-\frac12\right)\pi\right]\sin(\nu
\pi)\cosh\left[\nu\left(F-\frac12\right)u\right]}{\cosh(\nu u)-\cos(\nu
\pi)}\\
\left.- \frac{2\pi}{\nu}\sum_{p=1}^{\left[\!\left| {\nu}/2 \right|\!\right]}
\exp[-2mr\sin(p\pi/\nu)] \,\frac{\sin[(2F-1)p\pi]}{\sin^2(p\pi/\nu)} + \frac{\pi}{2N} \left(-1\right)^{N}\sin\left(2NF \pi \right) {\rm e}^{-2mr}
\, \delta_{\nu, \, 2N}\right\}\\
-\frac{\nu e}{\pi^2}\int\limits_{r}^\infty dr' \int\limits_m^\infty
\frac{dq\, q^2}{\sqrt{q^2-m^2} }\,\,
C\!\!\left(\frac{q}m\right)K_{\frac12+\nu\left(F-\frac12\right)}(qr') K_{\frac12-\nu\left(F-\frac12\right)}(qr'),
\end{multline}
and
\begin{multline}\label{c10}
\Phi_{\rm I}=\frac{e}{2m}\left\{\frac16 \left(\nu^2-1\right)
\left(F-\frac12\right)\right.\\
\left.-\left[\frac1\pi \int\limits_1^\infty
\frac{dt}{t\sqrt{t^2-1}}\,C(t)+\frac13\left(F-\frac12\right)\right]\left[\frac14-\nu^2\left(F-\frac12\right)^2\right]
\right\},
\end{multline}
where
\begin{multline}\label{c11}
C(t)= \frac{1}{\pi}\cos\left[\nu\left(F-\frac12\right)\pi\right] \\
\times \frac{\left(\frac{t}2\right)^{\nu\left(2F-1\right)}
\frac{\Gamma\left[\frac12-\nu\left(F-\frac12\right)\right]}{\Gamma\left[\frac12+\nu\left(F-\frac12\right)\right]}
\tan\left(\frac{\Theta}{2}+\frac\pi4 \right) - \left(\frac{t}2\right)^{-\nu\left(2F-1\right)}
\frac{\Gamma\left[\frac12+\nu\left(F-\frac12\right)\right]}{\Gamma\left[\frac12-\nu\left(F-\frac12\right)\right]}
\cot\left(\frac{\Theta}{2}+\frac\pi4 \right) }{\left(\frac{t}2\right)^{\nu\left(2F-1\right)}
\frac{\Gamma\left[\frac12-\nu\left(F-\frac12\right)\right]}{\Gamma\left[\frac12+\nu\left(F-\frac12\right)\right]}
\tan\left(\frac{\Theta}{2}+\frac\pi4 \right) + \frac2{t} +
\left(\frac{t}2\right)^{-\nu\left(2F-1\right)} \frac{\Gamma\left[\frac12+\nu\left(F-\frac12\right)\right]}{\Gamma\left[\frac12-\nu\left(F-\frac12\right)\right]}
\cot\left(\frac{\Theta}{2}+\frac\pi4 \right) }.
\end{multline}
Under the condition of minimal irregularity, see \eqref{3.18}, we
obtain
\begin{multline}\label{c12}
\left.j_\varphi(r)\right|_{F\neq
\frac12}=\frac{m}{(2\pi)^2}\,\left\{ {\rm
sgn}\!\!\left(\!F-\frac12\right)\int\limits_0^\infty
\frac{du}{\cosh(u/2)}\left[1+\frac1{2mr \cosh(u/2)}\right]{\rm
e}^{-2mr
\cosh(u/2)}\right. \\
\times \frac{\cos\left[\nu\left(F-\frac12\right)\pi\right]
\cosh\left[\nu\left(\left|F-\frac12\right|-1\right)u\right]-
\cos\left[\nu\left(\left|F-\frac12\right|-1\right)\pi\right]\cosh\left[\nu\left(F-\frac12\right)u\right]}{\cosh(\nu
u)-\cos(\nu \pi)} \\
+ \frac{2\pi}{\nu}\sum_{p=1}^{\left[\!\left| {\nu}/2
\right|\!\right]} \left[1 +\frac{1}{2mr\sin(p\pi/\nu)}\right]\,
\exp[-2mr\sin(p\pi/\nu)]
\,\frac{\sin[(2F-1)p\pi]}{\sin(p\pi/\nu)}\\
\left. - \frac{\pi}{2N} \left(-1\right)^{N}\sin\left(2NF \pi \right) \left(1 +\frac{1}{2mr}\right){\rm e}^{-2mr} \, \delta_{\nu, \, 2N}\right\},
\end{multline}
\begin{equation}\label{c13}
\left.j_\varphi(r)\right|_{F= \frac12}= -
\frac{\sin\theta}{2\pi^2}\int_m^\infty \frac{dq\,q^2}{\sqrt{q^2-m^2}
}\, \frac{{\rm e}^{-2qr}}{q+m\cos\theta},
\end{equation}
\begin{multline}\label{c14}
\left.B_{\rm I}(r)\right|_{F\neq\frac12}=\frac{\nu
e}{2(2\pi)^2}\frac1r\,\left\{ {\rm sgn}\!\!\left(\!F-\frac12\right)
\int\limits_0^\infty \frac{du}{\cosh^2 (u/2)}\,{\rm e}^{-2mr
\cosh(u/2)}\right. \\
\times \frac{\cos\left[\nu\left(F-\frac12\right)\pi\right]
\cosh\left[\nu\left(\left|F-\frac12\right|-1\right)u\right]-
\cos\left[\nu\left(\left|F-\frac12\right|-1\right)\pi\right]\cosh\left[\nu\left(F-\frac12\right)u\right]}{\cosh(\nu
u)-\cos(\nu \pi)}\\
\left.+ \frac{2\pi}{\nu}\sum_{p=1}^{\left[\!\left| {\nu}/2
\right|\!\right]} \exp[-2mr\sin(p\pi/\nu)]
\,\frac{\sin[(2F-1)p\pi]}{\sin^2(p\pi/\nu)} - \frac{\pi}{2N} \left(-1\right)^{N}\sin\left(2NF \pi \right) {\rm e}^{-2mr}
\, \delta_{\nu, \, 2N}\right\},
\end{multline}
\begin{equation}\label{c15}
\left.B_{\rm I}(r)\right|_{F = \frac12}= -\frac{\nu e
\sin\theta}{2\pi^2}\int_m^\infty
\frac{dq\,q^2}{\sqrt{q^2-m^2}}\frac{\Gamma(0,2qr)}{q+m \cos\theta},
\end{equation}
\begin{equation}\label{c16}
\left.\Phi_{\rm I}\right|_{F\neq \frac12}= - \frac{e}{6m}\,
\left[F-\frac12 - \frac12{\rm
sgn}\!\!\left(\!F-\frac12\right)\right]\Biggl\{\frac34
-\nu^2 \left[\frac14 - \left|F -\frac12\right| - F(1-F)\right] \Biggr\},
\end{equation}
and
\begin{equation}\label{c17}
\left.\Phi_{\rm I}\right|_{F= \frac12}= -\frac{e}{4\pi
m}\arctan\left(\tan\frac\theta2\right).
\end{equation}


\setcounter{equation}{0}
\renewcommand{\theequation}{D.\arabic{equation}}
\section*{Appendix D: case of massless quantum spinor matter}

We present here the results for the case of massless quantum
spinor matter in the background of the ANO vortex of nonzero transverse size.

In the case of  $\nu \geq 1$ and $\frac12\left(1-\frac1\nu\right) < F <
\frac12\left(1+\frac1\nu\right)$ or $\frac12 \leq \nu< 1$ and
$\frac12\left(\frac1\nu-1\right) < F < \frac12\left(3-\frac1\nu
\right)$, we obtain
\begin{multline}\label{d1}
\left.j_\varphi(r)\right|_{F<\frac12,\theta\neq
-\frac\pi2}\\
=\frac{1}{2(2\pi)^2}\frac1r
\left\{\frac{2\pi}\nu\sum_{p=1}^{\left[\!\left| {\nu}/2
\right|\!\right]} \frac{\sin[(2F-1)p\pi]}{\sin^2(p\pi/\nu)}-
\frac{\pi}{2N} (-1)^{N}\sin(2N F \pi)
\, \delta_{\nu, \, 2N} -
\int\limits_0^\infty \frac{du}{\cosh^2(u/2)} \right.\\ \left. \times
\frac{\cos\left[\nu\left(F-\frac12\right)\pi\right]
\cosh\left[\nu\left(F+\frac12 \right)u\right] -
\cos[\nu\left(F+\frac12 \right)\pi)]
\cosh\left[\nu\left(F-\frac12\right)u\right]}{\cosh(\nu u)-\cos(\nu
\pi)}\right\}\\
-\frac{r}{\pi^2}\int\limits_0^\infty dq\,q\left[\sum_{l=0}^\infty
\left.C^{(\wedge)}_{\nu\left(l-F+\frac12\right)+\frac12}(qr_0)\right|_{m=0}K_{\nu\left(l-F+\frac12\right)+\frac12}(qr)
K_{\nu\left(l-F+\frac12\right)-\frac12}(qr)\right.\\
-\left.\sum_{l=1}^\infty
\left.C^{(\vee)}_{\nu\left(l+F-\frac12\right)+\frac12}(qr_0)\right|_{m=0}K_{\nu\left(l+F-\frac12\right)+\frac12}(qr)
K_{\nu\left(l+F-\frac12\right)-\frac12}(qr)\right],
\end{multline}
\begin{multline}\label{d2}
\left.j_\varphi(r)\right|_{F>\frac12,\theta\neq
\frac\pi2}\\
=\frac{1}{2(2\pi)^2}\frac1r
\left\{\frac{2\pi}\nu\sum_{p=1}^{\left[\!\left| {\nu}/2
\right|\!\right]} \frac{\sin[(2F-1)p\pi]}{\sin^2(p\pi/\nu)}- \frac{\pi}{2N} (-1)^{N}\sin(2N F \pi)
\, \delta_{\nu, \, 2N} +
\int\limits_0^\infty \frac{du}{\cosh^2(u/2)} \right.\\ \left. \times
\frac{\cos\left[\nu\left(F-\frac12\right)\pi\right]
\cosh\left[\nu\left(F-\frac32 \right)u\right] -
\cos[\nu\left(F-\frac32 \right)\pi)]
\cosh\left[\nu\left(F-\frac12\right)u\right]}{\cosh(\nu u)-\cos(\nu
\pi)}\right\}\\
-\frac{r}{\pi^2}\int\limits_0^\infty dq\,q\left[\sum_{l=1}^\infty
\left.C^{(\wedge)}_{\nu\left(l-F+\frac12\right)+\frac12}(qr_0)\right|_{m=0}K_{\nu\left(l-F+\frac12\right)+\frac12}(qr)
K_{\nu\left(l-F+\frac12\right)-\frac12}(qr)\right.\\
-\left.\sum_{l=0}^\infty
\left.C^{(\vee)}_{\nu\left(l+F-\frac12\right)+\frac12}(qr_0)\right|_{m=0}K_{\nu\left(l+F-\frac12\right)+\frac12}(qr)
K_{\nu\left(l+F-\frac12\right)-\frac12}(qr)\right],
\end{multline}
\begin{multline}\label{d3}
\left.j_\varphi(r)\right|_{F\neq\frac12,\theta=\pm
\frac\pi2}\\
=\frac{1}{2(2\pi)^2}\frac1r
\left\{\frac{2\pi}\nu\sum_{p=1}^{\left[\!\left| {\nu}/2
\right|\!\right]} \frac{\sin[(2F-1)p\pi]}{\sin^2(p\pi/\nu)}- \frac{\pi}{2N} (-1)^{N}\sin(2N F \pi)
\, \delta_{\nu, \, 2N} \mp
\int\limits_0^\infty \frac{du}{\cosh^2(u/2)} \right.\\ \left. \times
\frac{\cos\left[\nu\left(F-\frac12\right)\pi\right]
\cosh\left[\nu\left(F-\frac12\pm1 \right)u\right] -
\cos[\nu\left(F-\frac12\pm1 \right)\pi)]
\cosh\left[\nu\left(F-\frac12\right)u\right]}{\cosh(\nu u)-\cos(\nu
\pi)}\right\}\\
\mp\frac{r}{\pi^2}\int\limits_0^\infty
dq\,q\left\{\frac{I_{\frac12\mp\nu\left(F-\frac12\right)}(qr_0)}{K_{\frac12\mp\nu\left(F-\frac12\right)}(qr_0)}
K_{\frac12+\nu\left(F-\frac12\right)}(qr)
K_{\frac12-\nu\left(F-\frac12\right)}(qr)\right.\\
+\sum_{l=1}^\infty\left[
\frac{I_{\nu\left(l-F+\frac12\right)\pm\frac12}(qr_0)}{K_{\nu\left(l-F+\frac12\right)\pm\frac12}(qr_0)}
K_{\nu\left(l-F+\frac12\right)+\frac12}(qr)
K_{\nu\left(l-F+\frac12\right)-\frac12}(qr)\right.\\
\left.\left.+\frac{I_{\nu\left(l+F-\frac12\right)\mp\frac12}(qr_0)}{K_{\nu\left(l+F-\frac12\right)\mp\frac12}(qr_0)}
K_{\nu\left(l+F-\frac12\right)+\frac12}(qr)
K_{\nu\left(l+F-\frac12\right)-\frac12}(qr)\right]\right\},
\end{multline}
and
\begin{equation}\label{d4}
\left.j_\varphi(r)\right|_{F=\frac12}=-\frac{\sin\theta}{(2\pi)^2}\left[\frac1{r-r_0}+8r\int\limits_0^\infty
dq\,q \sum_{l=1}^\infty \left.\tilde C_{\nu
l+\frac12}(qr_0)\right|_{m=0} K_{\nu l+\frac12}(qr)K_{\nu
l-\frac12}(qr)\right].
\end{equation}
It should be noted that the current is invariant under transformation $\theta\rightarrow \pi - \theta$. Thus the current is continuous in $\theta$, and its values at $\theta=0$ and $\theta=\pi$ coincide, in particular,
\begin{equation}\label{d5}
\left.j_\varphi(r)\right|_{F=\frac12, \, \theta=0}=\left.j_\varphi(r)\right|_{F=\frac12, \, \theta=\pi}=0.
\end{equation}
Since a piece of $j_\varphi(r)$ is proportional to $r^{-1}$, the corresponding piece of $B_{\rm I}(r)$ is also proportional to $r^{-1}$. Consequently, flux $\Phi_{\rm I}$, see \eqref{1.21}, is given by an integral that is linearly divergent at $r \rightarrow \infty$. Therefore, we have no choice but to introduce cutoff  $r_{\rm max} > r$ and the restricted flux,
\begin{equation}\label{d6}
\Phi_{{\rm I} (r_{\rm max})}=\frac{2\pi}{\nu} \int\limits_{r_0}^{r_{\rm max}} dr\,r
B_{\rm I}(r),
\end{equation}
where, as a consequence of \eqref{d1}-\eqref{d4},
\begin{multline}\label{d7}
\left.B_{\rm I}(r)\right|_{F<\frac12,\theta\neq
-\frac\pi2}
=\frac{\nu e}{2(2\pi)^2}\left(\frac1r
- \frac1r_{\rm max}\right)\left\{\frac{2\pi}\nu\sum_{p=1}^{\left[\!\left| {\nu}/2
\right|\!\right]} \frac{\sin[(2F-1)p\pi]}{\sin^2(p\pi/\nu)} \right. \\
- \frac{\pi}{2N} (-1)^{N}\sin(2N F \pi)
\, \delta_{\nu, \, 2N} -
\int\limits_0^\infty \frac{du}{\cosh^2(u/2)} \\ \left. \times
\frac{\cos\left[\nu\left(F-\frac12\right)\pi\right]
\cosh\left[\nu\left(F+\frac12 \right)u\right] -
\cos[\nu\left(F+\frac12 \right)\pi)]
\cosh\left[\nu\left(F-\frac12\right)u\right]}{\cosh(\nu u)-\cos(\nu
\pi)}\right\}\\
-\frac{\nu e}{\pi^2}\int\limits_{r}^{r_{\rm max}} dr'\int\limits_0^\infty
dq\,q\left[\sum_{l=0}^\infty
\left.C^{(\wedge)}_{\nu\left(l-F+\frac12\right)+\frac12}(qr_0)\right|_{m=0}K_{\nu\left(l-F+\frac12\right)+\frac12}(qr')
K_{\nu\left(l-F+\frac12\right)-\frac12}(qr')\right.\\
-\left.\sum_{l=1}^\infty
\left.C^{(\vee)}_{\nu\left(l+F-\frac12\right)+\frac12}(qr_0)\right|_{m=0}K_{\nu\left(l+F-\frac12\right)+\frac12}(qr')
K_{\nu\left(l+F-\frac12\right)-\frac12}(qr')\right],
\end{multline}
\begin{multline}\label{d8}
\left.B_{\rm I}(r)\right|_{F>\frac12,\theta\neq \frac\pi2}
=\frac{\nu
e}{2(2\pi)^2}\left(\frac1r
- \frac1r_{\rm max}\right) \left\{\frac{2\pi}\nu\sum_{p=1}^{\left[\!\left|
{\nu}/2 \right|\!\right]} \frac{\sin[(2F-1)p\pi]}{\sin^2(p\pi/\nu)}
\right. \\
- \frac{\pi}{2N} (-1)^{N}\sin(2N F \pi)
\, \delta_{\nu, \, 2N}+
\int\limits_0^\infty \frac{du}{\cosh^2(u/2)} \\ \left. \times
\frac{\cos\left[\nu\left(F-\frac12\right)\pi\right]
\cosh\left[\nu\left(F-\frac32 \right)u\right] -
\cos[\nu\left(F-\frac32 \right)\pi)]
\cosh\left[\nu\left(F-\frac12\right)u\right]}{\cosh(\nu u)-\cos(\nu
\pi)}\right\}\\
-\frac{\nu e}{\pi^2}\int\limits_{r}^{r_{\rm max}} dr'\int\limits_0^\infty
dq\,q\left[\sum_{l=1}^\infty
\left.C^{(\wedge)}_{\nu\left(l-F+\frac12\right)+\frac12}(qr_0)\right|_{m=0}K_{\nu\left(l-F+\frac12\right)+\frac12}(qr')
K_{\nu\left(l-F+\frac12\right)-\frac12}(qr')\right.\\
-\left.\sum_{l=0}^\infty
\left.C^{(\vee)}_{\nu\left(l+F-\frac12\right)+\frac12}(qr_0)\right|_{m=0}K_{\nu\left(l+F-\frac12\right)+\frac12}(qr')
K_{\nu\left(l+F-\frac12\right)-\frac12}(qr')\right],
\end{multline}
\begin{multline}\label{d9}
\left.B_{\rm I}(r)\right|_{F\neq\frac12,\theta=\pm
\frac\pi2}
=\frac{\nu e}{2(2\pi)^2}\left(\frac1r
- \frac1r_{\rm max}\right)
\left\{\frac{2\pi}\nu\sum_{p=1}^{\left[\!\left| {\nu}/2
\right|\!\right]} \frac{\sin[(2F-1)p\pi]}{\sin^2(p\pi/\nu)} \right. \\
 - \frac{\pi}{2N} (-1)^{N}\sin(2N F \pi)
\, \delta_{\nu, \, 2N} \mp
\int\limits_0^\infty \frac{du}{\cosh^2(u/2)} \\ \left. \times
\frac{\cos\left[\nu\left(F-\frac12\right)\pi\right]
\cosh\left[\nu\left(F-\frac12\pm1 \right)u\right] -
\cos[\nu\left(F-\frac12\pm1 \right)\pi)]
\cosh\left[\nu\left(F-\frac12\right)u\right]}{\cosh(\nu u)-\cos(\nu
\pi)}\right\}\\
\mp\frac{\nu e}{\pi^2}\int\limits_{r}^{r_{\rm max}} dr'\int\limits_0^\infty
dq\,q\left\{\frac{I_{\frac12\mp\nu\left(F-\frac12\right)}(qr_0)}{K_{\frac12\mp\nu\left(F-\frac12\right)}(qr_0)}
K_{\frac12+\nu\left(F-\frac12\right)}(qr')
K_{\frac12-\nu\left(F-\frac12\right)}(qr')\right.\\
+\sum_{l=1}^\infty\left[
\frac{I_{\nu\left(l-F+\frac12\right)\pm\frac12}(qr_0)}{K_{\nu\left(l-F+\frac12\right)\pm\frac12}(qr_0)}
K_{\nu\left(l-F+\frac12\right)+\frac12}(qr')
K_{\nu\left(l-F+\frac12\right)-\frac12}(qr')\right.\\
\left.\left.+\frac{I_{\nu\left(l+F-\frac12\right)\mp\frac12}(qr_0)}{K_{\nu\left(l+F-\frac12\right)\mp\frac12}(qr_0)}
K_{\nu\left(l+F-\frac12\right)+\frac12}(qr')
K_{\nu\left(l+F-\frac12\right)-\frac12}(qr')\right]\right\},
\end{multline}
and
\begin{multline}\label{d10}
\left.B_{\rm I}(r)\right|_{F=\frac12}=\frac{\nu e \sin\theta}{(2\pi)^2}\!\left[\!\frac1{r_0}\ln\left(1\!-\!\frac{r_0}{r}\right) - \frac1{r_0}\ln\left(1\!-\!\frac{r_0}{r_{\rm max}}\right)\right. \\
\left. - 8\!\int\limits_{r}^{r_{\rm max}}\!
dr'\! \int\limits_0^\infty\! dq\,q \sum_{l=1}^\infty \left.\tilde
C_{\nu l+\frac12}(qr_0)\right|_{m=0}\! K_{\nu l+\frac12}(qr')K_{\nu
l-\frac12}(qr')\right]\!.
\end{multline}

In the case of $\nu>1$ and $0<F< \frac12(1-\nu)$, $j_\varphi(r)$ is
given by the right-hand side of \eqref{d1} and $B_{\rm I}(r)$ is
given by the right-hand side of \eqref{d7}. In the case of $\nu>1$
and $\frac12(1+\nu)<F< 1$, $j_\varphi(r)$ is given by the right-hand
side of \eqref{d2} and $B_{\rm I}(r)$ is given by the right-hand side of \eqref{d8}.


Turning now to flux $\Phi_{{\rm I} (r_{\rm max})}$ \eqref{d6}, we
numerically calculate quantity $$\omega(\theta)=\lim_{r \rightarrow
r_0} \nu r j_\varphi (r) \left(\frac{r-r_0}{r_0}\right)^2 $$ and
compare it in Fig. 10 with the appropriate quantity in the case of
the massive spinor field (see Section 6). Note that $\omega(\theta)$ in
the massless case is strictly symmetric with respect to point
$\theta=\pi/2$ (location of the inverted peak in Fig. 10), whereas
$\omega(\theta)$ in the massive case is not symmetric, although this
asymmetry is so slight that it is not visible in Fig. 1 and Fig. 10. Note
also that coefficients
$\left.C^{(\wedge)}_{\rho}(qr_0)\right|_{\theta=\pm \pi/2}$,
$\left.C^{(\vee)}_{\rho}(qr_0)\right|_{\theta=\pm \pi/2}$, and
$\left.\tilde C_{\rho}(qr_0)\right|_{\theta=\pm \pi/2}$ in the
massive case coincide with those in the massless case, and the
differences in the values of $\omega(\theta)$ are due to different
measures of integration in the massive and massless cases. As
follows from the behavior of $\omega(\theta)$, the induced vacuum
magnetic flux, either $\Phi_{{\rm I}}$ \eqref{1.21} in the massive
case or $\Phi_{{\rm I} (r_{\rm max})}$ \eqref{d6} in the massless
case, is finite at $\theta=0$ and $\theta=\pi$ only, with coinciding
values at these points in the latter case. Thus we obtain
\begin{equation}\label{d11}
 \left.\Phi_{{\rm I} (r_{\rm max})}\right|_{\theta = \frac{\pi}{2} \mp \frac{\pi}{2}}
= 0, \quad F = 1/2
\end{equation}

\begin{figure}[t]
\begin{center}
\includegraphics[width=95mm]{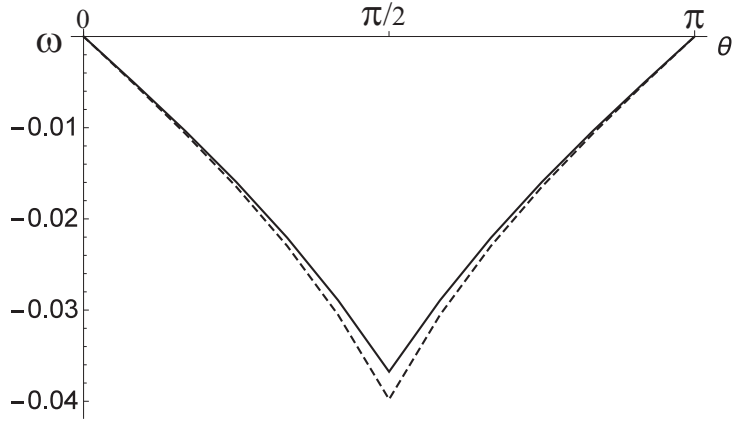}
\end{center} \caption{$\omega(\theta)$ in the case of the
massless spinor field (solid line) and in the case of the massive spinor
field (dashed line).}\label{Triangle2}
\end{figure}

\noindent and
\begin{multline}\label{d12}
\left.\Phi_{{\rm I} (r_{\rm max})}\right|_{\theta = \frac{\pi}{2} \mp \frac{\pi}{2}}
=e\,\frac{\left(r_{\rm max}-r_0\right)^2}{r_{\rm max}}
\left\{\frac{1}{4\nu}\sum_{p=1}^{\left[\!\left| {\nu}/2
\right|\!\right]} \frac{\sin[(2F-1)p\pi]}{\sin^2(p\pi/\nu)} \right.\\
- \frac{1}{16N} (-1)^{N}\sin(2N F \pi)
\, \delta_{\nu, \, 2N} + {\rm sgn}\left(F-\frac12\right)
\frac{1}{8\pi}\int\limits_0^\infty \frac{du}{\cosh^2(u/2)} \\
\left. \times \frac{\cos\left[\nu\left(F-\frac12\right)\pi)\right]
\cosh\left[\nu\left(\left|F-\frac12\right|-1\right)u\right]-
\cos\left[\nu\left(\left|F-\frac12\right|-1\right)\pi\right]\cosh\left[\nu\left(F-\frac12\right)u\right]}{\cosh(\nu u)-\cos(\nu
\pi)}\right\}\\
+ \frac{e}{2\pi} r_0 \int\limits_{0}^\infty
 dv\, \Biggl\{\frac12\Biggl[C^{(0)}_{\frac12+\nu\left(F-\frac12\right)}(v)-C^{(0)}_{\frac12-\nu\left(F-\frac12\right)}(v)\Biggr.\Biggr. \\
\Biggl. +{\rm
sgn}\left(F-\frac12\right)\Biggl(C^{(0)}_{\frac12+\nu\left(F-\frac12\right)}(v)+C^{(0)}_{\frac12-\nu\left(F-\frac12\right)}(v)\Biggr)\Biggr]
D^{(0)}_{\frac12+\nu\left|F-\frac12\right|}
\left(v;\frac{r_0}{r_{\rm max}}\right)
\\
\Biggl. + \sum_{l=1}^\infty \Biggl[C^{(0)}_{\nu \left(l+F-\frac12\right)+\frac12}(v) D^{(0)}_{\nu
\left(l+F-\frac12\right)+\frac12}\left(v;\frac{r_0}{r_{\rm max}}\right)
 - C^{(0)}_{\nu \left(l-F+\frac12\right)+\frac12}(v)
D^{(0)}_{\nu \left(l-F+\frac12\right)+\frac12}\left(v;\frac{r_0}{r_{\rm max}}\right) \Biggr]\Biggr\}, \\ F \neq 1/2,
\end{multline}
where
\begin{equation}\label{d13}
C^{(0)}_\rho(v)=\left[
I_\rho(v)K_\rho(v)  -
I_{\rho-1}(v)K_{\rho-1}(v)
\right] \left[
K^2_\rho(v) + K^2_{\rho-1}(v)\right]^{-1}
\end{equation}
and
\begin{multline}\label{d14}
D^{(0)}_{\rho}(v;w) = \rho K_\rho^2(v)-(\rho-1)K_{\rho+1}(v)K_{\rho-1}(v)
+ v \left[ K_{\rho}(v)\frac{d}{d \rho} K_{\rho-1}(v) \right.\\
\left. -
K_{\rho-1}(v)\frac{d}{d \rho} K_{\rho}(v)\right] - w^{-2}\left[\rho K_\rho^2\left(\frac{v}{w}\right)-(\rho-1)K_{\rho+1}\left(\frac{v}{w}\right)K_{\rho-1}\left(\frac{v}{w}\right)\right]\\
- \frac{v}{w} \left[ K_{\rho}\left(\frac{v}{w}\right)\frac{d}{d \rho} K_{\rho-1}\left(\frac{v}{w}\right) -
K_{\rho-1}\left(\frac{v}{w}\right)\frac{d}{d \rho} K_{\rho}\left(\frac{v}{w}\right) \right].
\end{multline}
Retaining only the terms that are maximally divergent in the limit of
$r_{\rm max}\rightarrow \infty$, we get
\begin{multline}\label{d15}
\left.\Phi_{{\rm I} (r_{\rm max})}\right|_{\theta = \frac{\pi}{2} \mp \frac{\pi}{2}}
=e \, r_{\rm max}
\left\{\frac{1}{4\nu}\sum_{p=1}^{\left[\!\left| {\nu}/2
\right|\!\right]} \frac{\sin[(2F-1)p\pi]}{\sin^2(p\pi/\nu)} \right.\\
- \frac{1}{16N} (-1)^{N}\sin(2N F \pi)
\, \delta_{\nu, \, 2N} + {\rm sgn}\left(F-\frac12\right)
\frac{1}{8\pi}\int\limits_0^\infty \frac{du}{\cosh^2(u/2)} \\
\left. \times \frac{\cos\left[\nu\left(F-\frac12\right)\pi)\right]
\cosh\left[\nu\left(\left|F-\frac12\right|-1\right)u\right]-
\cos\left[\nu\left(\left|F-\frac12\right|-1\right)\pi\right]\cosh\left[\nu\left(F-\frac12\right)u\right]}{\cosh(\nu u)-\cos(\nu \pi)}\right\}\\
 + O\left(e \, r_0\right), \quad F \neq 1/2.
\end{multline}
Thus, we obtain the following relation between current
$\left.j_\varphi(r)\right|_{\theta = \frac{\pi}{2} \mp \frac{\pi}{2}}$ and magnetic field strength $\left.B_{\rm I}(r)\right|_{\theta = \frac{\pi}{2} \mp \frac{\pi}{2}}$ in the physically sensible case of $r_{\rm max} \gg r_0$:
\begin{equation}\label{d16}
\nu e \left.j_\varphi(r)\right|_{\theta = \frac{\pi}{2} \mp \frac{\pi}{2}}=\frac {r_{\rm max}}{r_{\rm max} - r} \left.B_{\rm I}(r)\right|_{\theta = \frac{\pi}{2} \mp \frac{\pi}{2}}=\frac{\nu}{\pi r_{\rm max} r}\left.\Phi_{{\rm I} (r_{\rm max})}\right|_{\theta = \frac{\pi}{2} \mp \frac{\pi}{2}}, \quad  r \gg r_0,
\end{equation}
where flux $\left.\Phi_{{\rm I} (r_{\rm max})}\right|_{\theta = \frac{\pi}{2} \mp \frac{\pi}{2}}$ is given by \eqref{d15}.

In particular, we get in the case of $\nu=1$
\begin{equation}\label{d17}
\left.\Phi_{{\rm I} (r_{\rm max})}\right|_{\nu=1,\,\theta=\frac{\pi}2\mp
\frac{\pi}{2}} \\
= \frac{e}4\, r_{\rm max} \tan(F\pi) \left|F-\frac12\right|
\left(\left|F-\frac12\right|-1\right)+O(er_0)
\end{equation}
and
\begin{multline}\label{d18}
\left. e j_\varphi(r)\right|_{\nu=1,\,\theta=\frac{\pi}2\mp \frac{\pi}{2}}
=\frac{r_{\rm max}}{r_{\rm max}-r}\,\left. B_{I}(r)\right|_{\nu=1,\,\theta=\frac{\pi}2\mp \frac{\pi}{2}}
= \frac{e}{4\pi r} \tan(F\pi) \left|F-\frac12\right|
\left(\left|F-\frac12\right|-1\right), \\
r\gg r_0.
\end{multline}
The last relation for the current was obtained in \cite{Si9e} in the $r_0 = 0$ case under the condition of minimal irregularity with requirements of the charge conjugation invariance and continuity in $\theta$
(see (10.6) in this reference where the definition of the current differs by an extra $r^{-1}$). Note a discontinuity at $F=1/2$, which is independent of $\nu$,
\begin{equation}\label{d19}
\lim_{F\rightarrow (1/2)_{\pm}}\left. e
j_\varphi(r)\right|_{\theta \neq \pm
\frac{\pi}{2}}=\pm\frac{e}{4\pi^2 r}, \quad r\gg r_0.
\end{equation}
This is distinct from the case of quantum scalar matter under the Dirichlet boundary condition, when the current that is induced in the vacuum by the infinitely thin vortex is continuous and vanishing at $F=1/2$ \cite{SiB1,SiB2,SiV9}, see the appropriate expression from these references at $m=0$ and $\nu=1$:
\begin{equation}\label{d20}
\left. e j_\varphi(r)\right|_{{\rm scalar},\,{\rm Dirichlet}}= - \frac{e}{4\pi r}\tan(F\pi) \left(F-\frac12\right)^2.
\end{equation}

\end{document}